\documentclass[twocolumn]{aastex63}
\shorttitle{Quantifying Feedback from Narrow Line Region Outflows}
\shortauthors{Revalski et al.}

\usepackage{float}
\usepackage{color}
\usepackage{comment}
\usepackage{graphics}
\usepackage{epstopdf}
\usepackage{microtype}
\usepackage{subfigure}
\usepackage[normalem]{ulem}
\usepackage{natbib}
\graphicspath{{./}{figures/}}
\newcommand{\lbol}{\ensuremath{L_{bol}}}
\newcommand{\mbh}{\ensuremath{M_{BH}}}
\newcommand{\ledd}{\ensuremath{\lbol/L_{Edd}}}
\newcommand{\othree}{[O~III]~}
\newcommand{\hst}{{\it HST}~}

\newcommand{\kms}{~km~s$^{-1}$~}
\newcommand{\zsun}{\ensuremath{~Z_{\odot}~}}
\definecolor{malachite}{rgb}{0.04, 0.85, 0.32}

\usepackage{soul,xcolor}
\setstcolor{red}

\begin{document}

\title{\vspace{-1em}Quantifying Feedback from Narrow Line Region Outflows in Nearby Active Galaxies. III.\\Results for the Seyfert 2 Galaxies Markarian~3, Markarian~78, and NGC~1068\footnote{Based on observations made with the NASA/ESA Hubble Space Telescope, obtained from the Data Archive at the Space Telescope Science Institute, which is operated by the Association of Universities for Research in Astronomy, Inc., under NASA contract NAS 5-26555. These observations are associated with program No. 5140, 5754, 7404, 7573, and 8480.}\footnote{Based in part on observations obtained with the Apache Point Observatory 3.5-meter telescope, which is owned and operated by the Astrophysical Research Consortium.}}








\correspondingauthor{Mitchell Revalski}
\email{mrevalski@stsci.edu}

\author[0000-0002-4917-7873]{Mitchell Revalski}
\affiliation{Space Telescope Science Institute, 3700 San Martin Drive, Baltimore, MD 21218, USA}

\author[0000-0001-8658-2723]{Beena Meena}
\affiliation{Department of Physics and Astronomy, Georgia State University, 25 Park Place, Suite 605, Atlanta, GA 30303, USA}

\author[0000-0001-5099-8700]{Francisco Martinez}
\affiliation{Center for Relativistic Astrophysics, School of Physics, Georgia Institute of Technology,
Atlanta, GA 30332, USA}

\author[0000-0001-5862-2150]{Garrett E. Polack}
\affiliation{Department of Physics and Astronomy, Georgia State University, 25 Park Place, Suite 605, Atlanta, GA 30303, USA}

\author[0000-0002-6465-3639]{D. Michael Crenshaw}
\affiliation{Department of Physics and Astronomy, Georgia State University, 25 Park Place, Suite 605, Atlanta, GA 30303, USA}

\author[0000-0002-6928-9848]{Steven B. Kraemer}
\affiliation{Institute for Astrophysics and Computational Sciences, Department of Physics, The Catholic University of America, Washington, DC 20064, USA}

\author[0000-0002-8837-8803]{Nicholas R. Collins}
\affiliation{Telophase Corporation at NASA's Goddard Space Flight Center, Code 667, Greenbelt, MD 20771, USA}

\author[0000-0002-3365-8875]{Travis C. Fischer}
\affiliation{AURA for ESA, Space Telescope Science Institute, 3700 San Martin Drive, Baltimore, MD 21218, USA}

\author[0000-0003-2450-3246]{Henrique R. Schmitt}
\affiliation{Naval Research Laboratory, Washington, DC 20375, USA}

\author[0000-0002-2617-5517]{Judy Schmidt}
\affiliation{Designer/Developer for the Astrophysics Source Code Library}

\author[0000-0002-2203-7889]{W. Peter Maksym}
\affiliation{Harvard-Smithsonian Center for Astrophysics, 60 Garden Street, Cambridge, MA 02138, USA}

\author[0000-0002-9946-4731]{Marc Rafelski}
\affiliation{Space Telescope Science Institute, 3700 San Martin Drive, Baltimore, MD 21218, USA}
\affiliation{Department of Physics and Astronomy, Johns Hopkins University, Baltimore, MD 21218, USA}

\begin{abstract}
Outflows of ionized gas driven by active galactic nuclei (AGN) may significantly impact the evolution of their host galaxies. However, determining the energetics of these outflows is difficult with spatially unresolved observations that are subject to strong global selection effects. We present part of an ongoing study using {\it Hubble Space Telescope} ({\it HST}) and Apache Point Observatory (APO) spectroscopy and imaging to derive spatially-resolved mass outflow rates and energetics for narrow line region (NLR) outflows in nearby AGN that are based on multi-component photoionization models to account for spatial variations in the gas ionization, density, abundances, and dust content. This expanded analysis adds Mrk~3, Mrk~78, and NGC~1068, doubling the sample in \cite{Revalski2019}. We find that the outflows contain total ionized gas masses of $M \approx 10^{5.5} - 10^{7.5}$ $M_{\odot}$ and reach peak velocities of $v \approx 800 - 2000$ km s$^{-1}$. The outflows reach maximum mass outflow rates of $\dot M_{out} \approx 3 - 12$ $M_{\odot}$ yr$^{-1}$ and encompass total kinetic energies of $E \approx 10^{54} - 10^{56}$ erg. The outflows extend to radial distances of $r \approx 0.1 - 3$ kpc from the nucleus, with the gas masses, outflow energetics, and radial extents positively correlated with AGN luminosity. The outflow rates are consistent with in situ ionization and acceleration where gas is radiatively driven at multiple radii. These radial variations indicate that spatially-resolved observations are essential for localizing AGN feedback and determining the most accurate outflow parameters.
\end{abstract}


\section{Introduction}

\subsection{Feedback from Outflows in Active Galaxies}

Outflows of ionized and molecular gas may play an important role in the coevolution of active galactic nuclei (AGN) and their host galaxies by regulating supermassive black hole (SMBH) accretion rates and evacuating reservoirs of potential star-forming gas from galaxy bulges \citep{Ciotti2001, Hopkins2005, Kormendy2013, Heckman2014, Fiore2017, Cresci2018, Harrison2018, Storchi-Bergmann2019, Laha2020, Veilleux2020}. Outflows are observed over a range of spatial scales, and those connecting the sub-parsec central engine to the kiloparsec scale galaxy environment can be found in the narrow-line region (NLR), which is composed of ionized gas $\sim$1 -- 1000+ parsecs (pcs) from the SMBH with densities of $n_\mathrm{H} \approx 10^2$ -- $10^6$~cm$^{-3}$ \citep{Peterson1997}. These outflows are of particular interest because they extend from the smallest scales that can be spatially-resolved in nearby galaxies (pcs from the SMBH) to bulge-galaxy scales where they may affect galactic evolution. We can determine whether or not NLR outflows are providing significant feedback to their host galaxies through quantifying their impact by measuring the outflowing mass ($M$) and velocity ($v$) over a spatial extent ($\delta r$). These parameters are then used to calculate properties including mass outflow rates ($\dot M = M v / \delta r$), kinetic energies ($ E = 1/2 M v^2$), kinetic energy flow rates ($\dot E = 1/2 \dot M v^2$), momenta ($p = Mv$), and momenta flow rates ($\dot p = \dot M v$).

Determining these quantities accurately for individual AGN has generally faced two obstacles. First, spatially unresolved observations only allow these properties to be determined globally, averaging over the spatial extent of the outflow to approximate the energetics with a single mass, velocity, and radial extent. Second, several methods for estimating the mass of the ionized gas that involve different assumptions yield a range of mass estimates for the same galaxies. While global techniques allow the mass to be quickly estimated for a large number of targets with available data, the underlying assumptions have not been critically examined for a large sample and in some cases are subject to systematics that overestimate the energetics by $\sim$1 -- 3 dex \citep{Karouzos2016, Bischetti2017, Venturi2020}. Recently, these systematic uncertainties are being better understood by utilizing spatially-resolved imaging and spectroscopy to map how the outflows change as a function of distance from the nucleus \citep{Durre2018, Durre2019, Venturi2018, Comeron2021, Garcia-Bernete2021, TrindadeFalcao2021}.

To tackle the second issue of determining accurate gas masses, we have developed a technique using multi-component photoionization models that match the emission line spectra, which tightly constrains the gas densities and allows us to calculate the mass of the ionized outflows with high precision \citep{Collins2009, Crenshaw2015, Revalski2019}. We have an ongoing program to quantify the energetic impact of spatially-resolved NLR outflows that was launched in an initial investigation by \cite{Crenshaw2015} focused on NGC~4151. This bright, prototypical Seyfert 1 galaxy displays outflow velocities up to $\sim$800\kms with an outflow gas mass of $3\times10^5$ M$_{\odot}$ that reaches a peak mass outflow rate of $\sim$3~M$_{\odot}$ yr$^{-1}$. This outflow rate is higher than the mass accretion rate onto the SMBH and the outflow rate seen for UV/X-ray absorbers at smaller radii \citep{Crenshaw2012}, indicating the potential importance of NLR outflows as a feedback mechanism.

We expanded this work to higher luminosities by conducting a similar analysis for Mrk~573 \citep{Revalski2018a} and Mrk~34 \citep{Revalski2018b}. These galaxies display more extended and energetic outflows than NGC~4151, highlighting the need for a systematic study of nearby AGN across a range of luminosity, SMBH mass, galaxy type, and environment (e.g. \citealp{Rojas2020, Yesuf2020}). In this study, we expand upon our earlier investigation with analyses of the nearby Seyfert galaxies Mrk~3, Mrk~78, and NGC~1068. These AGN were selected because they display clear signatures of outflow \citep{Fischer2013, Fischer2014} and have archival \hst spectroscopy and imaging required to implement our modeling technique. We complete the analysis for Mrk~78 in this paper, and draw modeling results from \cite{Ruiz2001, Collins2005, Collins2009} for Mrk~3 and \cite{Kraemer2000a, Kraemer2000b} for NGC~1068 to calculate the outflow energetics. We describe the observations (\S2), analysis (\S3), modeling (\S4), and calculations (\S5) for these new targets, as well as the results (\S6), discussion (\S7), and conclusions (\S8) for the entire sample.

\begin{deluxetable*}{lccccccccc}
\vspace{-0.5em}
\setlength{\tabcolsep}{0.085in} 
\tablecaption{Physical Properties of the Active Galaxy Sample}
\tablehead{
\colhead{Catalog} & \colhead{Redshift} & \colhead{Distance} &\colhead{Scale} & \colhead{Inclination} & \colhead{$\log$(\lbol)} & \colhead{$\log$(\mbh)} & \colhead{\ledd} & \colhead{References} & \colhead{Analysis \vspace{-0.5em}}\\
\colhead{Name} & \colhead{(21 cm)} & \colhead{(Mpc)} &\colhead{(pc/$\arcsec$)} & \colhead{(deg)} & \colhead{(erg s$^{-1}$)} & \colhead{($M_{\odot}$)} & \colhead{(unitless)} & \colhead{(Cols. 5,6,7)} & \colhead{Refs. \vspace{-0.5em}}\\
\colhead{(1)} & \colhead{(2)} & \colhead{(3)} &\colhead{(4)} & \colhead{(5)} & \colhead{(6)} & \colhead{(7)} & \colhead{(8)} & \colhead{(9)} & \colhead{(10)}
}
\startdata
NGC 4151 & 0.0033 & 13.3 & 67.4 & 20 & 43.9 & 7.6 & 0.01 & 1, 2, 3 & 12 \\
NGC 1068 & 0.0038 & 16.0 & 77.6 & 40 & 45.0 & 7.2 & 0.50 & 4, 5, 5 & $\star$ \\
Mrk 3    & 0.0135 & 56.6 & 274.5 & 64$^\alpha$ & 45.3 & 8.7 & 0.04 & 6, 6, 5 & $\star$ \\
Mrk 573   & 0.0172 & 72.0 & 349.1 & 38 & 45.5 & 7.3 & 0.75 & 7, 8, 5 & 8 \\
Mrk 78   & 0.0372 & 154.2 & 747.4 & 55 & 45.9 & 7.9 & 0.79 & 9, $\star$, 5 & $\star$ \\
Mrk 34     & 0.0505 & 207.9 & 1007.7 & 41 & 46.2 & 7.5 & 3.98 & 10, 10, 11 & 10 \\
\enddata
\tablecomments{Columns are (1) target name, (2) 21~cm redshift from the NASA/IPAC Extragalactic Database, (3) Hubble distance and (4) spatial scale assuming H$_0$ = 71 km s$^{-1}$ Mpc$^{-1}$, (5) host galaxy inclination, (6) bolometric luminosity estimated from [O~III] imaging, (7) black hole mass, and (8) the corresponding Eddington ratio (\ledd) calculated using $L_{Edd} = 1.26 \times 10^{38} \left(M/M_{\odot}\right)$ erg~s$^{-1}$. Column (9) gives the references for columns~5~--~7 and column (10) provides the reference to our mass outflow modeling. References are: (1) \citealp{Das2005} (2) \citealp{Crenshaw2012}, (3) \citealp{Bentz2006}, (4) \citealp{Das2006}, (5) \citealp{Woo2002}, (6) \citealp{Collins2009}, (7) \citealp{Fischer2017}, (8) \citealp{Revalski2018a}, (9) \citealp{Schmitt2000}, (10) \citealp{Revalski2018b}, (11) \citealp{Oh2011}, (12) \citealp{Crenshaw2015}, and ($\star$) this work. $^\alpha$The host galaxy inclination is 33$\degr$, but the outflows occupy a gas disk at an inclination of 64$\degr$ as described in \cite{Gnilka2020}.}
\label{sample}
\vspace{-2em}
\end{deluxetable*}

\subsection{Characteristics of the Sample}

Mrk~3, Mrk~78, and NGC~1068 have been the subject of multiwavelength investigations in the radio \citep{Ulvestad1984}, optical \citep{Fischer2013}, UV \citep{Ferland1986}, and X-ray \citep{Awaki1991}. As shown in Figure~\ref{structure}, Mrk~3 is a Seyfert~2 galaxy with an S0 classification and a backward S-shaped NLR that is produced by external fueling as cold gas from a nearby companion galaxy is ionized within the AGN radiation field (see \citealp{Gnilka2020} and references therein for an extensive review). Mrk~78 is a Seyfert~2 galaxy with an SB classification and intertwined radio and optical features obscured by a thick dust lane \citep{Adams1973, DeRobertis1987, Pedlar1989, Capetti1994, RamosAlmeida2006, Jackson2007, Fischer2011, Liu2017}. NGC~1068 is the nearest Seyfert~2 galaxy and has been studied extensively to understand the physical processes at work in AGN at the smallest resolvable spatial scales (e.g. \citealp{Antonucci1985, Pogge1993, Jaffe2004, Garcia-Burillo2014, May2014, Kraemer2015, May2017}). \hst color-composite images of these AGN are presented in Figure~\ref{structure}, and their physical properties are provided in Table~\ref{sample}.

\section{Observations}

\subsection{Hubble Space Telescope}

The archival {\it Hubble Space Telescope} spectroscopy and \othree imaging used in this study were obtained with the Space Telescope Imaging Spectrograph (STIS), Wide Field and Planetary Camera 2 (WFPC2), and the Faint Object Camera (FOC). We retrieved calibrated data (DOI:\dataset[10.17909/t9-4581-8p50]{\doi{10.17909/t9-4581-8p50}}) from the Mikulski Archive at the Space Telescope Science Institute (MAST) and combined multiple dithered spectroscopic exposures using the Interactive Data Language (IDL) and the Image Reduction and Analysis Facility (IRAF, \citealp{Tody1986, Tody1993}) for imaging.

\begin{figure*}[htb!]
\centering
\includegraphics[width=0.45\textwidth]{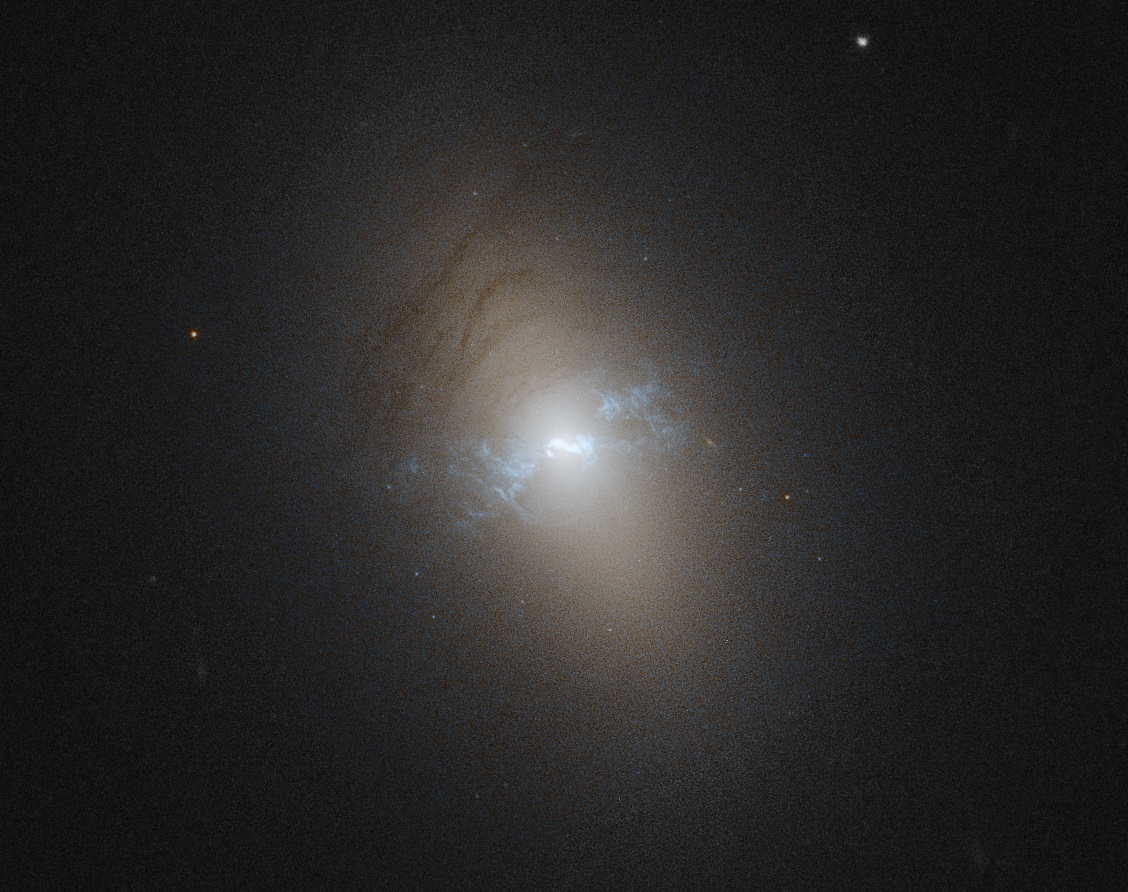}
\includegraphics[width=0.45\textwidth]{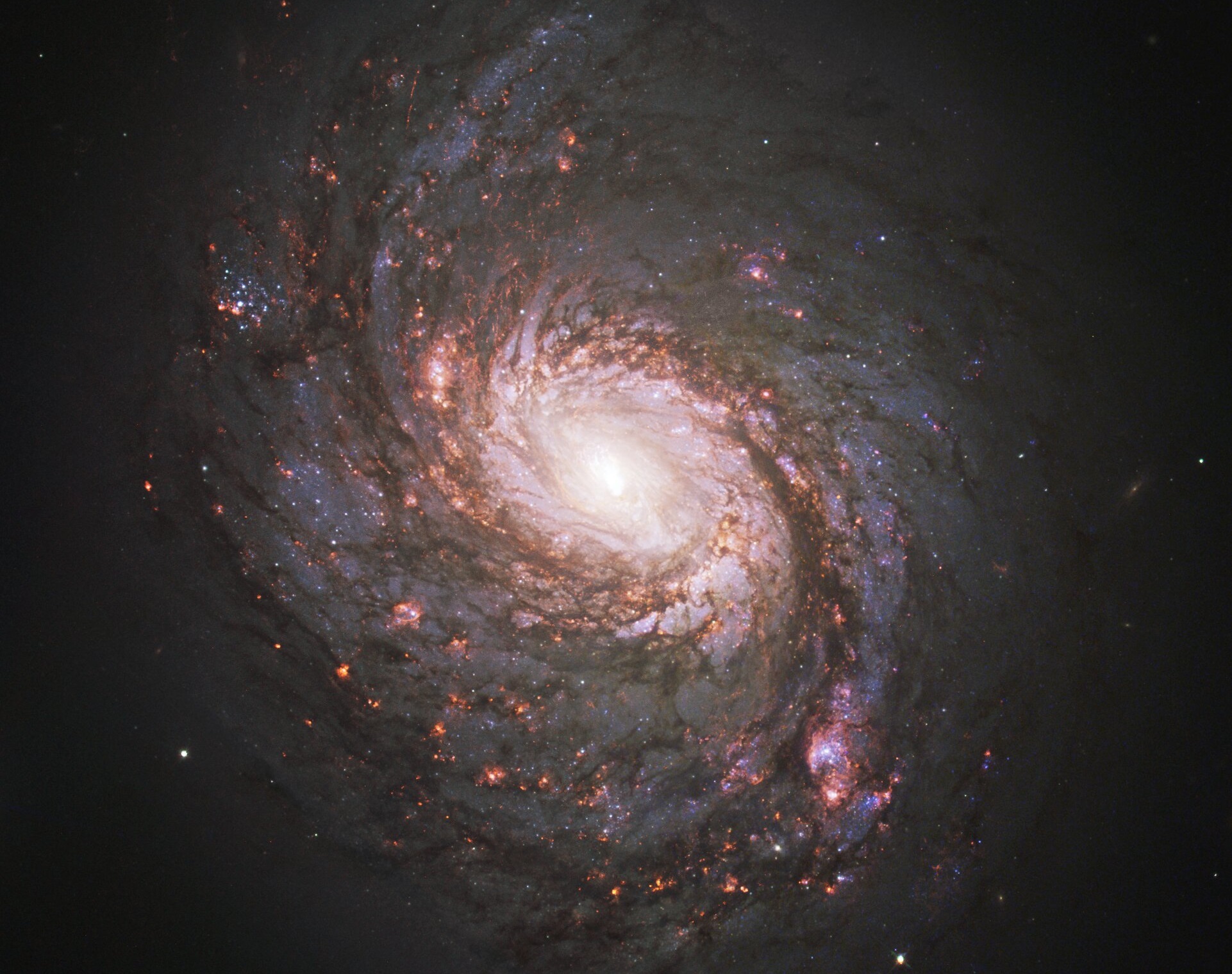}
\includegraphics[width=0.906\textwidth]{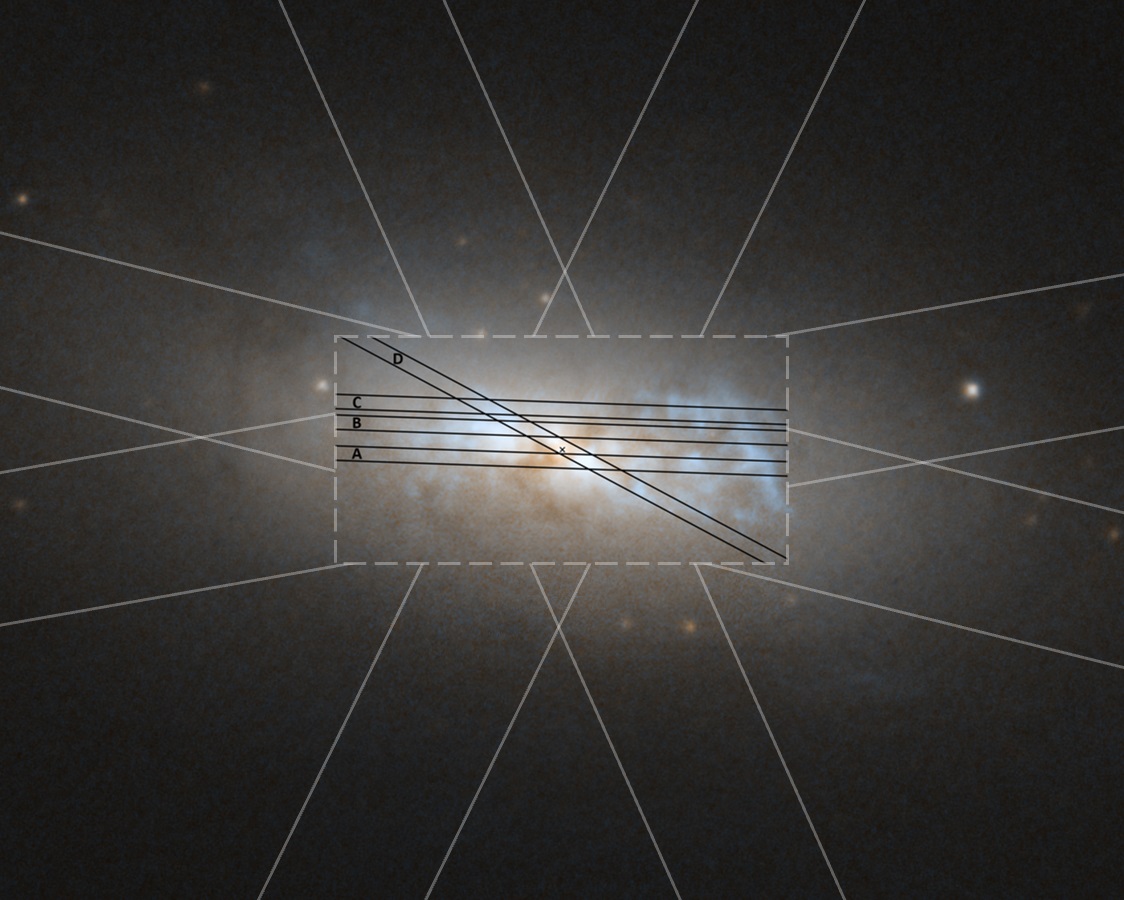}
\caption{\hst color-composite images of Mrk~3 (top-left, $56\arcsec \times 45\arcsec$, WFPC2/WF F814W, WFPC2/PC F606W), NGC~1068 (top-right, $144\arcsec \times 114\arcsec$, ACS/WFC F658N, F814W, WFPC2 F606W, F450W), and Mrk 78 (bottom, $15\arcsec \times 12\arcsec$, ACS/WFC F814W, STIS/CCD 50CCD, FOC/96 F502M). North is up and east is to the left in all panels, and high-resolution images are available at \url{https://www.flickr.com/photos/geckzilla/}. The image of Mrk~78 has a $6\arcsec \times 3\arcsec$ inset rectangle, and the solid lines delineate the locations and slit widths of the observations, with \hst STIS (0$\farcs$2 wide slit) in black and APO DIS (2$\farcs$0 wide slit) in gray. The \hst and APO slits spatially overlap and are truncated at the dashed rectangle for visibility. The `$\times$' marks the location of the optical and infrared continuum peak \citep{RamosAlmeida2006, Fischer2011}. Bright white-blue emission near the nuclei shows AGN-ionized gas in the NLR and ENLR. Diffuse white is primarily stellar continuum emission and the dark lanes and patches are due to dust. The red clumps in NGC~1068 are H~II regions created by star formation outside of the nucleus.}
\label{structure}
\end{figure*}

Mrk~78 was observed extensively with \hst STIS using the G140L, G430L, G430M, and G750M gratings with a $52\arcsec \times 0\farcs2$ slit to investigate jet-gas interactions within the NLR (Program ID 7404, PI: M. Whittle; \citealt{Whittle2004, Whittle2005, Rosario2007}). As shown in Figure~\ref{structure}, the observations consist of four long-slit pointings labeled A, B, C, and D. Slits A -- C are parallel and spatially offset from one another at a position angle (PA) of $88\degr$, while slit D intersects these near the nucleus at a PA of $66\degr$.

In our analysis, we focus on the medium-dispersion G750M observations to trace the ionized gas kinematics, as well as the low-dispersion G430L observations that contain diagnostic emission lines required for photoionization modeling. Details of the observations are provided in Table~\ref{data}, and extracted spectra for each slit are shown in Figure~\ref{spectra}. For \othree imaging, we use FOC observations through the F502M filter with F550M for continuum subtraction (Program ID 5140, PI: F. Macchetto).

Mrk~3 and NGC~1068 have been well-studied, and we use portions of our previous investigations in this analysis. The spectroscopic results for Mrk~3 are based on \hst STIS long-slit observations using the G140L, G230L, G430L, and G750L gratings at a PA of $71\degr$ with the $52\arcsec \times 0\farcs1$ slit (Program ID 8480, PI: S.~Kraemer) that are described in \cite{Collins2005}. NGC~1068 was observed with the same gratings along a PA of $202\degr$ and the $52\arcsec \times 0\farcs1$ slit (Program ID 7573, PI: S.~Kraemer), with details of the observations presented in \cite{Crenshaw2000a}. For \othree imaging, we use an FOC observation of Mrk~3 through the F502M filter and F550M for continuum subtraction (Program ID 5140, PI: F.~Macchetto). NGC~1068 was observed with WFPC2/PC using the F502N filter with F547M for continuum subtraction (Program ID 5754, PI: H.~Ford). Details for the images of Mrk~3, Mrk~78, and NGC~1068 are provided in Table~\ref{data}.

\subsection{Apache Point Observatory}

We observed Mrk~78 with the Dual Imaging Spectrograph (DIS) on the Apache Point Observatory (APO) 3.5-meter telescope to obtain deep spectroscopy of the ionized gas outside of the narrow \hst slits. The DIS gathers spectra in blue and red channels simultaneously, enabling us to characterize the kinematics and physical conditions of gas in the extended narrow-line region (ENLR) on scales $> 1\arcsec$, using a variety of emission lines surrounding H$\beta$ and H$\alpha$. We collected spectra using a $2\arcsec$-wide slit at PAs of $24\degr$, $76\degr$, $100\degr$, and $152\degr$ to sample near the major and minor axes, with the photometric major axis of the host galaxy at a PA of $\sim$84$\degr$ \citep{Schmitt2000}. Details of the observations and instrument are presented in Table~\ref{data}.

\movetabledown=1.75in
\begin{rotatetable*}
\begin{deluxetable*}{cccccccccccccccc}
\tabletypesize{\footnotesize}
\setlength{\tabcolsep}{0.03in}
\tablecaption{Summary of Observations}
\tablehead{
\colhead{Target} & \colhead{Observing} & \colhead{Instrument} & \colhead{Proposal} & \colhead{Observation} & \colhead{Date} & \colhead{Exposure} & \colhead{Grating} & \colhead{Slit} & \colhead{Spectral} & \colhead{Wavelength} & \colhead{Spatial} & \colhead{Position} & \colhead{Spatial} & \colhead{Mean} &\colhead{Mean}\\
\colhead{Name} & \colhead{Facility} & \colhead{Name} & \colhead{ID} & \colhead{ID} & \colhead{(UT)} & \colhead{Time} & \colhead{or Filter} & \colhead{ID} & \colhead{Dispersion} & \colhead{Range} & \colhead{Scale} & \colhead{Angle} &\colhead{Offset$^\star$} & \colhead{Airmass} &\colhead{Seeing}\\
\colhead{} & \colhead{} & \colhead{} & \colhead{} & \colhead{} & \colhead{(y-m-d)} & \colhead{(s)} & \colhead{} & \colhead{} & \colhead{(\AA~pix$^{-1}$)} & \colhead{(\AA)} & \colhead{($\arcsec$~pix$^{-1}$)} & \colhead{(deg)} &\colhead{($\arcsec$)} & \colhead{} &\colhead{($\arcsec$)}
}
\startdata
Mrk 78 & HST & STIS & 7404 & O4DJ02030 & 1998-02-28 & 2052 & G430L & A & 2.73 & 2900-5700 & 0.051 & 88.05 & 0.125 & ... & ...\\
Mrk 78 & HST & STIS & 7404 & O4DJ02060 & 1998-02-28 & 1730 & G430L & B & 2.73 & 2900-5700 & 0.051 & 88.05 &-0.27 & ... & ...\\
Mrk 78 & HST & STIS & 7404 & O4DJ02090 & 1998-03-01 & 2052 & G430L & C & 2.73 & 2900-5700 & 0.051 & 88.05 & -0.55 & ... & ...\\
Mrk 78 & HST & STIS & 7404 & O4DJ04010 & 1998-03-01 & 1643 & G430L & D & 2.73 & 2900-5700 & 0.051 & 61.56 & -0.05 & ... & ...\\
Mrk 78 & HST & STIS & 7404 & O4DJ02010 & 1998-02-28 & 1100 & G750M & A & 0.56 & 6480-7054 & 0.051 & 88.05 & 0.125 & ... & ...\\
Mrk 78 & HST & STIS & 7404 & 04DJ02040 & 1998-02-28 & 1199 & G750M & B & 0.56 & 6480-7054 & 0.051 & 88.05 & -0.27 & ... & ...\\
Mrk 78 & HST & STIS & 7404 & O4DJ02070 & 1998-02-28 & 1172 & G750M & C & 0.56 & 6480-7054 & 0.051 & 88.05 & -0.55 & ... & ...\\
Mrk 78 & HST & STIS & 7404 & O4DJ04030 & 1998-03-01 & 1320 & G750M & D & 0.56 & 6480-7054 & 0.051 & 61.56 & -0.05 & ... & ...\\
Mrk 78 & HST & STIS & 7404 & O4DJ01020 & 1997-11-16 & 120 & MIRVIS & ... & ... & 1640-10270 & 0.051 & ... & ... & ... & ...\\
Mrk 78 & HST & FOC & 5140 & X2580303T & 1994-03-19 & 800 & F502M & ... & ... & 4645-5389 & 0.014 & ... & ... & ... & ...\\
Mrk 78 & HST & FOC & 5140 & X2580304T & 1994-03-19 & 1196 & F550M & ... & ... & 5303-5726 & 0.014 & ... & ... & ... & ...\\
Mrk 78 & APO & DIS & ... & ... & 2016-01-02 & 2137 & B1200 & ... & 0.615 & 4257-5517 & 0.42 & 24 & ... & 1.36 & 1.48\\
Mrk 78 & APO & DIS & ... & ... & 2016-01-02 & 2137 & R1200 & ... & 0.580 & 6020-7180 & 0.40 & 24 & ... & 1.36 & 1.55\\
Mrk 78 & APO & DIS & ... & ... & 2014-10-25 & 2400 & B1200 & ... & 0.615 & 4760-6000 & 0.42 & 76 & ... & 1.20 & 1.52\\
Mrk 78 & APO & DIS & ... & ... & 2014-10-25 & 2400 & R1200 & ... & 0.580 & 6002-7162 & 0.40 & 76 & ... & 1.20 & 1.80\\
Mrk 78 & APO & DIS & ... & ... & 2015-02-19 & 2400 & B1200 & ... & 0.615 & 4481-5721 & 0.42 & 100 & ... & 1.18 & 1.48\\
Mrk 78 & APO & DIS & ... & ... & 2015-02-19 & 2400 & R1200 & ... & 0.580 & 6002-7162 & 0.40 & 100 & ... & 1.18 & 1.55\\
Mrk 78 & APO & DIS & ... & ... & 2015-12-03 & 2700 & B1200 & ... & 0.615 & 4278-5518 & 0.42 & 152 & ... & 1.49 & 1.52\\
Mrk 78 & APO & DIS & ... & ... & 2015-12-03 & 2700 & R1200 & ... & 0.580 & 6020-7180 & 0.40 & 152 & ... & 1.49 & 1.55\\
\hline
Mrk 3 & HST & FOC & 5140 & X2580103T & 1994-03-20 & 750 & F502M & ... & ... & 4645-5389 & 0.014 & ... & ... & ... & ...\\
Mrk 3 & HST & FOC & 5140 & X2580104T & 1994-03-20 & 1196 & F550M & ... & ... & 5303-5726 & 0.014 & ... & ... & ... & ...\\
\hline
NGC 1068 & HST & WFPC2/PC & 5754 & U2M30103T & 1995-01-17 & 300 & F502N & ... & ... & 4969-5044 & 0.045 & ... & ... & ... & ...\\
NGC 1068 & HST & WFPC2/PC & 5754 & U2M30104T & 1995-01-17 & 600 & F502N & ... & ... & 4969-5044 & 0.045 & ... & ... & ... & ...\\
NGC 1068 & HST & WFPC2/PC & 5754 & U2M30101T & 1995-01-17 & 140 & F547M & ... & ... & 5060-5885 & 0.045 & ... & ... & ... & ...\\
NGC 1068 & HST & WFPC2/PC & 5754 & U2M30102T & 1995-01-17 & 300 & F547M & ... & ... & 5060-5885 & 0.045 & ... & ... & ... & ...
\enddata
\vspace{-0.2em}
\begin{minipage}{8.54in}
\tablecomments{A summary of the observations and data used in this study. The columns list the observing facility, instrument, \hst proposal ID, MAST archive observation ID, observation date, exposure time, grating (for spectra) or filter (for imaging), the four {\it HST} slit names, spectral dispersion, wavelength range (for spectra) or bandpass (for imaging, defined as the range where the system throughput exceeds 1\%), spatial resolution, position angles of the slits, spatial offset from the continuum peak, airmass, and seeing. The seeing was calculated by measuring the full-width at half-maximum (FWHM) of the brightness profiles of the standard stars along the slit. All {\it HST} values are defined in their respective instrument handbooks \citep{McMaster2008, Riley2017}, with the exact STIS spatial scale quoted as 0.05078$\arcsec$~pix$^{-1}$. $^\star$Observations with non-zero values are spatially offset from the nucleus. The data may be obtained from MAST using the following DOI:\dataset[10.17909/t9-4581-8p50]{\doi{10.17909/t9-4581-8p50}}.}
\end{minipage}
\label{data}
\end{deluxetable*}
\end{rotatetable*}

We reduced the APO spectroscopy using IRAF \citep{Tody1986, Tody1993} and standard techniques including bias subtraction, image trimming, bad pixel replacement, flat-fielding, Laplacian edge cosmic-ray removal \citep{vanDokkum2001}, image combining, and sky-line subtraction. Wavelength calibration was completed using comparison arc lamp images, and velocities were corrected to heliocentric. Flux calibration was completed using Oke standard stars \citep{Oke1990} and the airmass at mid-exposure. Finally, the DIS dispersion and spatial axes are not perpendicular, so we fit a line to the galaxy continuum and resampled the data so measurements of emission lines from the same pixel rows correspond to the same spatial locations.

\begin{figure*}
\centering
\includegraphics[width=0.95\textwidth]{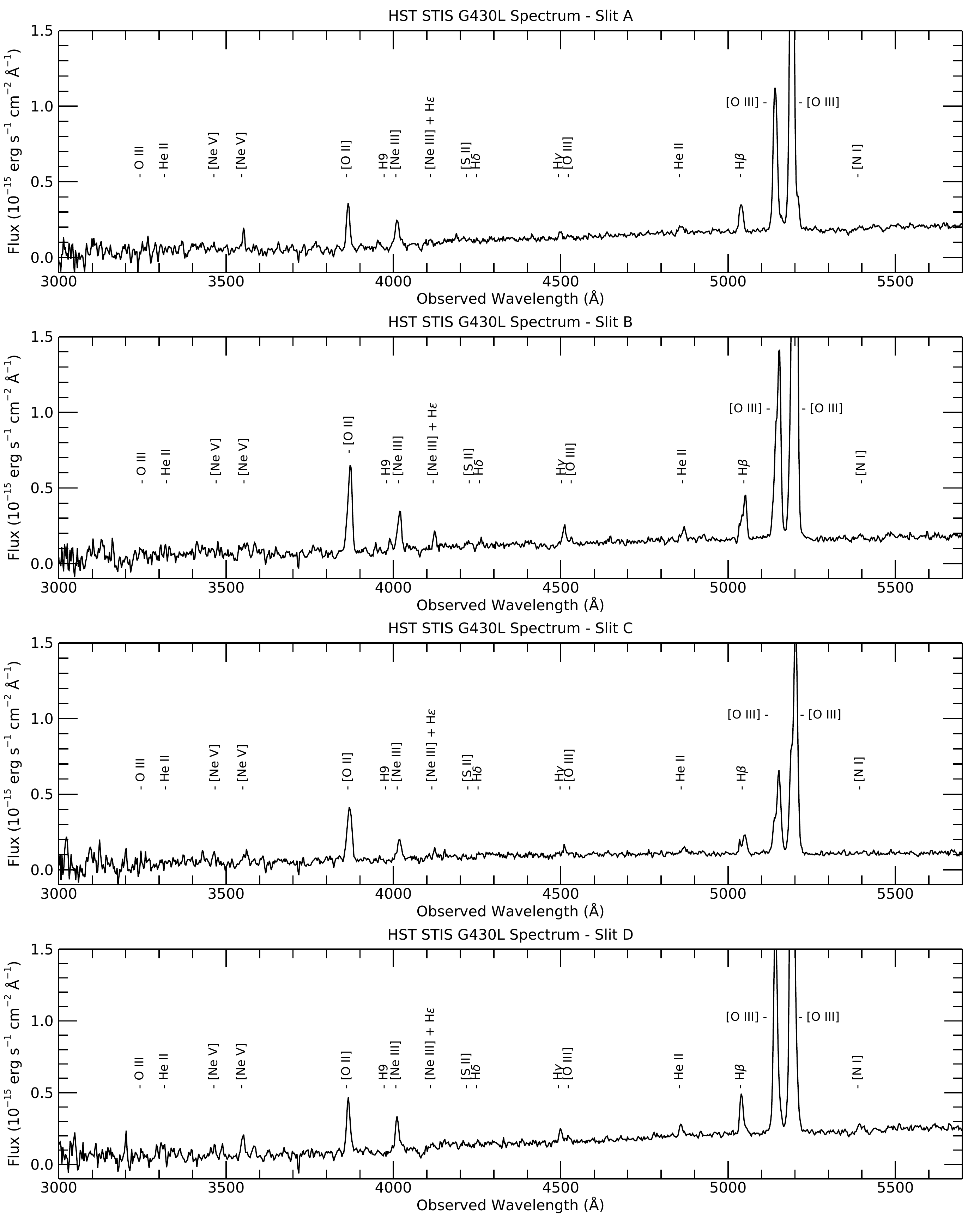}
\caption{Mrk~78 spectral traces of the nuclear emission spatially summed over $\sim1\farcs0$, with the positions of emission lines labeled. The spectra are shown at observed wavelengths and from top to bottom are \hst STIS G430L and G750M for slits A, B, C, and D.}
\label{spectra}
\end{figure*}
\addtocounter{figure}{-1}
\begin{figure*}
\centering
\includegraphics[width=0.95\textwidth]{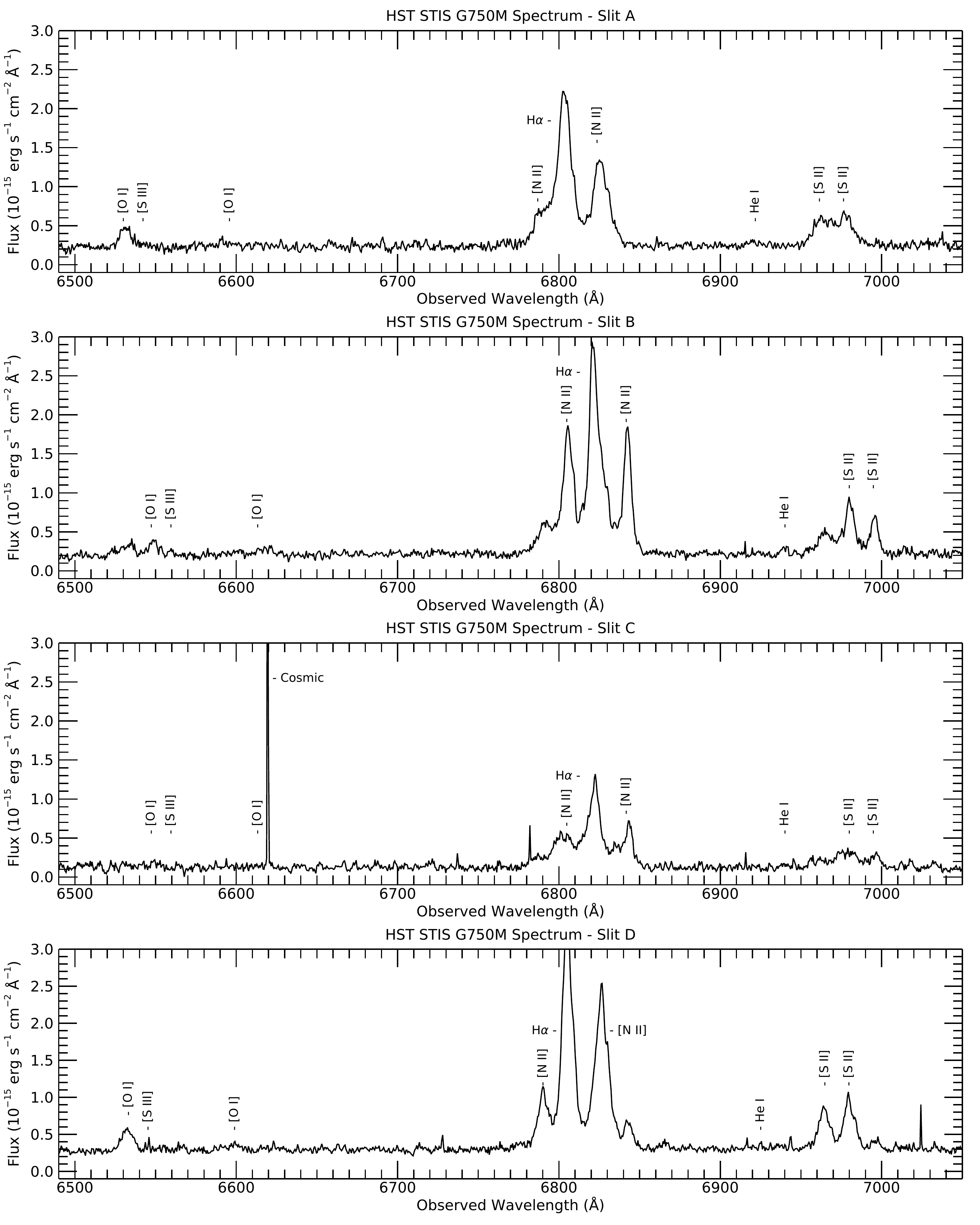}
\caption{{\it continued.}}
\end{figure*}

\begin{figure}
\centering
\includegraphics[width=0.49\textwidth]{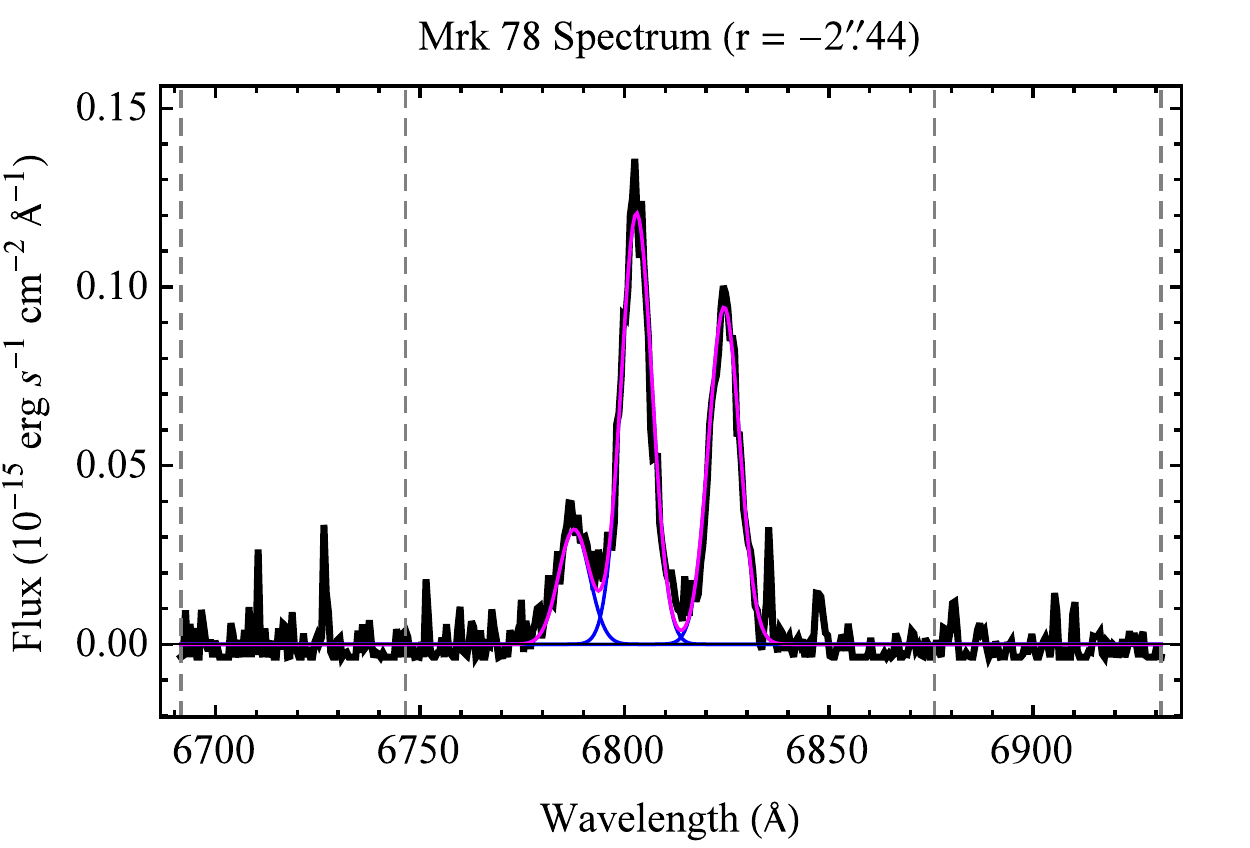}\vspace{0.05em}\\ 
\includegraphics[width=0.49\textwidth]{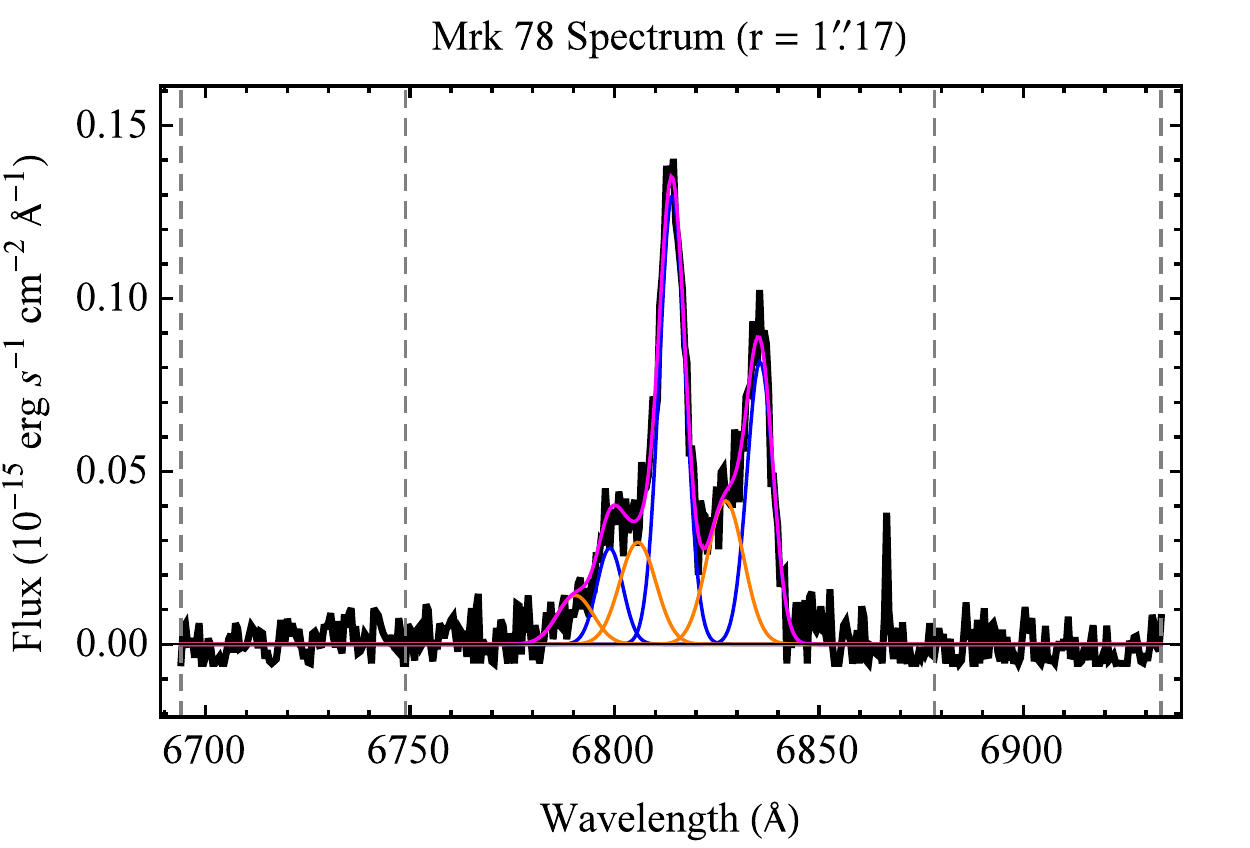}\vspace{0.05em}\\ 
\includegraphics[width=0.49\textwidth]{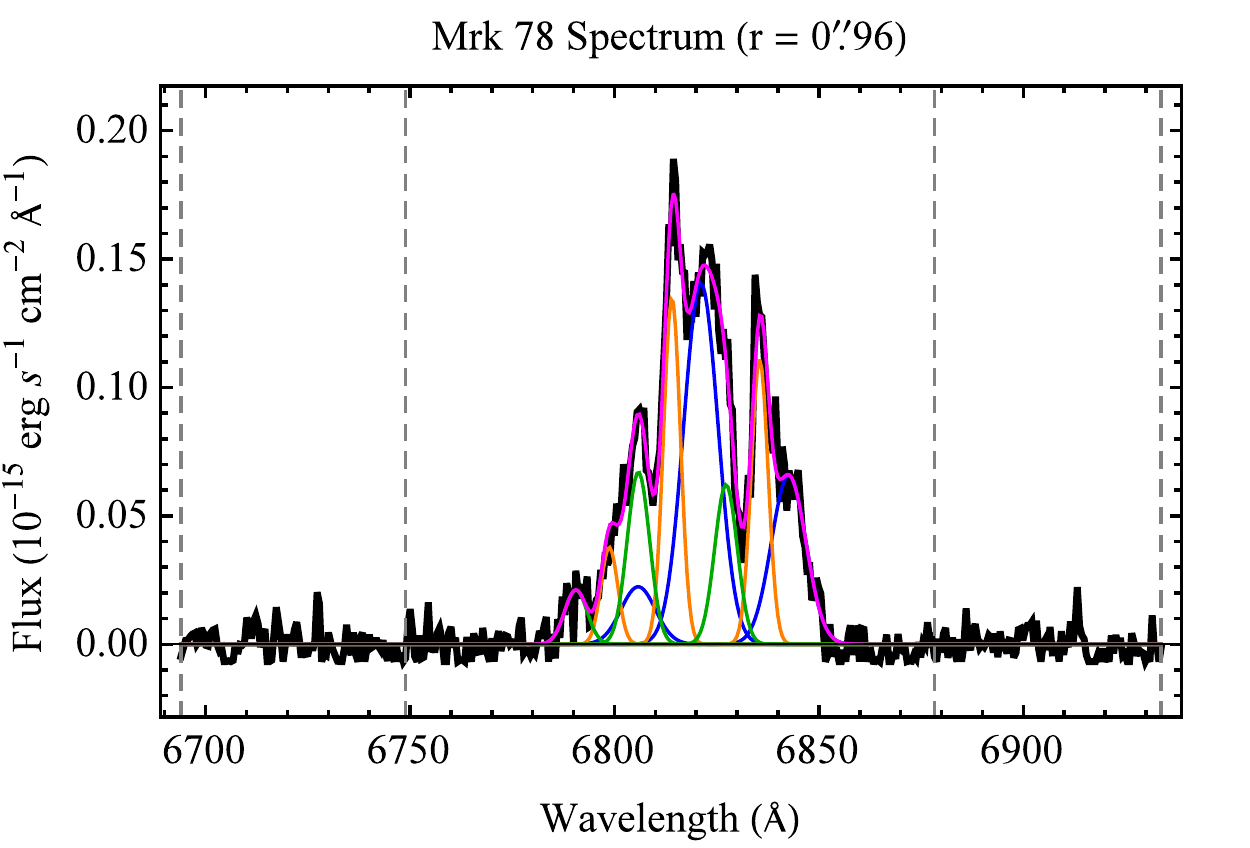} 
\caption{Examples of the H$\alpha$ + [N~II] Bayesian spectral decomposition process for Mrk~78 \hst STIS G750M observations, which determines whether one (top), two (middle), or three (bottom) meaningful kinematic components exist at each spatial location along the slits. The spectra are fit with an increasing number of components until the difference between competing models is less than a Bayesian criterion as described in the text. The data are shown in black, the continuum fit regions between gray dashed lines, and the first, second, and third components in order of decreasing peak flux are shown in blue, orange, and green, respectively. The sum of all Gaussian kinematic components is shown in magenta.}
\label{fitting}
\end{figure}

\section{Analysis}

\subsection{Spectral Fitting}

We fit Gaussian profiles to the emission lines in our spectra at each location along the slits to derive the spatially-resolved gas kinematics and emission line flux ratios required to generate photoionization models. We use a Bayesian fitting routine that we developed based on the Importance Nested Sampling algorithm in MultiNest \citep{Feroz2008, Feroz2009, Feroz2019, Buchner2014} that we have used in a variety of kinematic studies \citep{Fischer2018}. This procedure determines the number of meaningful kinematic components and characterizes each as a Gaussian with a variable velocity centroid, width, and height above the underlying continuum. The data are fit with an increasing number of kinematic components until the Bayesian likelihood criteria $ln(Z)$ is $< 5$ between models, at which point the simpler model with fewer components is selected. A detailed description of this process is given in \cite{Fischer2017}, and we present examples of one, two, and three component fits to the \hst data in Figure~\ref{fitting}.

We extracted spectra in $0\farcs2$ and $0\farcs42$ intervals along the slits for the \hst and APO data, respectively, and adopted a minimum peak-flux signal-to-noise (S/N) ratio of $> 2$ above the continuum for positive line detections. This over-samples the APO data as flux is shared between adjacent pixels due to the seeing; however, the radial gradients are preserved and we only derive the amplitude of the galaxy rotational velocity and the large-scale extent of the ionized gas using these ground-based data.

We ensure that the fits to each emission line sample the same kinematic components of the gas by using a spectral template technique that fixes the line centroids and widths based on freely-varying fits to the strong, velocity-resolved H$\alpha$ $\lambda$6563 \AA\ $+$ [N~II] $\lambda\lambda$6548, 6584 \AA\ emission lines in the \hst STIS G750M spectra. While the \othree $\lambda\lambda$4959, 5007 \AA~emission lines are brighter, we found that H$\alpha$ produced fits with smaller residuals than when using \othree in the lower dispersion G430L spectra and did not require independently fitting the G430M spectra. The line widths were constrained to minimum values of the line spread functions (LSFs) for each grating, measured from calibration lamp exposures, and maximum FWHM of 2000 km s$^{-1}$.

For all of the other emission lines, the fits are scaled from H$\alpha$ to preserve the same intrinsic velocity widths and centroids, and account for the instrument line spread functions between different gratings. The height of each Gaussian component is free to vary to enclose the total emission line flux, and the uncertainties are given by the residuals between the data and fits. This process allows us to accurately fit weak diagnostic emission lines that are important for comparison with photoionization models, but small differences in the intrinsic line widths may be neglected (see \S3.1 of \citealp{Revalski2018a}).

We further constrain the fitting process by fixing the relative height ratios of doublet lines to their theoretical values \citep{Osterbrock2006}. Specifically, \othree $\lambda \lambda$5007/4959 = 3.01, [O~I] $\lambda \lambda$6300/6363 = 3.0, and [N~II] $\lambda \lambda$6584/6548 = 2.95. We fit blended or closely spaced lines such as H$\alpha$ and [N~II], the [S~II] doublet, and others simultaneously, and fix the relative separations of these lines relative to H$\alpha$ to their laboratory values. We followed a similar procedure for the APO DIS spectroscopic data, using the strong \othree $\lambda$5007 \AA~emission line to trace the large-scale gas kinematics, because in this case the blue and red spectra have essentially the same wavelength resolution.

\subsection{Ionized Gas Kinematics}

\begin{figure*}[ht!]
\centering
\includegraphics[width=0.49\textwidth]{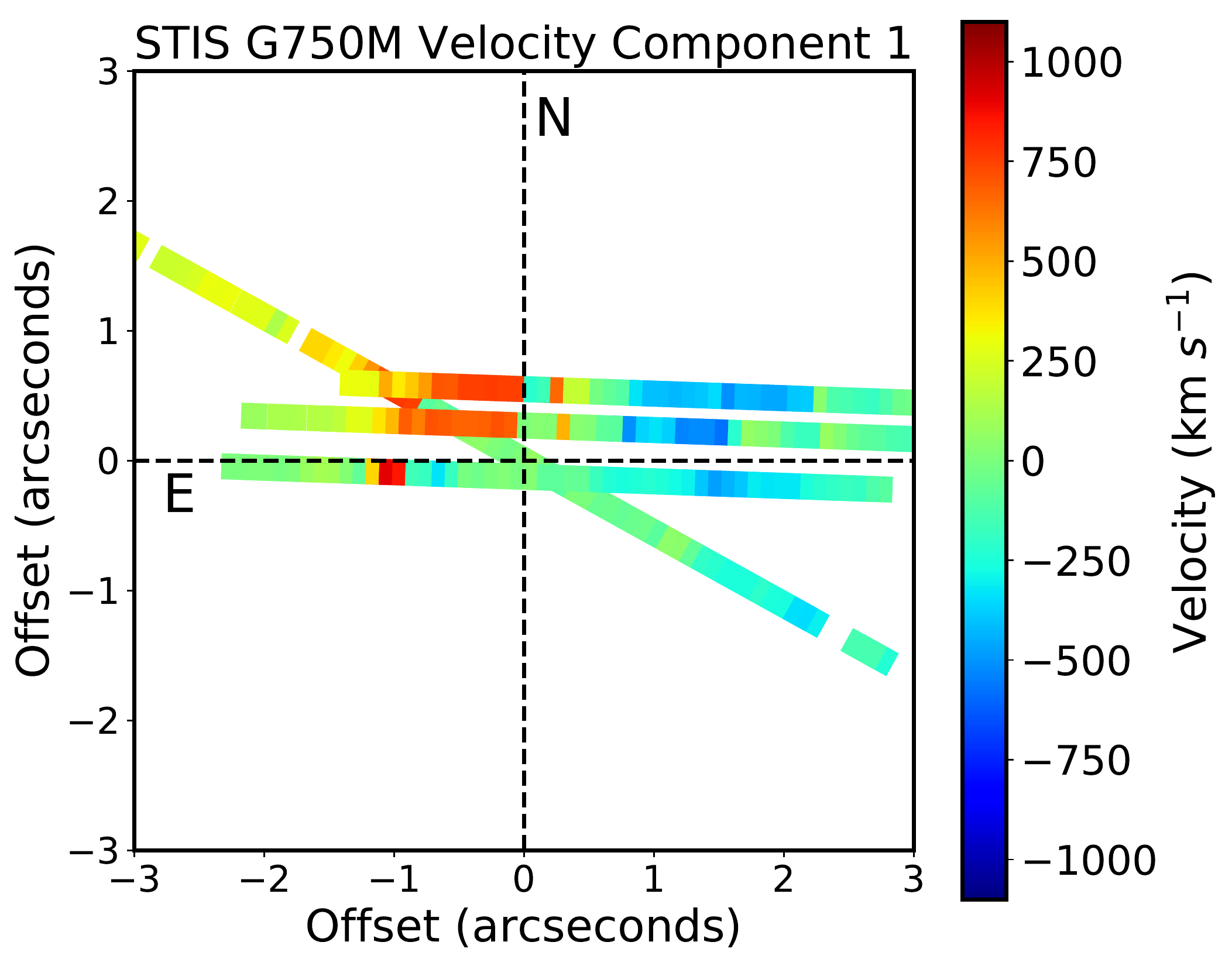}
\includegraphics[width=0.49\textwidth]{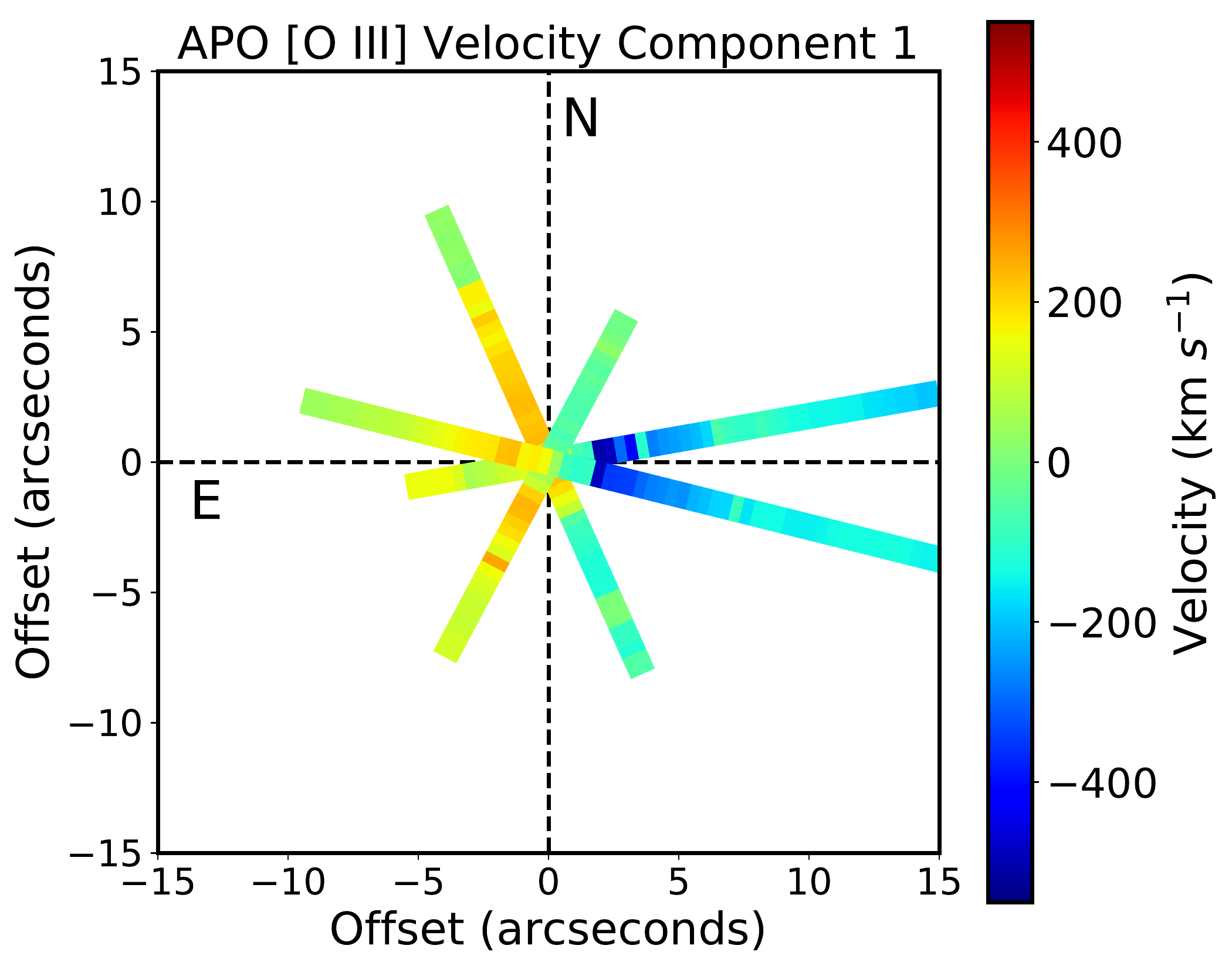}
\includegraphics[width=0.49\textwidth]{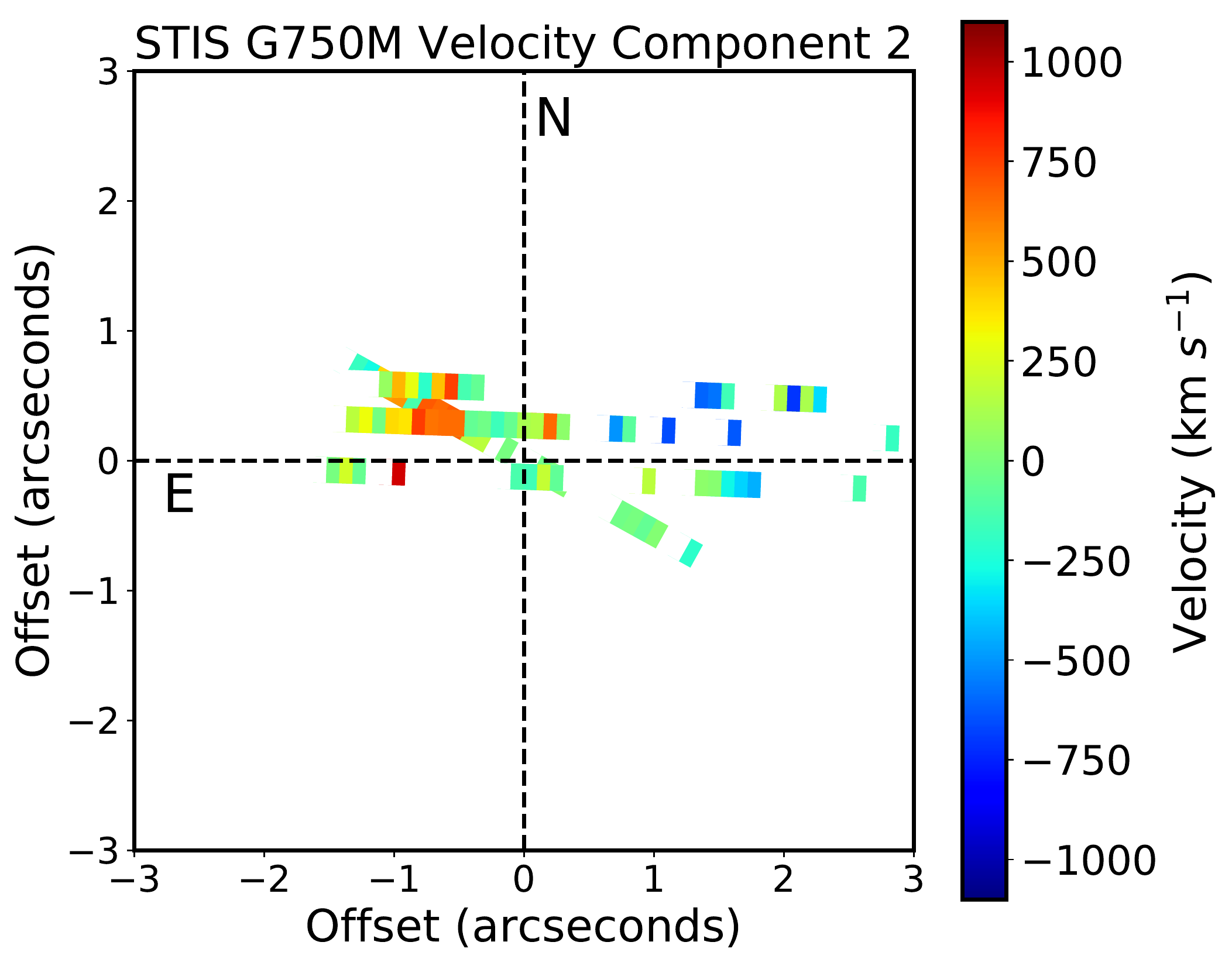}
\includegraphics[width=0.49\textwidth]{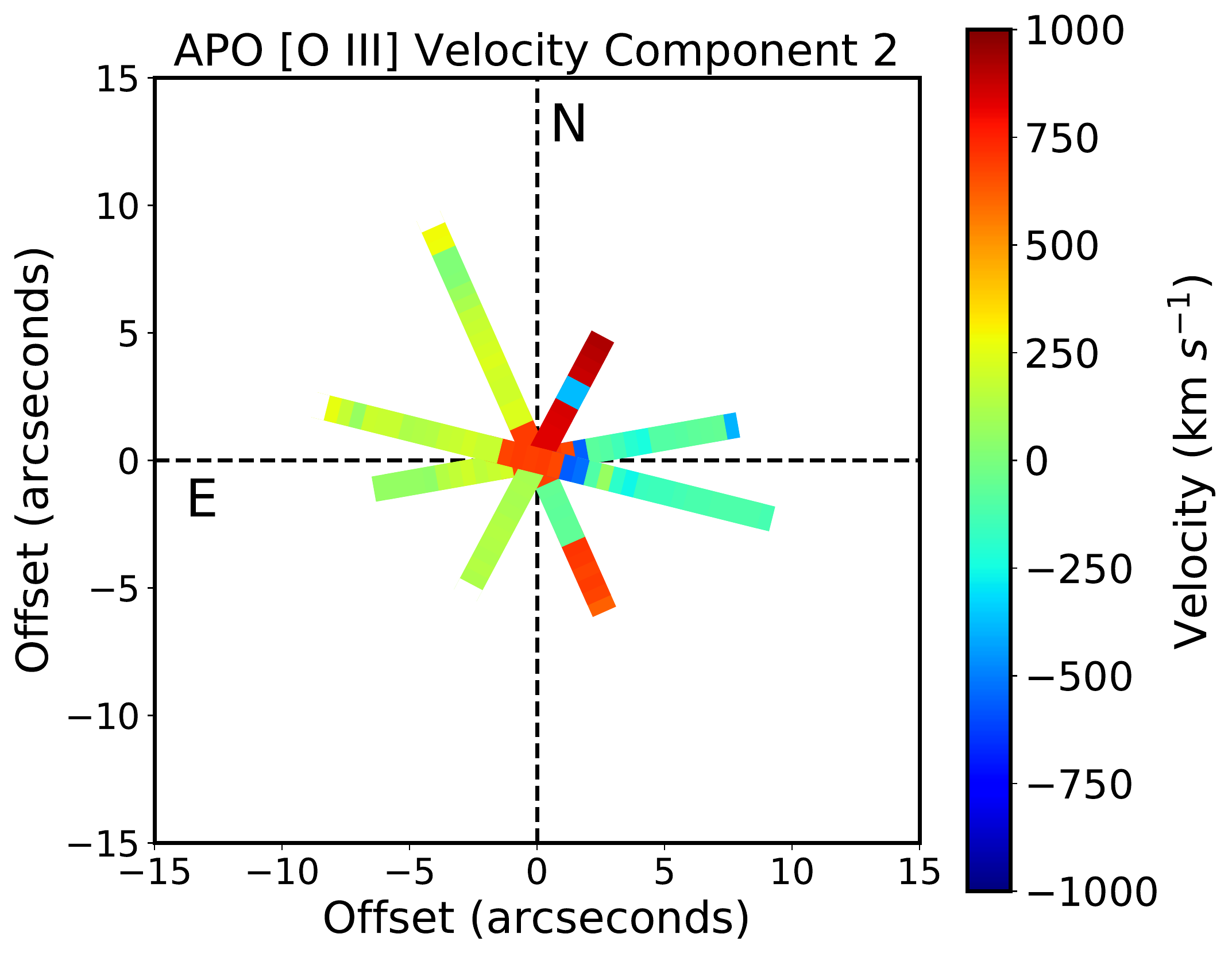}
\includegraphics[width=0.49\textwidth]{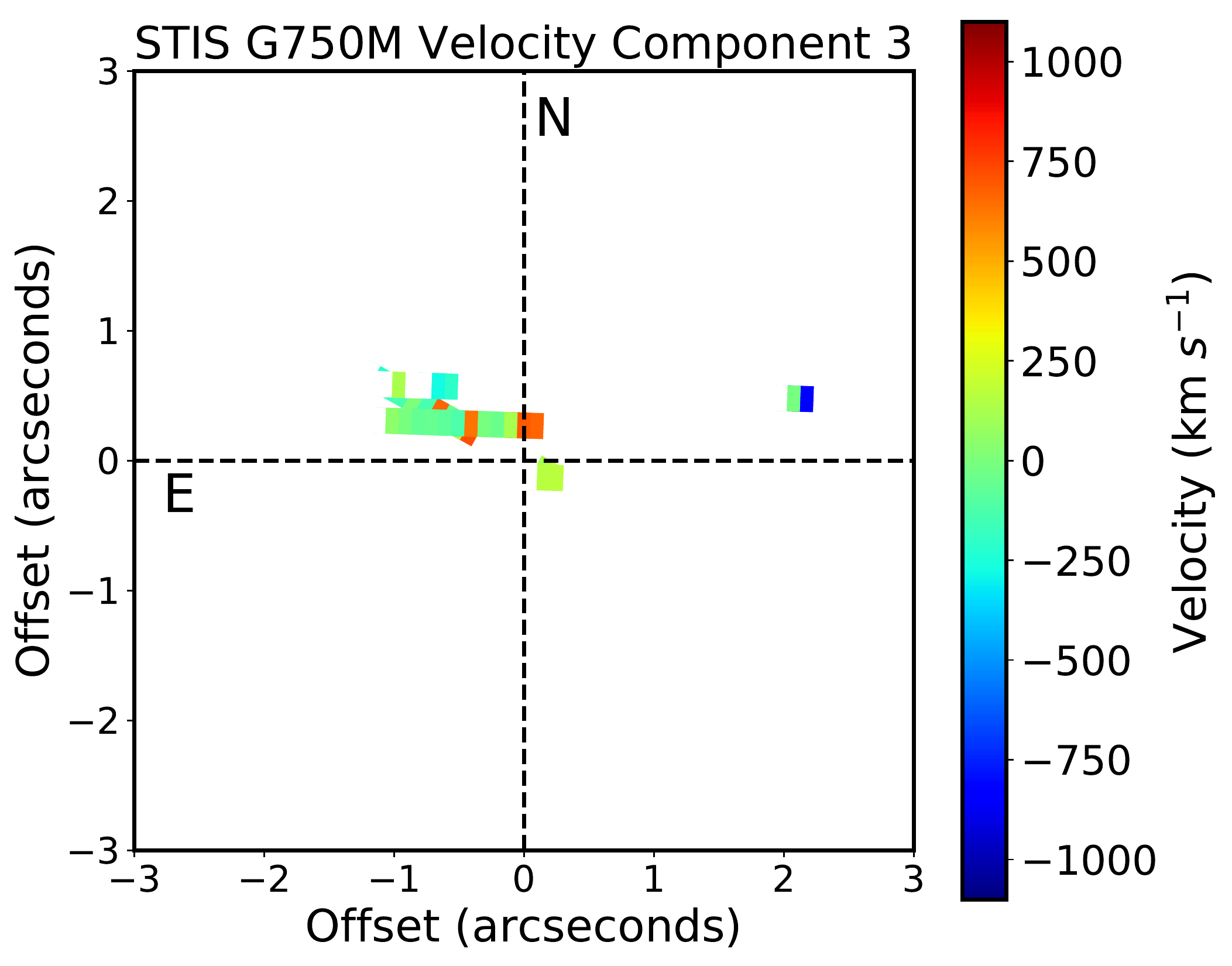}
\includegraphics[width=0.49\textwidth]{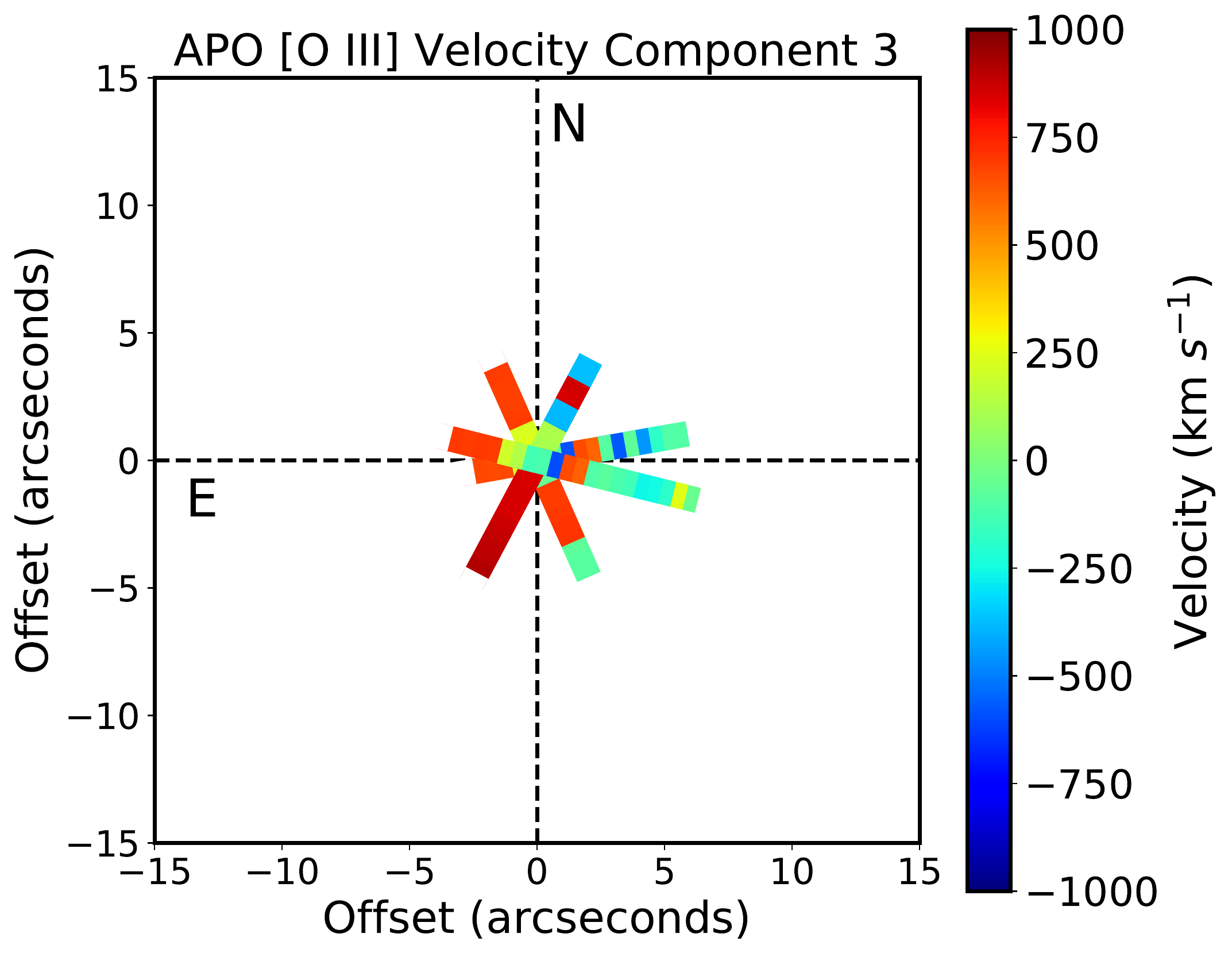}
\caption{Mrk~78 velocity maps of the ionized gas derived from fits to the \hst STIS H$\alpha$ (left) and APO DIS \othree (right) spectroscopy. The panels show the individual kinematic components for the highest (top), intermediate (middle), and lowest (bottom) flux components. The strong redshifts and blueshifts in the upper-left panel indicate the presence of ionized outflows, while the lower amplitude velocities in the upper-right panel (which has a smaller colorbar range) trace the larger scale rotation of gas in the extended NLR. The \hst panels span $\pm 3\arcsec$ and the slits are 0$\farcs$2 in width, while the APO panels span $\pm 15\arcsec$ and the slits are 2$\farcs$0 in width, but are represented by 1$\farcs$0 wide rectangles for visibility.}
\label{velmaps}
\end{figure*}

\begin{figure*}[ht!]
\centering
\subfigure{
\includegraphics[width=0.49\textwidth]{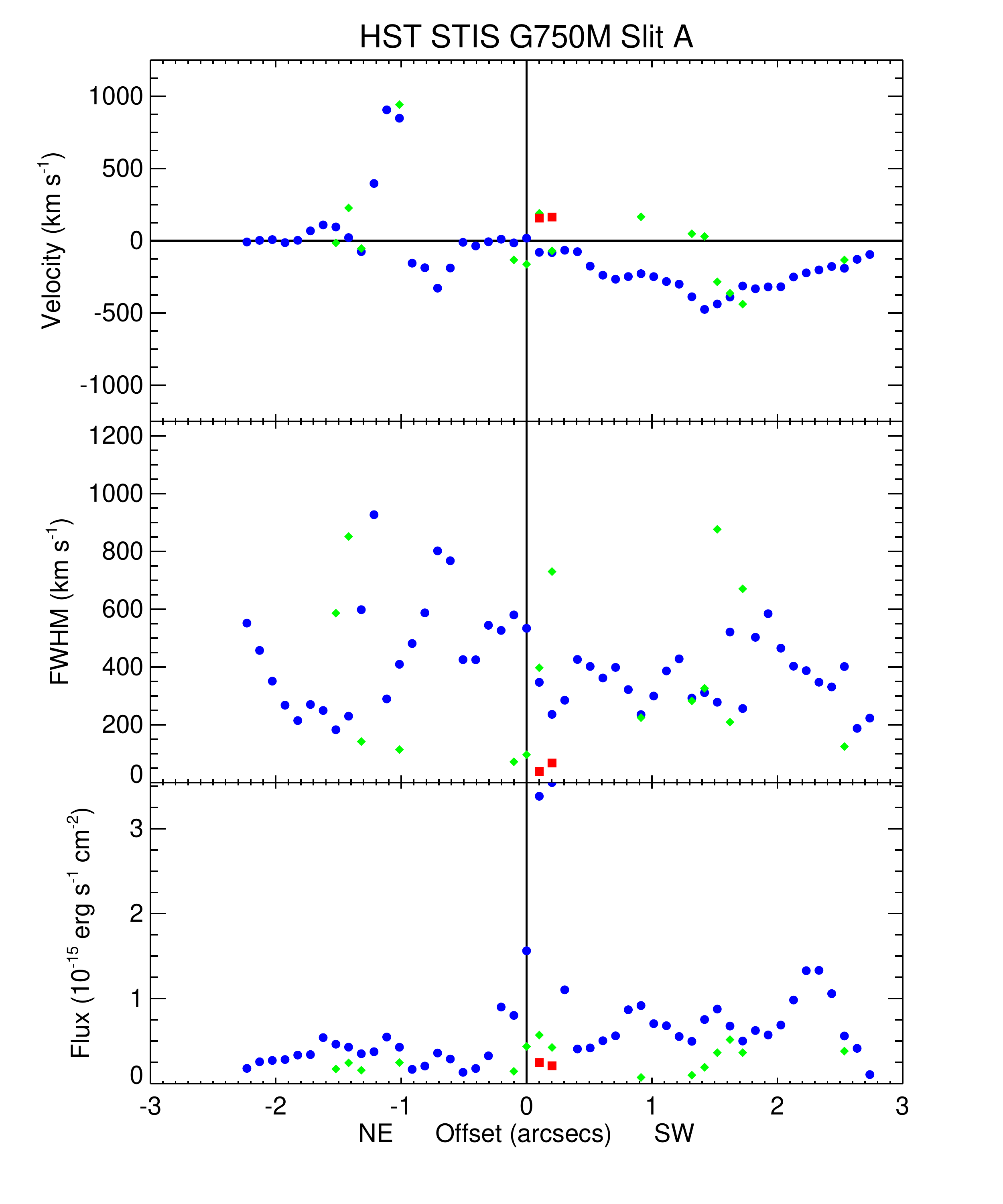}}
\subfigure{
\includegraphics[width=0.49\textwidth]{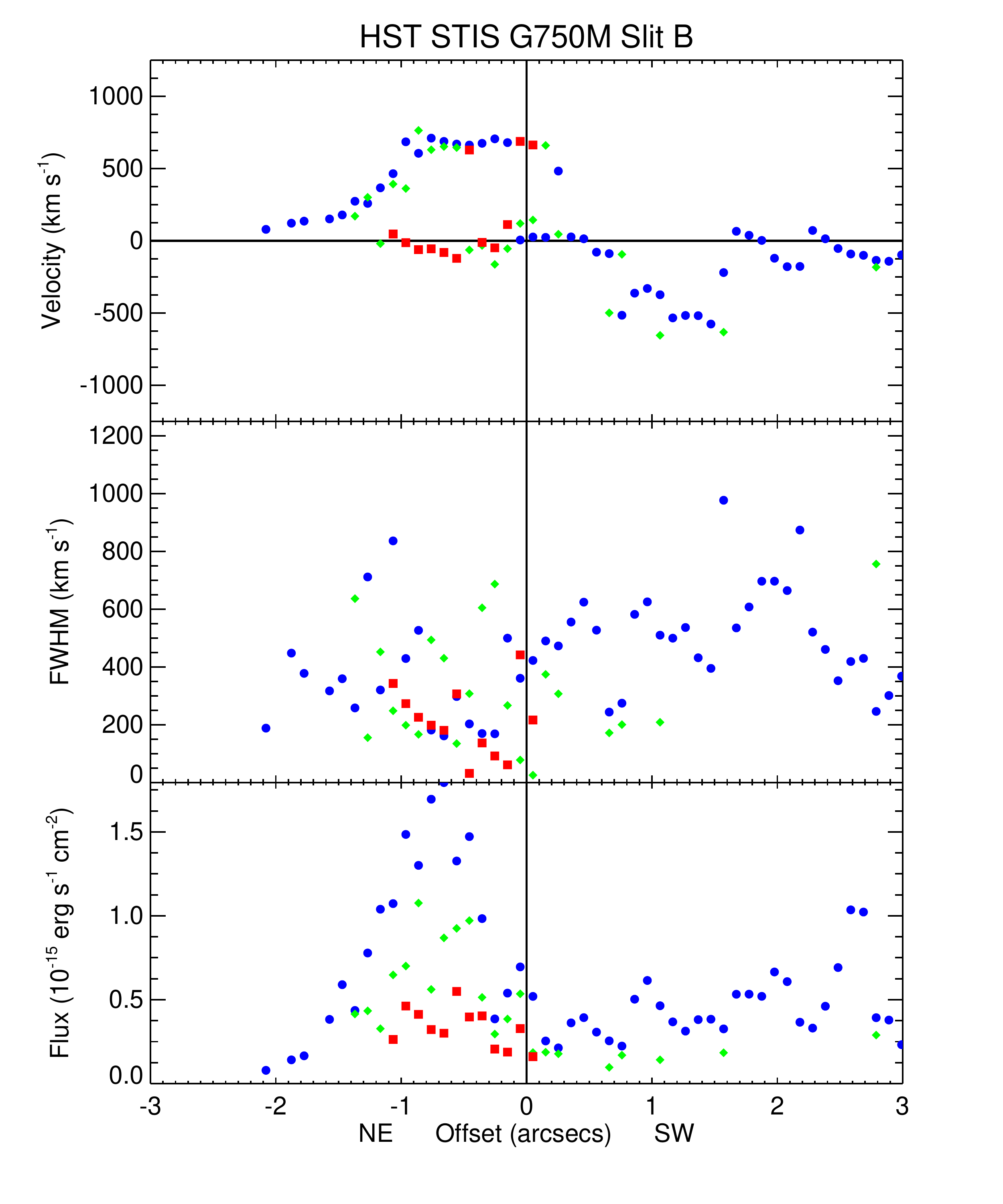}}
\subfigure{
\includegraphics[width=0.49\textwidth]{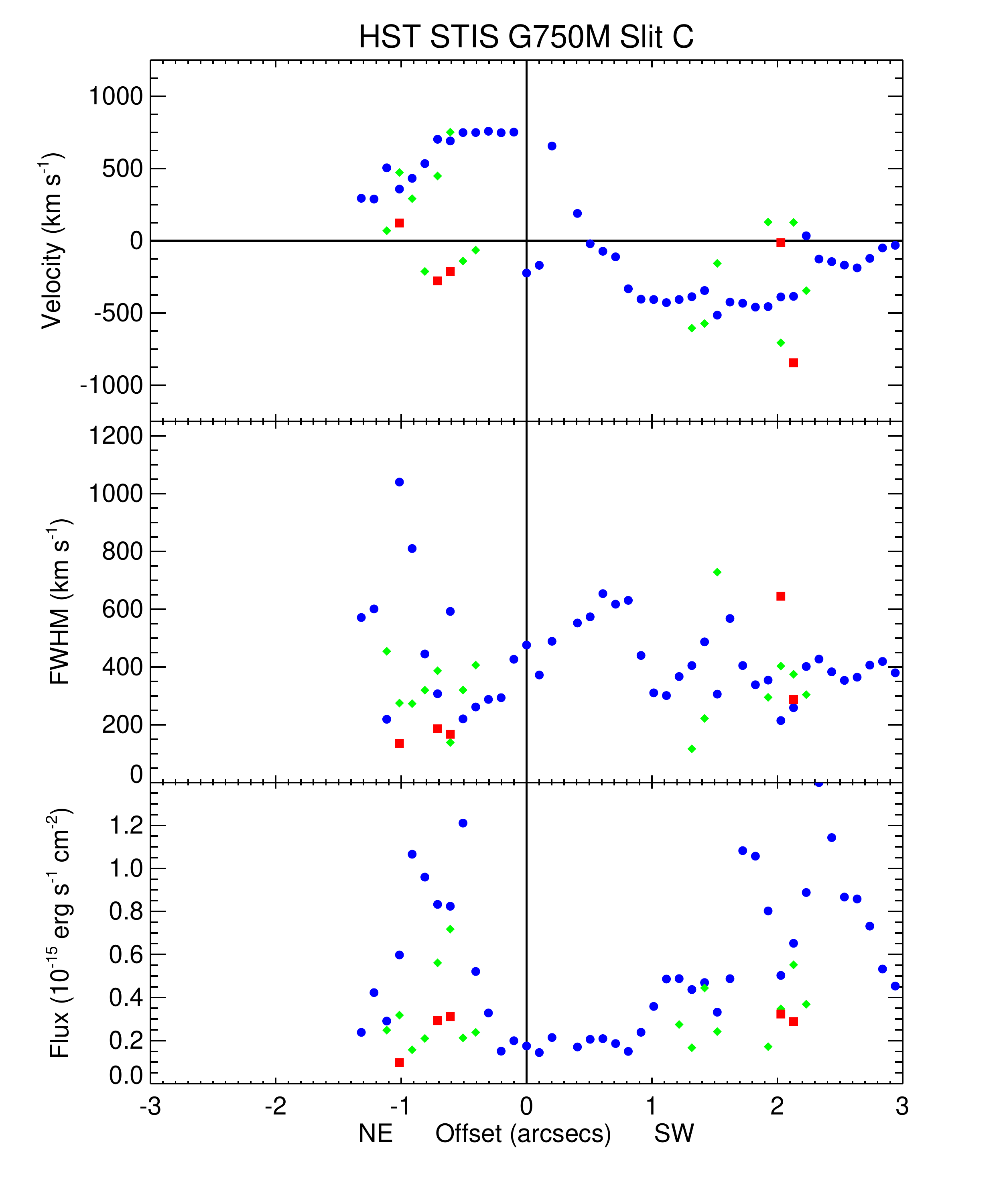}}
\subfigure{
\includegraphics[width=0.49\textwidth]{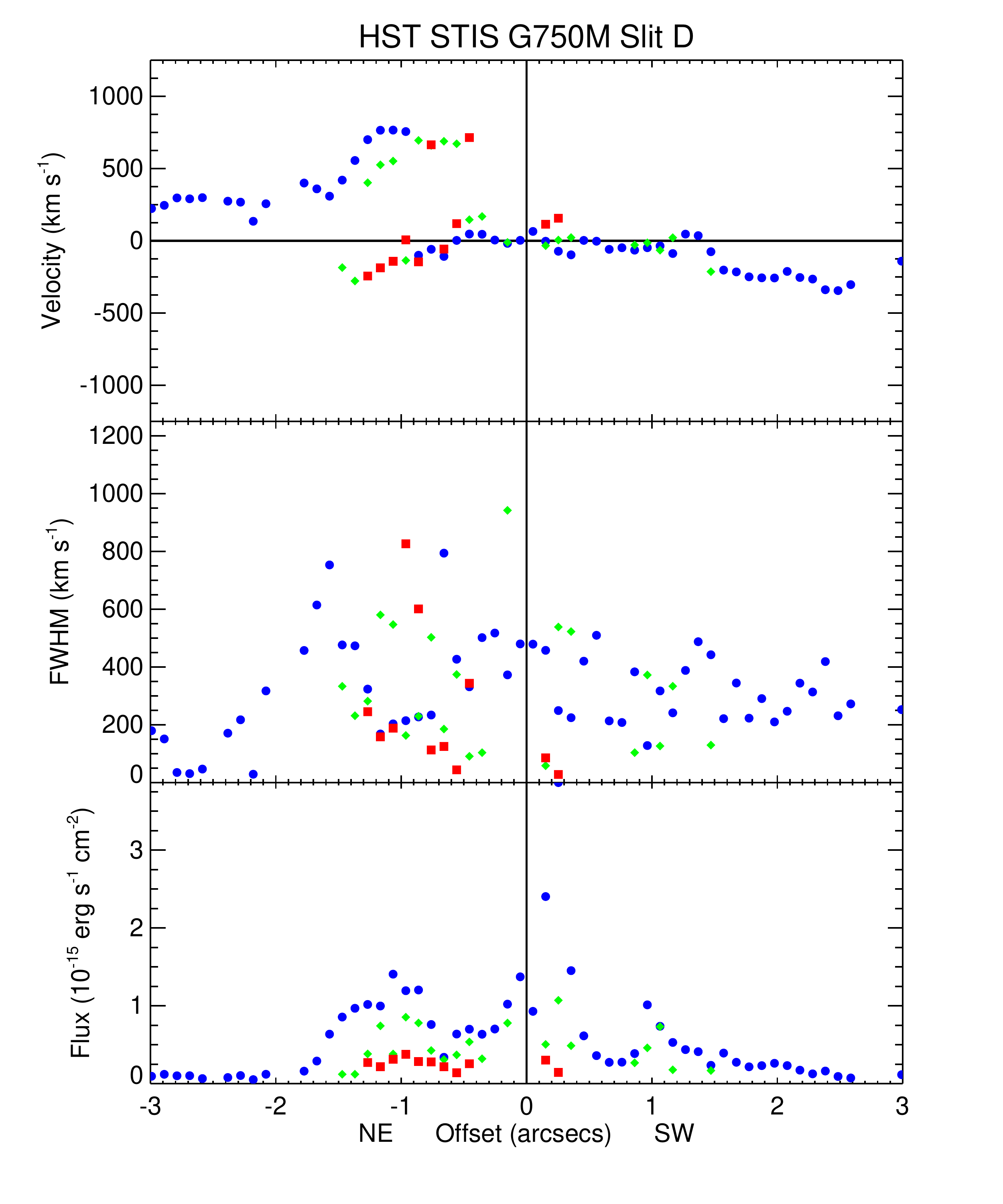}}
\vspace{-2em}
\caption{Mrk~78 observed velocity centroids (top), FWHM (middle), and integrated line fluxes (bottom) for the H$\alpha$ $\lambda$6563 emission line in each of the four HST STIS G750M spectral observations. The points are color-coded from strongest to weakest peak flux in the order: blue circles, green diamonds, and red squares. Slits A, B, and C are at a PA of 88$\degr$, while slit D is at 61$\degr$ east of north.}
\label{hstkinematics}
\end{figure*}

\begin{figure*}[ht!]
\centering
\subfigure{
\includegraphics[width=0.49\textwidth]{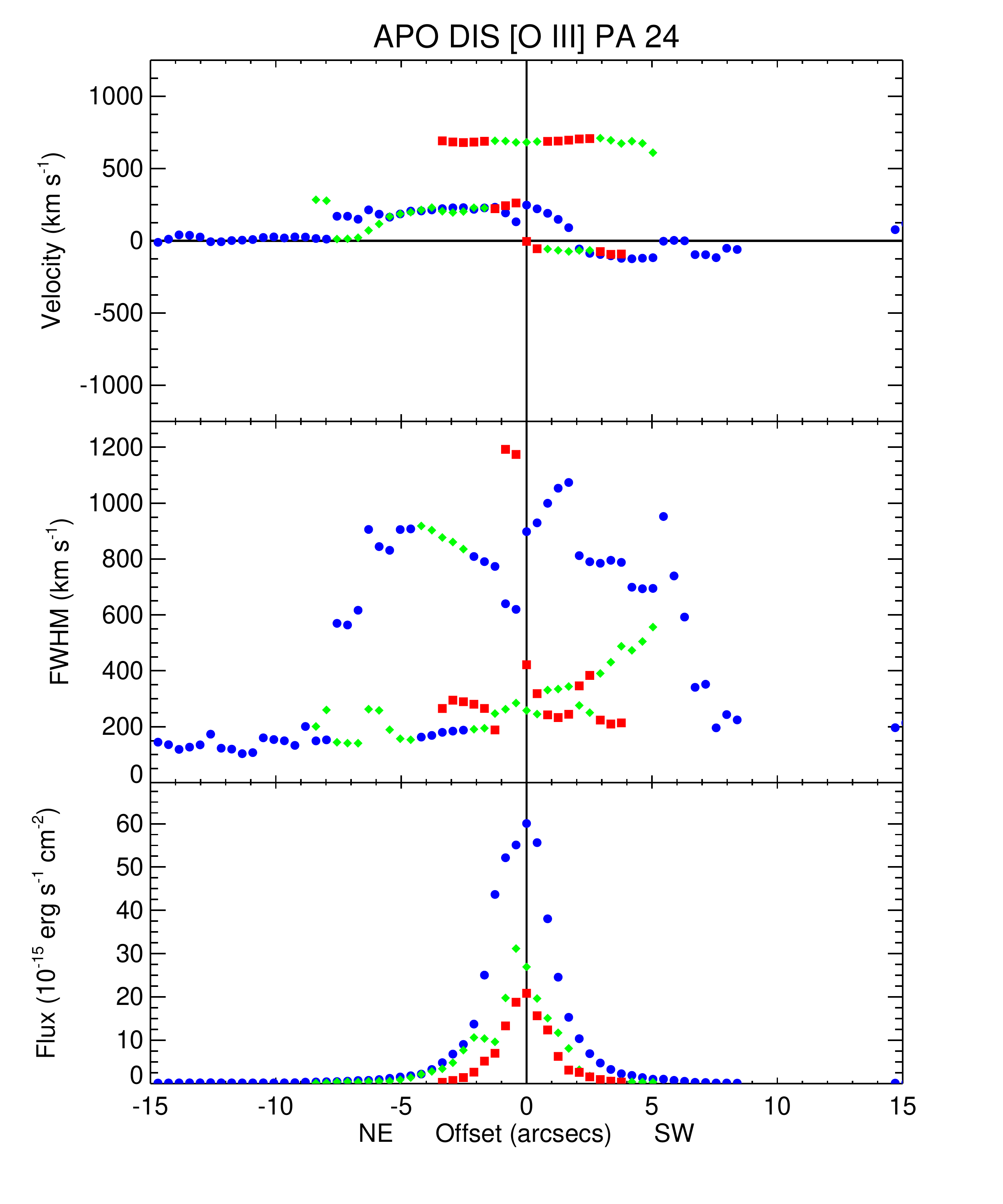}}
\subfigure{
\includegraphics[width=0.49\textwidth]{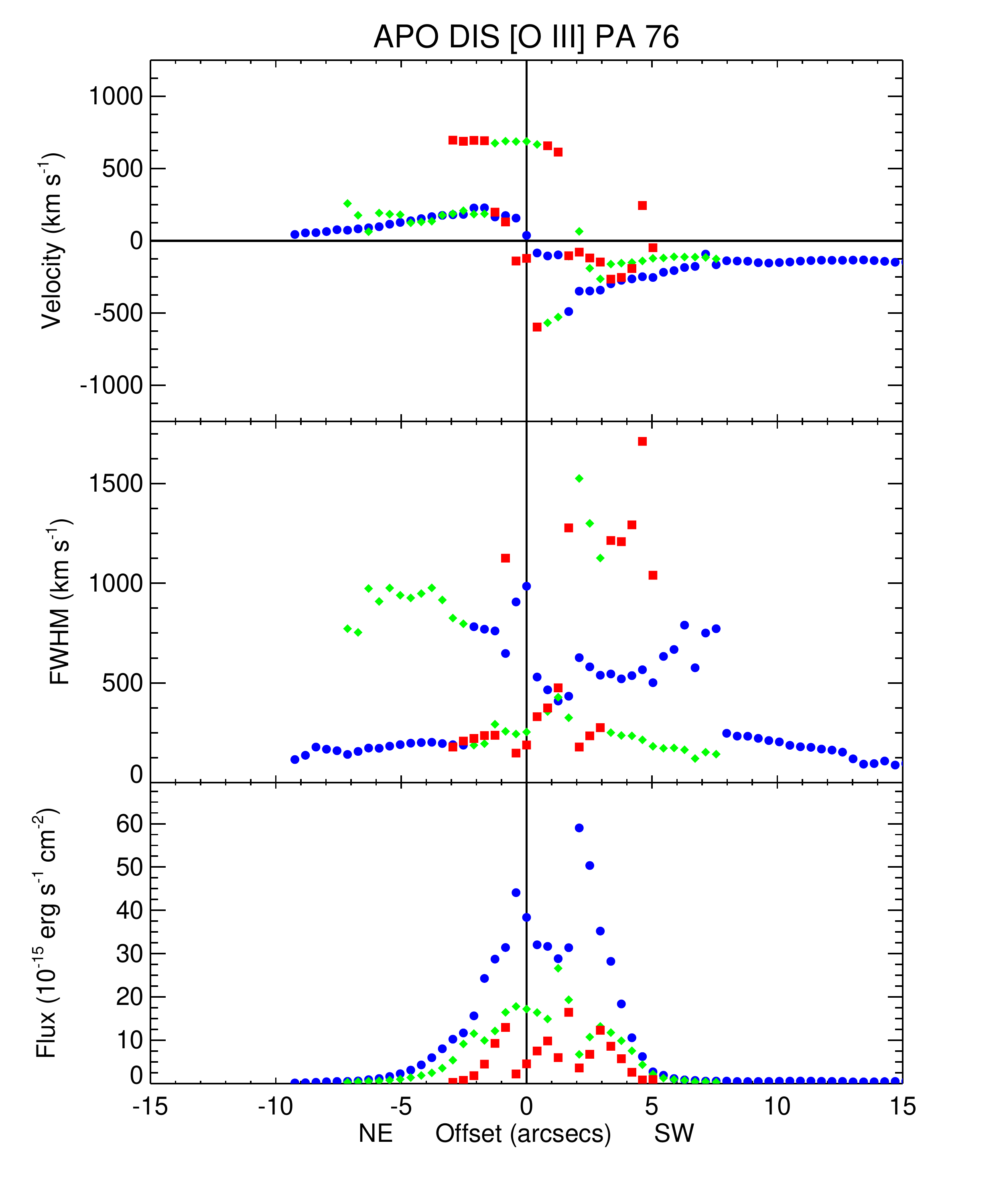}}
\subfigure{
\includegraphics[width=0.49\textwidth]{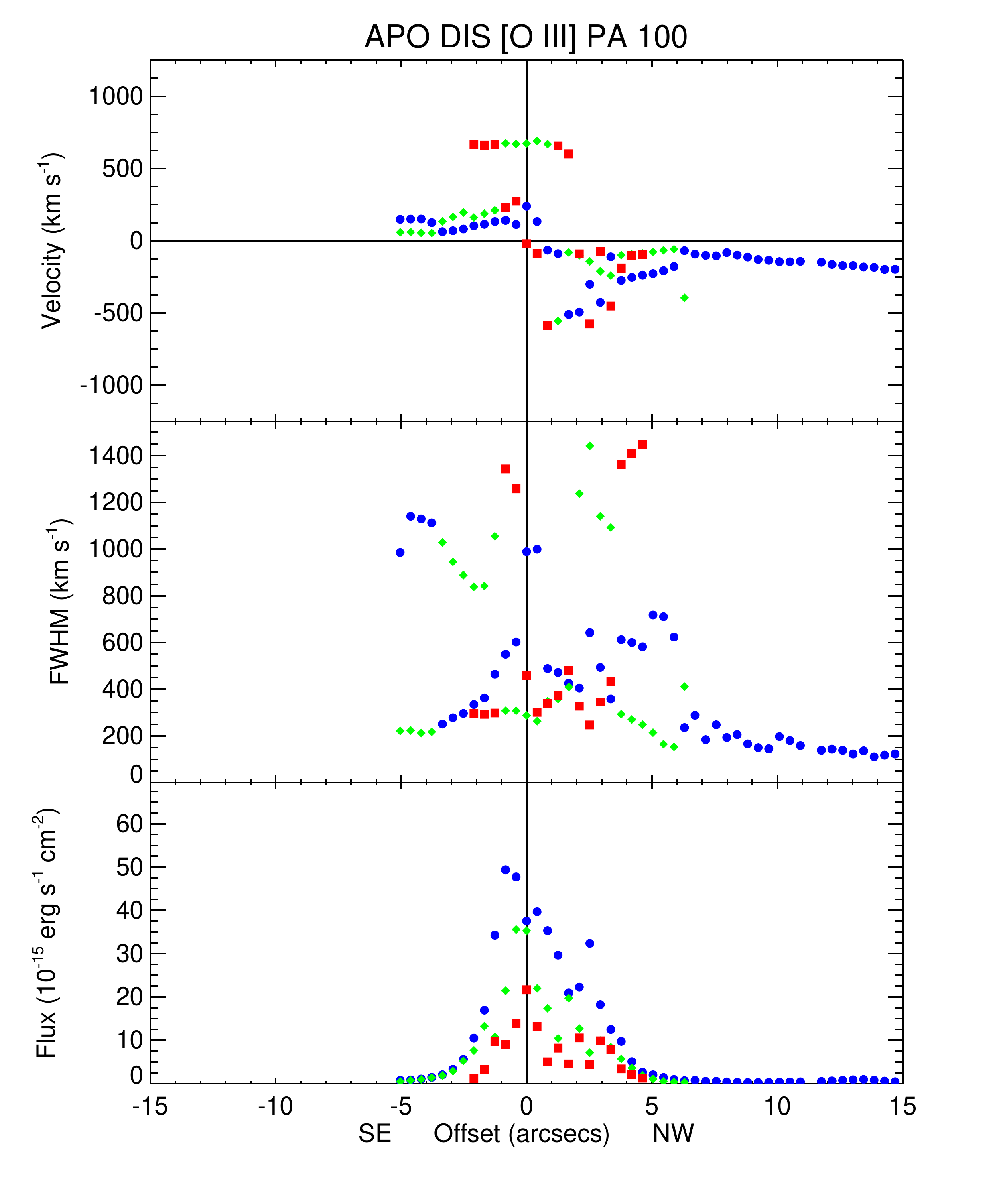}}
\subfigure{
\includegraphics[width=0.49\textwidth]{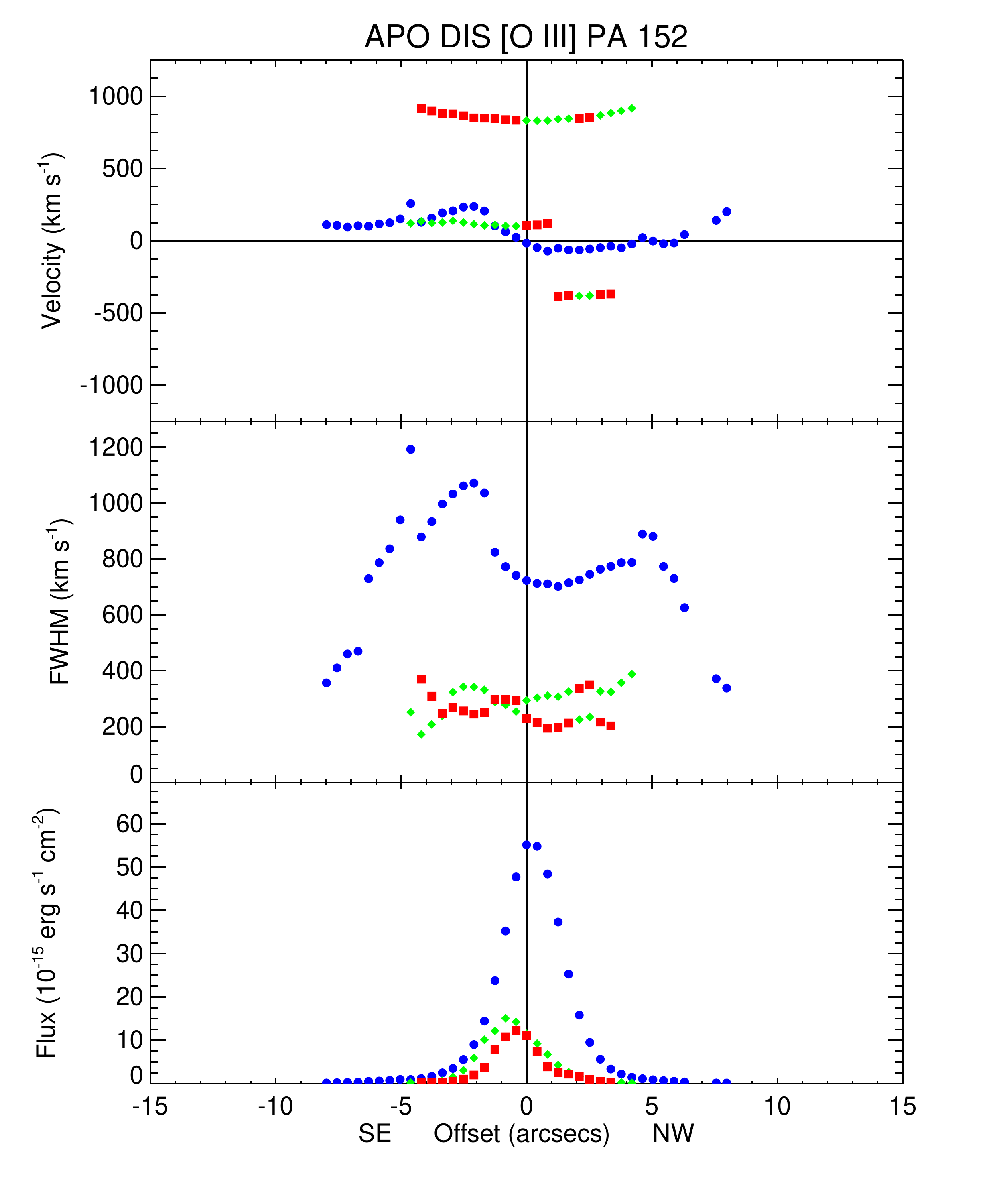}}
\vspace{-2em}
\caption{Mrk~78 observed velocity centroids (top), FWHM (middle), and integrated line fluxes (bottom) for the [O~III] $\lambda$5007 emission line in each of the four APO DIS long-slit observations. The points are color-coded from strongest to weakest peak flux in the order: blue circles, green diamonds, and red squares. Upper left to lower right are PA 24$\degr$, 76$\degr$, 100$\degr$, and 152$\degr$ east of north.}
\label{apokinematics}
\end{figure*}

In \cite{Fischer2011}, we used \hst STIS G430M observations of \othree to explore the NLR kinematics of Mrk~78. The four slit positions are labeled A, B, C, and D in Figure~\ref{structure}, and additional details are given in \cite{Fischer2011}. We expand the analysis for Mrk~78 by fitting the \hst STIS G750M observations to determine the H$\alpha$ kinematics for the same slit positions. We present velocity maps of the kinematic components in Figure~\ref{velmaps}, and plots of the gas velocity centroids, line widths, and fluxes for the \hst and APO observations in Figures~\ref{hstkinematics} and \ref{apokinematics}. There are a maximum of three kinematic components at each spatial location.

Overall, the \hst STIS H$\alpha$ kinematics presented in Figures~\ref{velmaps} and \ref{hstkinematics} show excellent agreement with the [O~III] kinematics in \cite{Fischer2011}. Variations in velocity, line width, and flux between the individual slits indicate the clumpy and inhomogeneous nature of the gas and outflows. High-velocity outflows reaching $\sim$1000\kms are observed in all slits to radial extents of $\sim$1$\arcsec$ ($\sim$750 pc), with moderate-velocity outflows up to $\sim$500 km s$^{-1}$ reaching out to $\sim$3$\arcsec$ ($\sim$2.2 kpc) from the nucleus. The large full-width at half-maximum (FWHM) values in excess of $\sim$250\kms are consistent with outflows dominating the kinematics to at least 3$\arcsec$ from the central SMBH.

The APO \othree kinematics in Figures~\ref{velmaps} and \ref{apokinematics} are dominated by rotational motion between $\sim$5$\arcsec$ and $\sim$15$\arcsec$ ($\sim$4 -- 11~kpc) with FWHM $<$ 250 km s$^{-1}$. The observed velocity amplitudes are $\sim$100\kms along PAs of 76$\degr$ and 100$\degr$, both of which are close to the photometric major axis of $\sim 84\degr$. The velocity centroids are blueshifted in the west and redshifted in the east, and based on the above photometric major axis and inclination of the host galaxy disk (55$\degr$; \citealp{Schmitt2000}), the deprojected velocity amplitude of the galaxy's rotation is approximately 125 km s$^{-1}$.

The higher velocities and FWHM of the brightest \othree component inside of 5$\arcsec$ to 7$\arcsec$ (depending on the PA) are indicative of the outflow seen in the STIS data, although the velocity amplitudes near the center are somewhat reduced due to averaging over a much larger projected area within the APO DIS slits. Interestingly, the APO observations are able to isolate a high-velocity (600 to 900 km s$^{-1}$), lower flux redshifted component out to $\sim$5$\arcsec$ ($\sim$4 kpc) from the nucleus, and a similar (300 to 500 km s$^{-1}$) low flux blueshifted component. These low-flux, high-velocity components indicate that weak outflows extend to radial distances of $\sim$4 kpc, beyond those detected in the \hst observations. This is caused by the narrower slits and shorter integration times of the \hst spectroscopy, highlighting the effect of sensitivity on the determination of outflow extents \citep{Kang2018}. While these components have high velocities, they are significantly lower in flux than the primary outflow and rotational components, indicating a smaller contribution to the outflow mass and energy budget.

We describe the complex kinematics of the ionized gas in the NLR of Mrk~3 in \cite{Gnilka2020} based on \hst STIS, APO DIS, and Gemini Near-Infrared Integral Field Spectrometer (NIFS) observations. As shown in Figure~7 of \cite{Gnilka2020}, the \hst STIS [O~III] and H$\alpha$ radial velocities peak at blueshifted and redshifted values of $\sim$700\kms close ($\sim$0$\farcs$2) to the central SMBH, and return to systemic values at a distance of $\sim$1$\farcs$2 ($\sim$330~pc), similar to the pattern seen in Mrk~78. The NIFS observations cover the entire NLR over a span of 3$\arcsec$ and show emission line knots with a range of blueshifted and redshifted velocities up to $-$1200 and $+$1500 km s$^{-1}$, respectively. At larger radii, the APO kinematics show a transition from outflow to rotation between 1$\farcs$2 and 4$\arcsec$, and a rotational component thereafter in the ENLR that reaches up to $\sim$20$\arcsec$ ($\sim$5.4~kpc) from the nucleus.

\clearpage

\begin{deluxetable*}{lccccccccc}[ht!]
\tabletypesize{\normalsize}
\setlength{\tabcolsep}{0.125in} 
\tablecaption{Biconical Outflow Model Parameters}
\tablehead{
\colhead{Name} & \colhead{PA} & \colhead{Incl.} &\colhead{H.O.A.} & \colhead{$r_{turn}$} & \colhead{$v_{max}$} & \colhead{$r_{max}$} & \colhead{$v_{deproj}$} & \colhead{$r_{deproj}$} & \colhead{Refs. \vspace{-0.5em}} \\
\colhead{} & \colhead{(deg)} & \colhead{(deg)} &\colhead{(deg)} & \colhead{(pc)} & \colhead{(km s$^{-1}$)} & \colhead{(pc)} & \colhead{(near/far)} & \colhead{(near/far)} & \colhead{\vspace{-0.5em}}\\
\colhead{(1)} & \colhead{(2)} & \colhead{(3)} &\colhead{(4)} & \colhead{(5)} & \colhead{(6)} & \colhead{(7)} & \colhead{(8)} & \colhead{(9)} & \colhead{(10)}
}
\startdata
Mrk~3      & 71 & 5 (NE)  & 20     & 55  & 1400 & 330 & 2.37/3.86 & 1.10/1.04  & 1, 2 \\
Mrk~78     & 65 & 30 (SW) & 10, 35 & 900 & 1200 & 3300 & 1.10/3.86 & 2.37/1.04 & 3 \\
NGC~1068   & 30 & 5 (NE)   & 35     & 148 & 1300 & 435 & 1.56/2.00 & 1.30/1.16  & 4 \\
\enddata
\tablecomments{Columns are (1) target name, (2) position angle of the bicone axis, (3) inclination angle of the bicone axis from the plane of the sky (direction closest to viewer), (4) model half opening angle, (5) turnover distance of maximum velocity, (6) maximum space velocity, (7) maximum outflow distance, (8) the velocity and (9) distance deprojection factors, and (10) references for the original kinematic models: (1)~\citealp{Ruiz2001}, (2)~\citealp{Crenshaw2010}, (3)~\citealp{Fischer2011}, and (4)~\citealp{Das2005}.}
\label{biconeparams}
\vspace{-0em}
\end{deluxetable*}

The kinematics of the NLR in NGC~1068 are described in detail by \cite{Das2005}, based on an analysis of \othree emission in multiple parallel \hst STIS observations using the G430M grating. The blueshifted and redshifted radial velocities of the ionized gas increase from near systemic to $\sim$1000 -- 1500\kms at a projected distance of 1$\farcs$9 ($\sim$150~pc) from the SMBH, followed by a decline to systemic velocity at around 5$\farcs6$ ($\sim$435~pc). At larger distances, the radial velocities are near systemic out to at least 8$\arcsec$ ($\sim$580~pc) and display correspondingly low FWHM that are indicative of rotational kinematics.

\subsection{Biconical Outflow Models}

To determine the mass outflow parameters, we need the true space velocities and distances of the ionized gas from the SMBHs for each AGN. We can obtain these by adopting kinematic models of the outflows whose geometries determine deprojection factors for the velocities and distances. We adopt biconical outflow models derived in our previous studies of Mrk~3 \citep{Ruiz2001, Crenshaw2010}, Mrk~78 \citep{Fischer2011}, and NGC~1068 \citep{Das2005}. In these models, the outflows increase from zero velocity at the SMBH to a maximum value ($v_{max}$) at a turnover radius ($r_{turn}$), and then decline to the systemic (rotational) value at a maximum distance ($r_{max}$). The symmetric 3D biconical models are then sampled along the same PAs as the \hst STIS slits, and the model parameters are varied until a match is obtained with the overall trend of radial velocities. The bicones are hollow along their central axes and are defined by minimum and maximum half-opening angles (HOAs). 

\begin{figure*}[ht!]
\centering
\includegraphics[width=0.42\textwidth, angle=90]{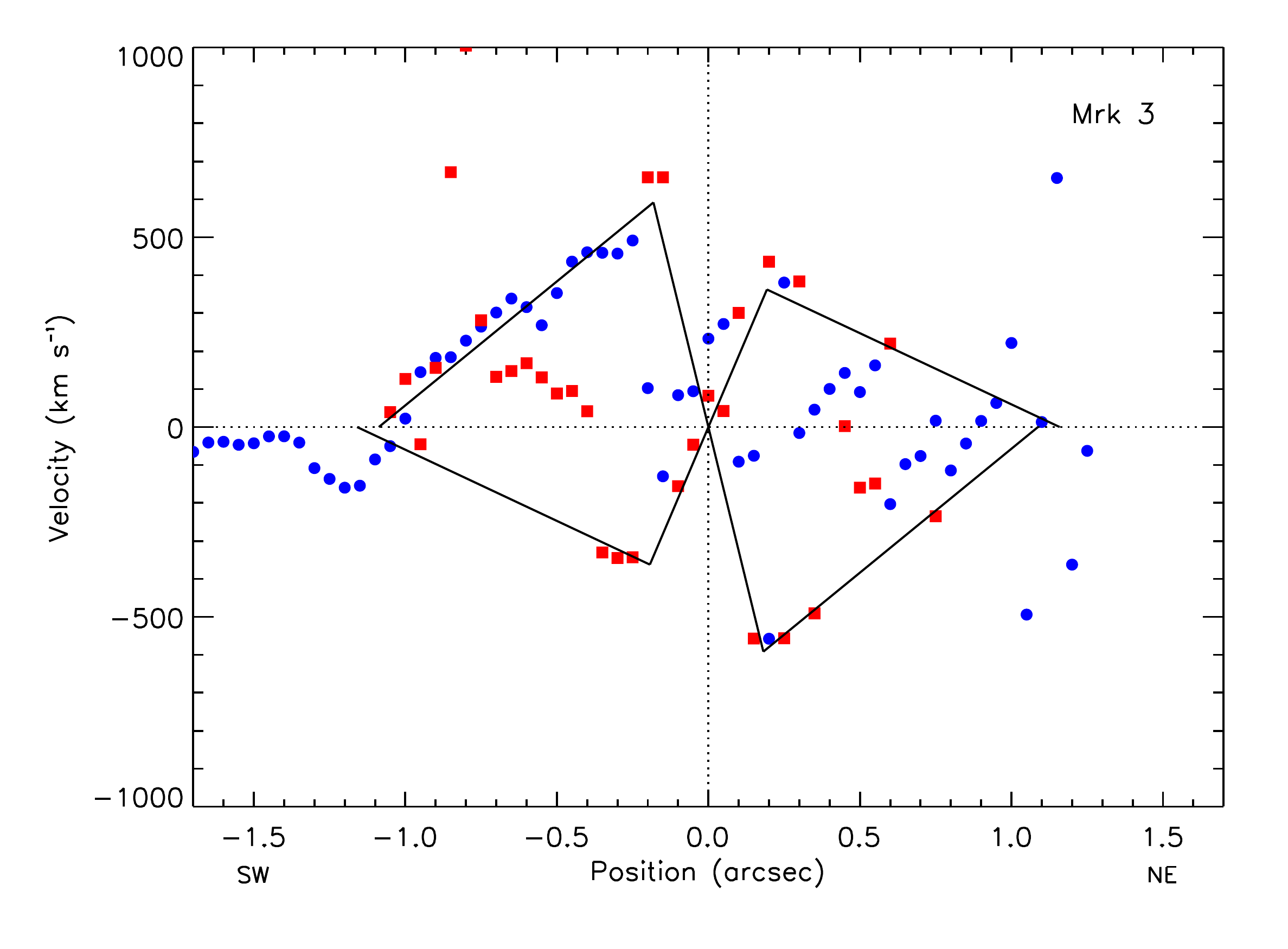}
\vspace{-2.2em}
\includegraphics[width=0.42\textwidth]{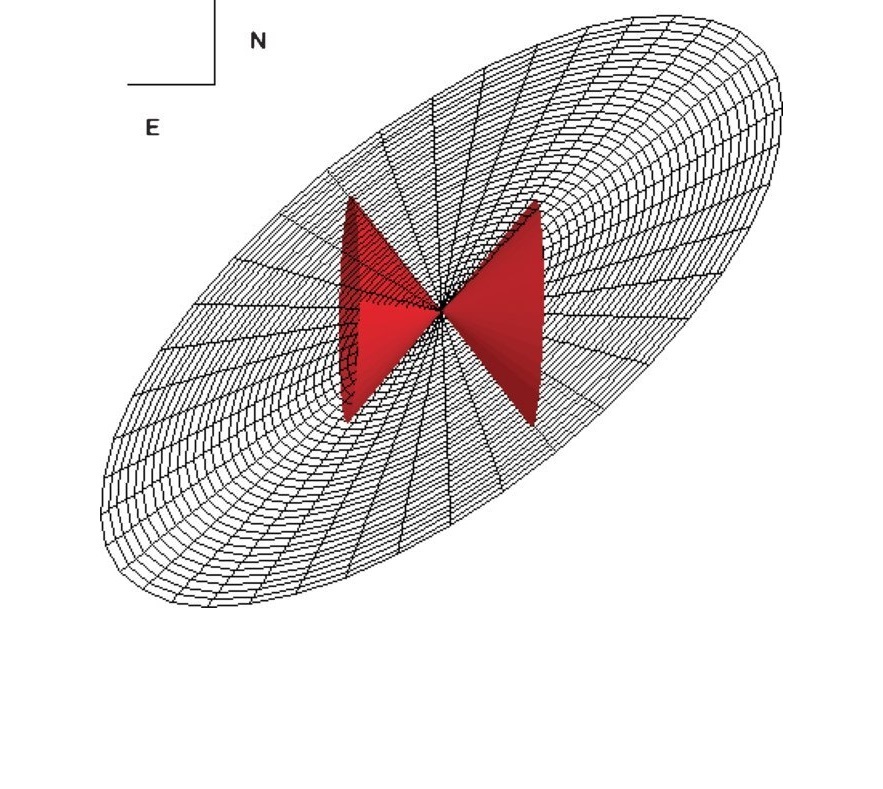}
\includegraphics[width=0.42\textwidth, angle=90]{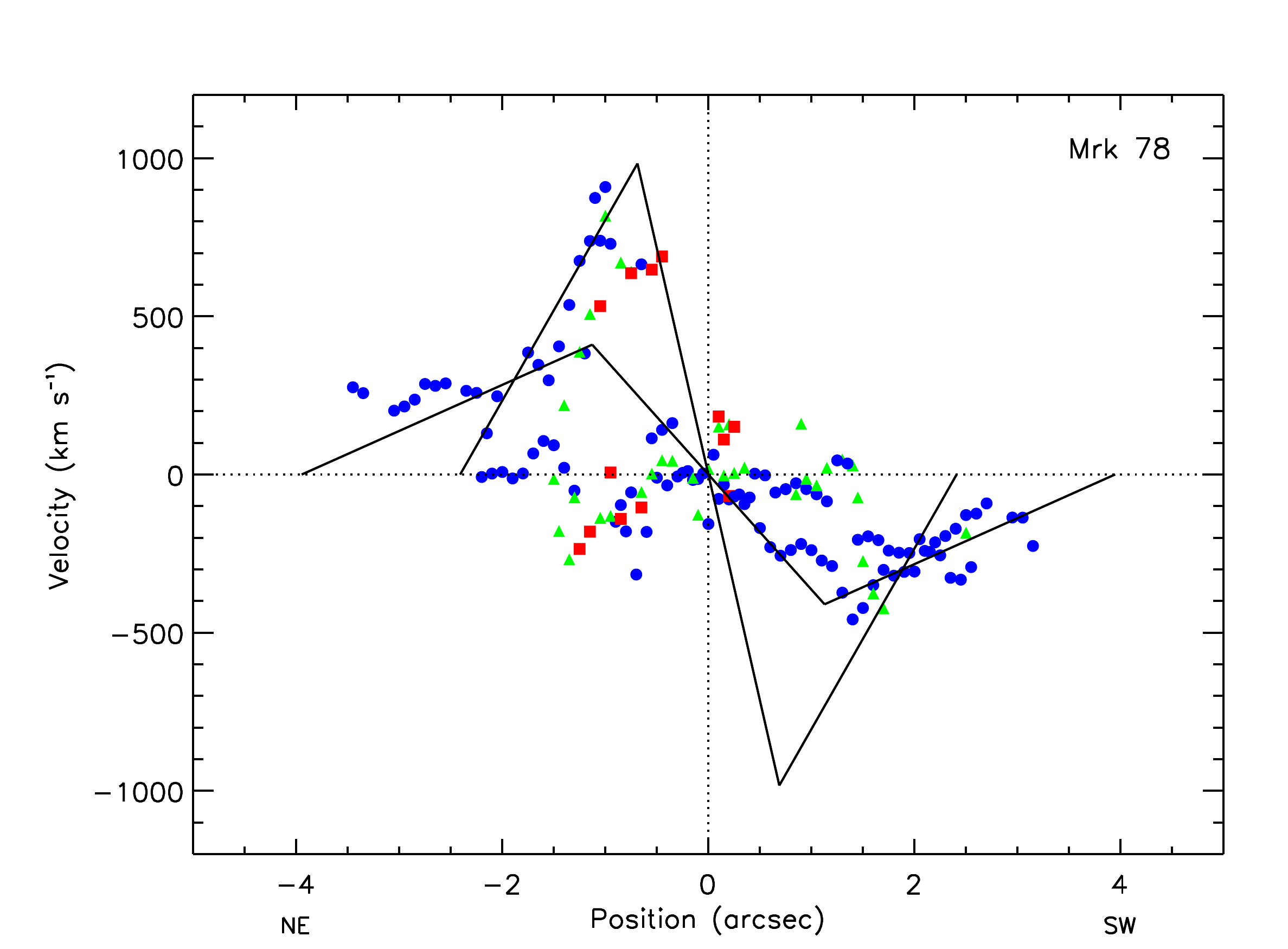}
\includegraphics[width=0.42\textwidth]{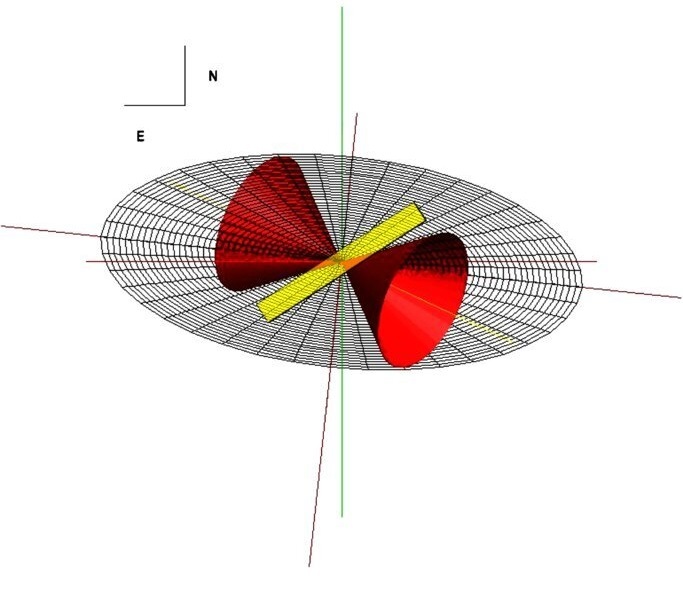}
\includegraphics[width=0.42\textwidth, angle=90]{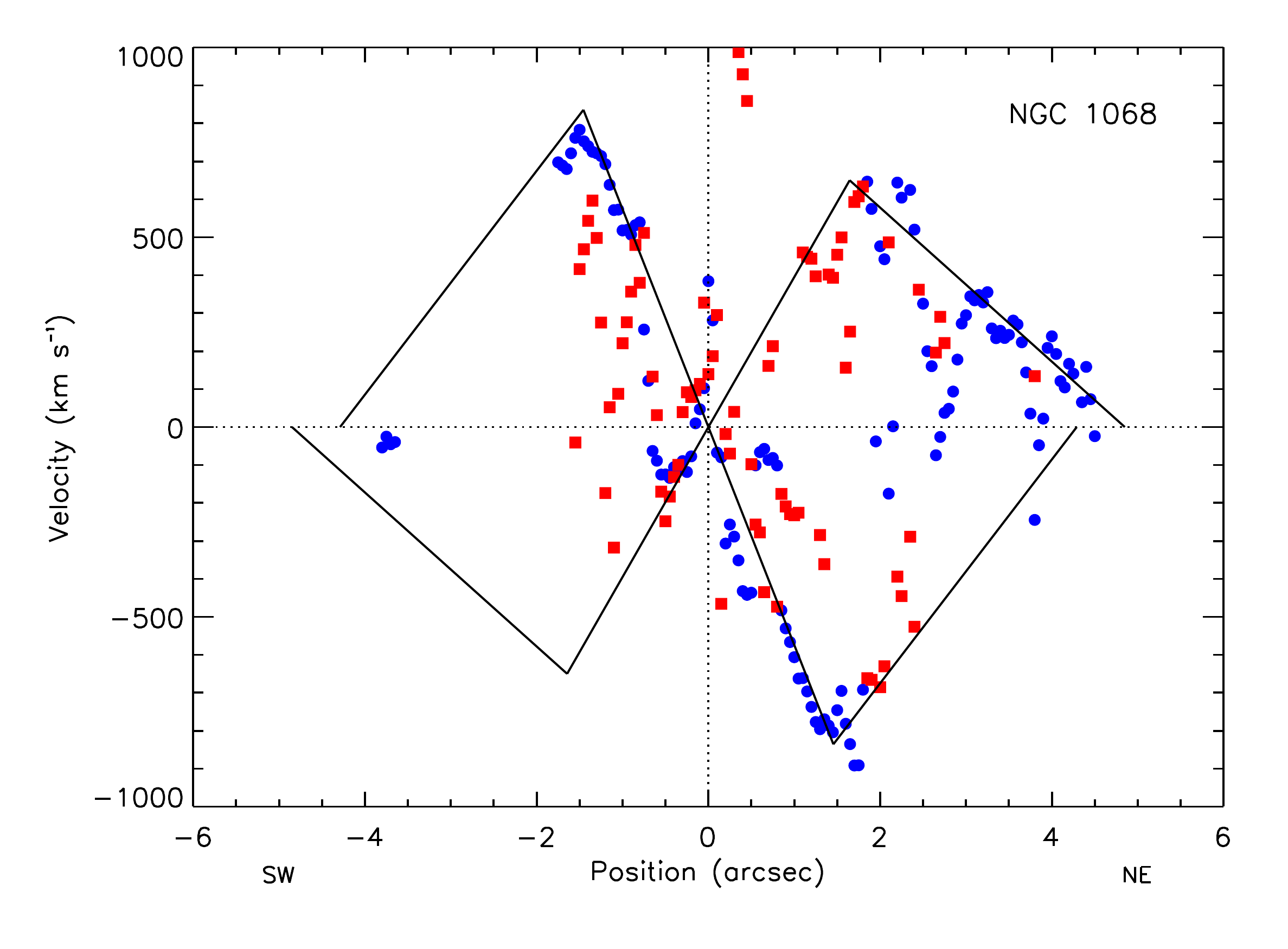}
\includegraphics[width=0.42\textwidth]{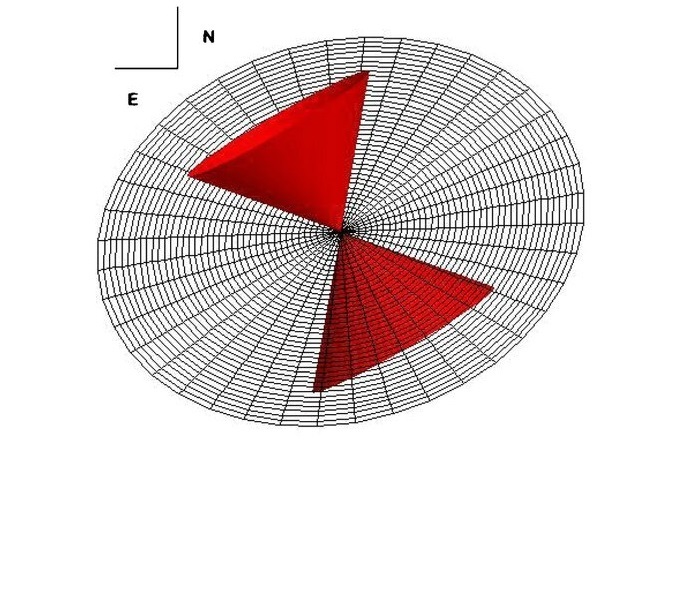}
\vspace{-2.8em}
\caption{Biconical outflow model fits (left, black lines) and geometries (right) for the observed gas kinematics of Mrk~3 (top), Mrk 78 (middle), and NGC~1068 (bottom). Radial velocities are color-coded as in Figures~\ref{hstkinematics} and \ref{apokinematics}. The geometric models highlight the outer opening angle of the ionized biconical outflows (red cones) and the host galaxy disk (black annuli) and have been reproduced from \cite{Crenshaw2010}, \cite{Fischer2011}, and \cite{Crenshaw2010b}. For Mrk~78, only slits A and D intersect the nucleus and are included in the figure, with slits B and C showing some points that partially match the more blue-shifted cone model line. Mrk~3 and NGC~1068 were observed with the low dispersion gratings and only required two kinematic components, with blue having the higher flux. Radial velocities are averaged within the photoionization model bins and matched to the nearest model line to obtain deprojection factors for the radial distances and velocities.}
\label{biconemodels}
\end{figure*}

In our previous studies of Mrk~573 \citep{Revalski2018a} and Mrk~34 \citep{Revalski2018b}, we adopted a variation of this model where the outflows within the ionizing bicone travel along the disk of the galaxy rather than the edges, which was based on evidence that they originate from ionized dust spirals in the host galaxy disk \citep{Fischer2017}. However, this variation is not appropriate for Mrk~3 or NGC~1068 because the nearly equal blueshifts and redshifts on either side of the SMBH are consistent with flows along the sides of a mostly hollow bicone (Figure~\ref{biconemodels}). In the case of Mrk~78, the STIS slit locations are close to the major axis of the galaxy disk (PA $=$ 84$\degr$) and the disk-flow model would result in observed radial velocities close to zero, in disagreement with the observations. Thus, we adopt our original biconical outflow models for this analysis.

For Mrk~3 and NGC~1068, we further simplify the models by choosing a single HOA between the minimum and maximum values that best fits the observed velocities, weighted by the high flux component, as shown by the straight lines in Figure~\ref{biconemodels}. Mrk~78 has a significantly higher inclination, and separate HOAs for the near and far sides of each cone provide a better fit. Knowing the input model values, we determine deprojection factors for the near and far sides of each bicone.

The outflow models are shown in Figure~\ref{biconemodels} and the geometric parameters are provided in Table~\ref{biconeparams}. Using these fits, we determine the deprojection factors for both velocity and distance from the center, average the radial velocities in each photoionization model bin (typically $0\farcs2 - 0\farcs3$ in width), and choose the nearest velocity law. This is straightforward for Mrk~3 and NGC~1068 because the redshifted and blueshifted points are well separated due to their low inclinations. For Mrk~78, the sides of each cone are difficult to separate, and we choose the low-amplitude velocity law for radii at $r >$ $2\arcsec$, the high-amplitude law for $r <$ $2\arcsec$, and a mean $r_{deproj}$ of 1.68.

\subsection{Emission Line Ratios}

We use the Gaussian fit parameters to calculate integrated emission line fluxes and their ratios relative to H$\beta$~$\lambda$4861. We sum the fluxes of the kinematic components at each radius for Mrk~78 because the biconical models are consistent with all of the gas outflowing in the STIS data, and the spectra generally display low S/N. Mrk~78 has four \hst STIS slits that show similar line ratios and we calculate a single set of flux-weighted average line ratios to simplify the photoionization modeling process. This prevents us from detecting differences in the physical conditions of the kinematic components, but is unavoidable given the low S/N of the spectra.

We calculate the flux-weighted average line ratios by first deriving the observed emission line ratios for each of the four \hst slits and correct them for reddening using the procedure described in \S3.3 of \cite{Revalski2018a}. We then deproject the distances along the slits to radial distances from the nucleus using the Pythagorean theorem,
\begin{equation}
D = \sqrt{(\delta N \times S)^2 + (\delta R)^2},
\end{equation}
\noindent
where $\delta N$ is the distance in pixels from the pixel closest to the nuclear continuum peak, $S$ is the \hst STIS plate scale of $0\farcs05078$ pixel$^{-1}$, and $\delta R$ is the offset distance of each slit from the nucleus provided in Table~\ref{data}. We then bin all of the measurements in $0\farcs2$ radial intervals together and calculate a single set of flux-weighted average line ratios and uncertainties for each radial distance.

These emission line ratios are given in Table~\ref{ratios} and the observed and reddening-corrected line ratios for the individual slits are available as online-only content. Unlike our previous targets, Mrk~78 displays a radial trend in the [O~III]/H$\beta$ ratios, decreasing by a factor of three from $\sim$18 to $\sim$6 in the \hst STIS observations. This indicates a change in the ionization state of the optical emission line gas across the spatial extent of the outflow.

In the nuclear region, emission lines with a range of ionization potentials (IPs) are detected, from neutral [O~I] to [Ne~V]. The number of detected lines and their fluxes decrease with increasing distance from the nucleus, which typically makes accurate modeling more difficult. This is alleviated by using a flux-weighted average of the four \hst STIS slits that intersect different emission line knots of various brightnesses. Despite the increased coverage afforded by four slits, the NLR emission is intrinsically weaker towards the northeast, resulting in larger uncertainties. The emission line ratios for Mrk~3 are given in Tables 2 -- 3 of \cite{Collins2005}, while for NGC~1068 they are provided in Tables 1 -- 2 of \cite{Kraemer2000b}. The measurements for these targets cover a similar swath of wavelengths, as well as emission lines in the UV portion of the spectrum.

\subsection{Emission Line Diagnostics}

\begin{figure*}
\centering
\includegraphics[width=0.50\textwidth]{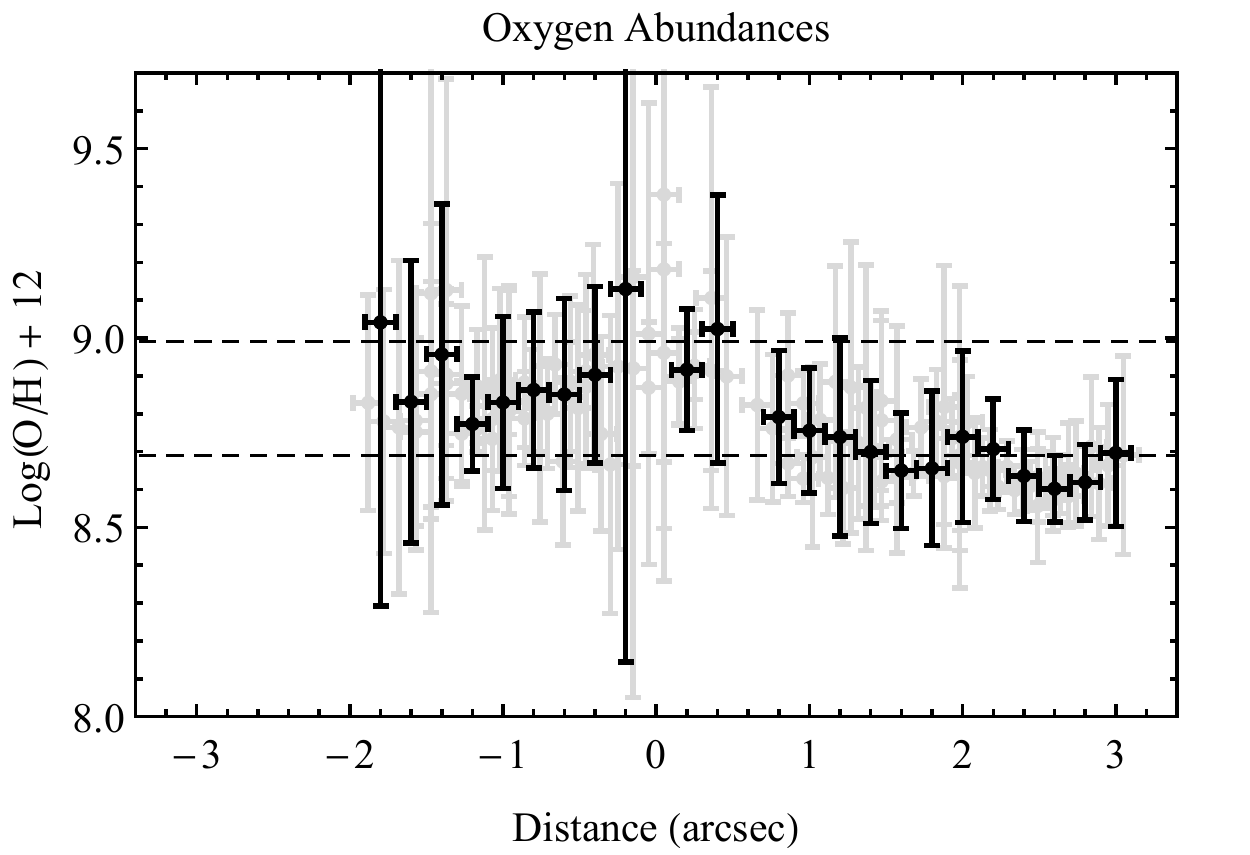}
\includegraphics[width=0.48\textwidth]{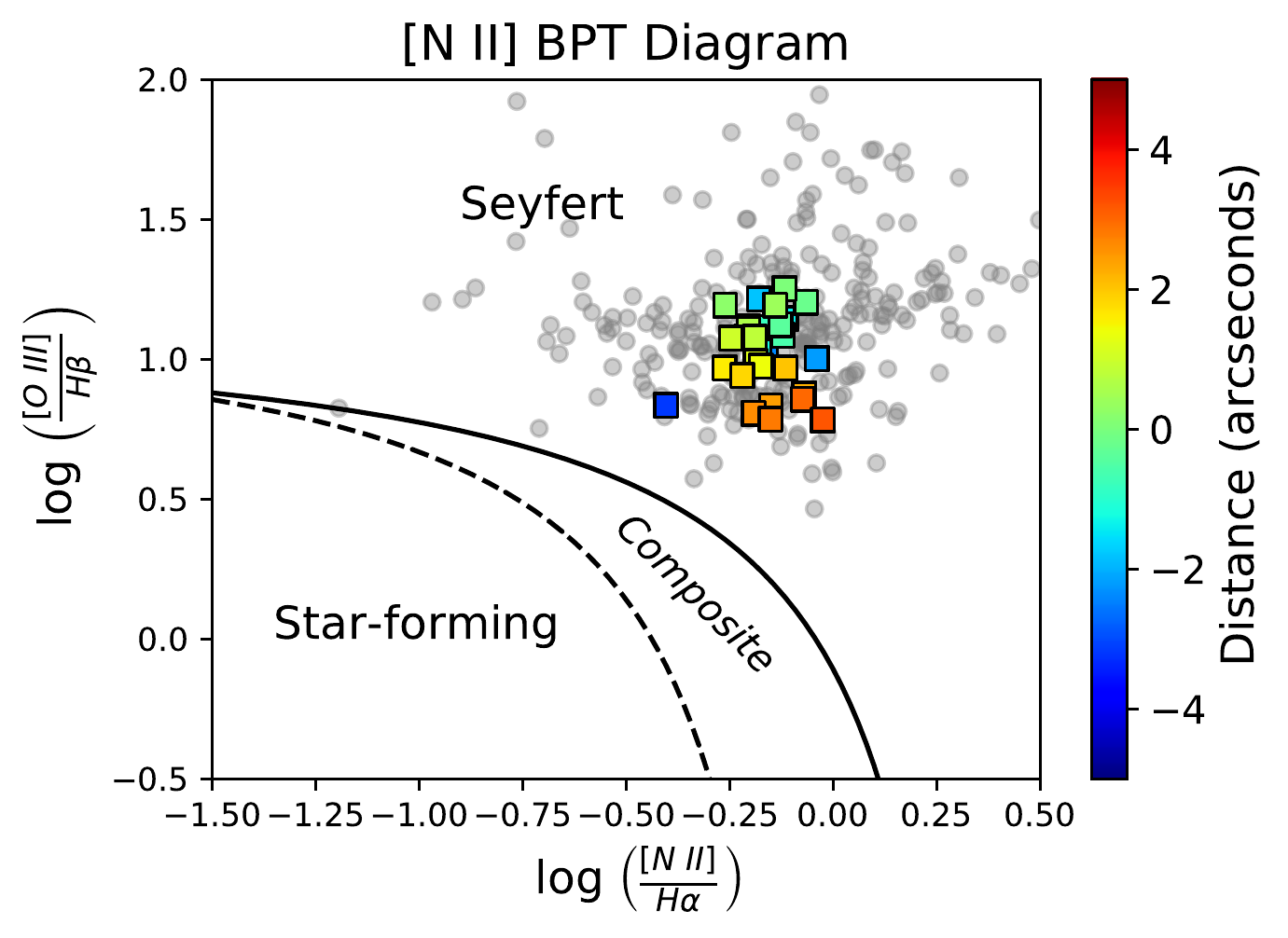}
\includegraphics[width=0.50\textwidth]{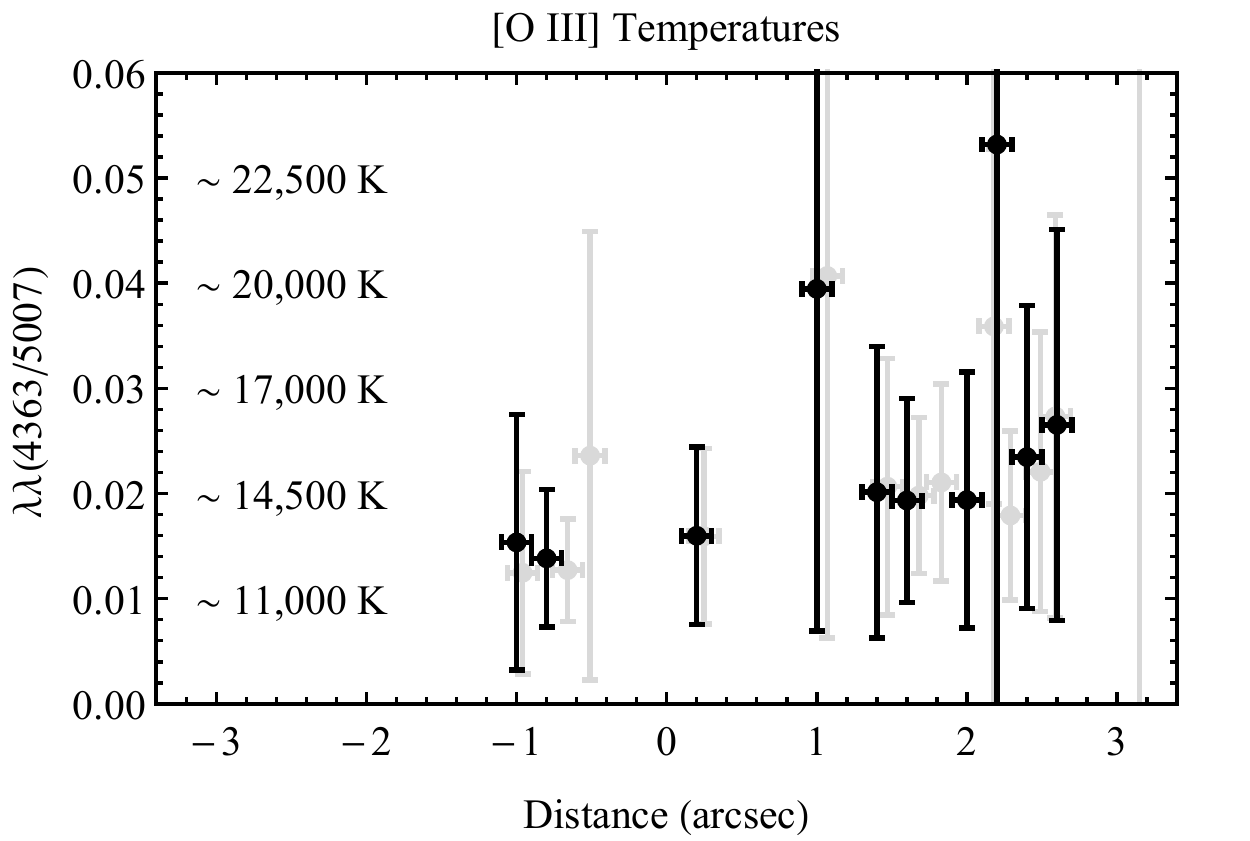}
\includegraphics[width=0.48\textwidth]{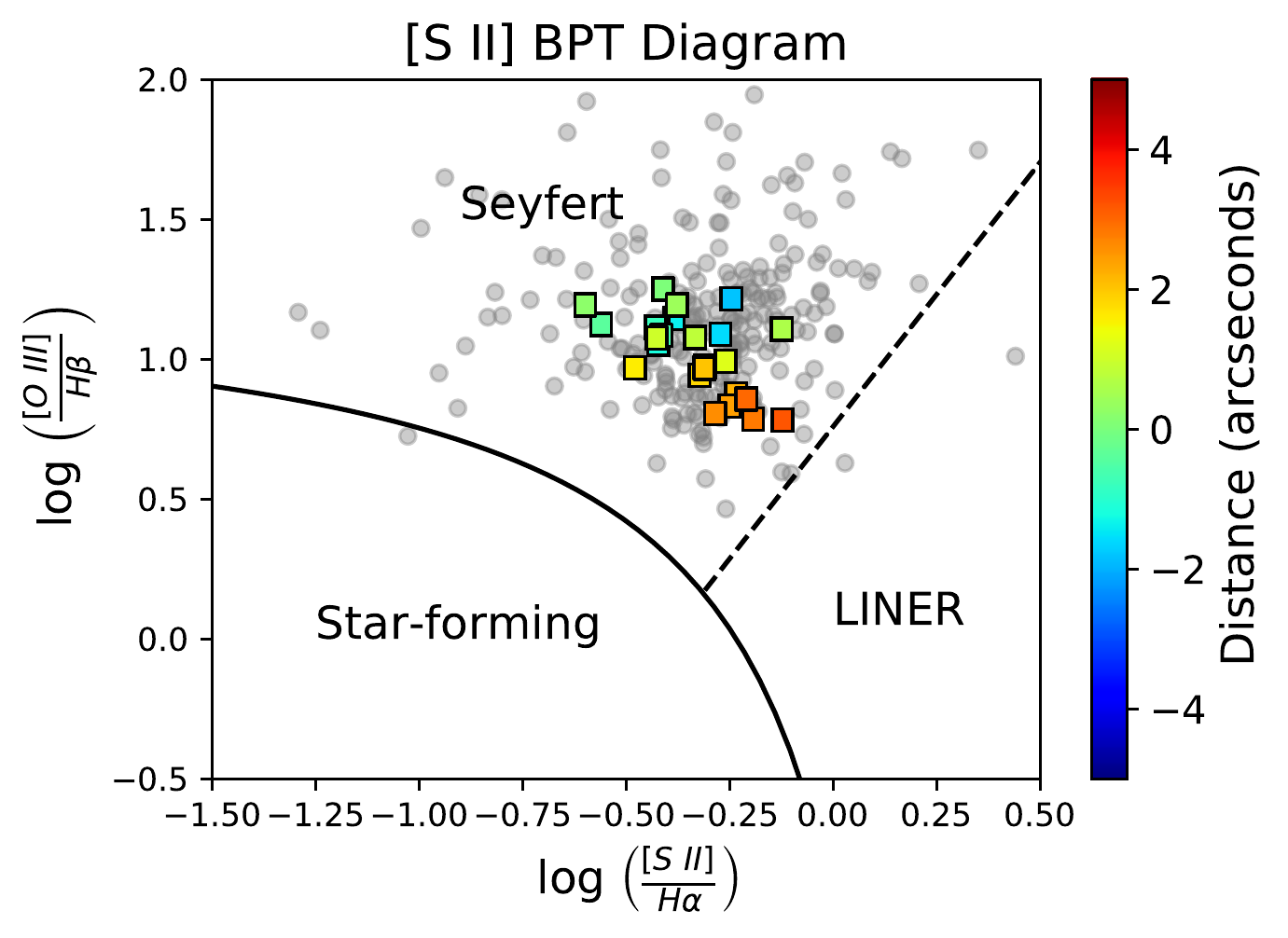}
\includegraphics[width=0.50\textwidth]{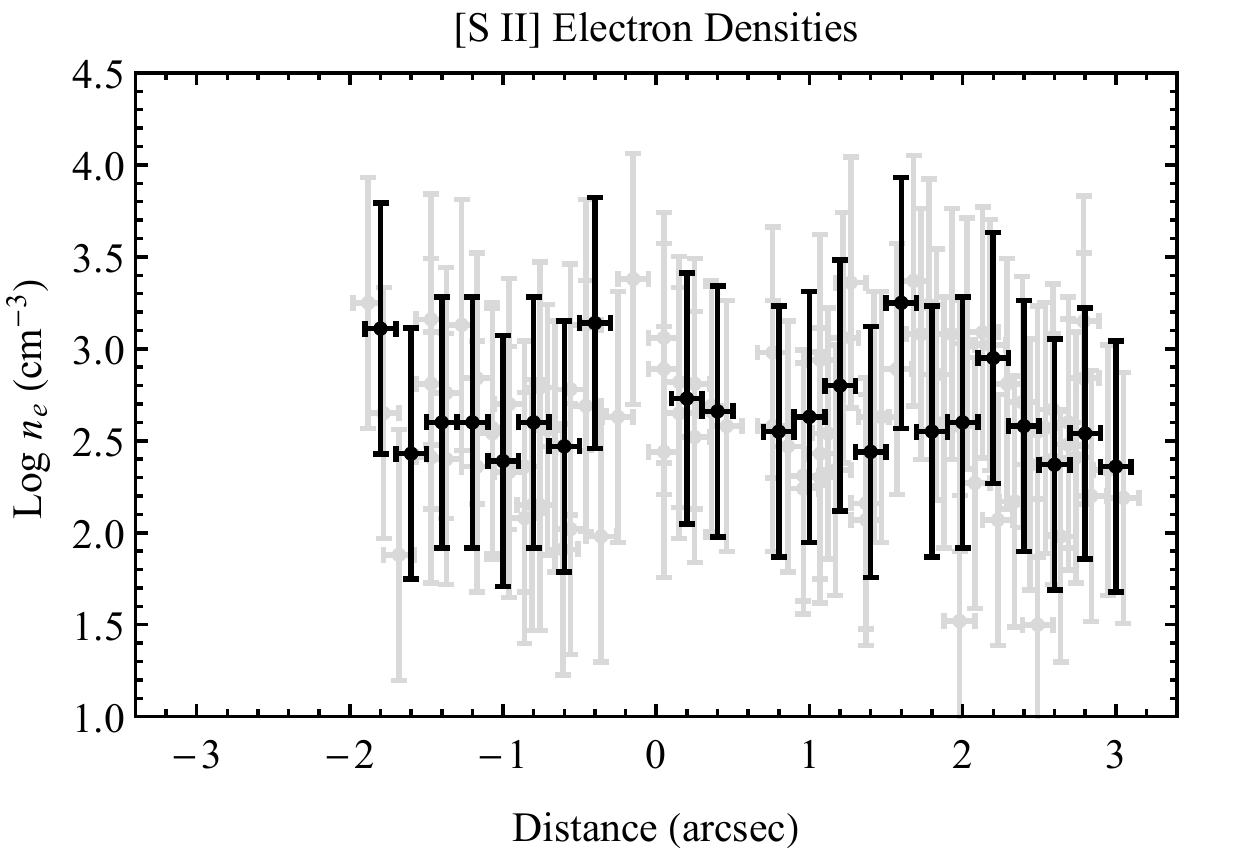}
\includegraphics[width=0.48\textwidth]{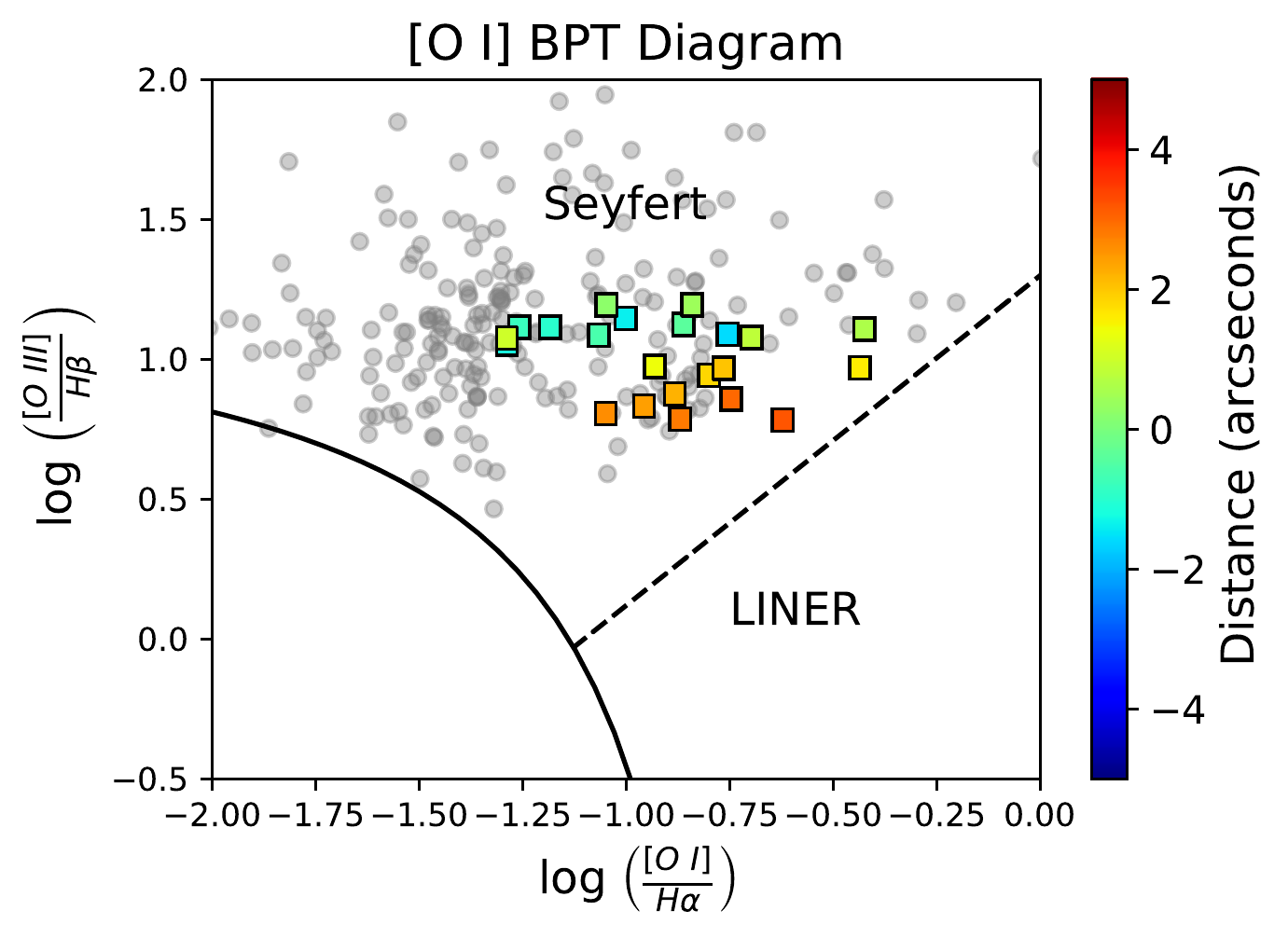}
\caption{Mrk~78 emission line diagnostic diagrams based on the line ratios in Table~\ref{ratios}. The left column displays the oxygen abundances (top), temperatures (middle), and densities (bottom) derived from the \hst STIS spectroscopy. The flux-weighted mean of all four slits at each deprojected distance is shown in black, and the dispersion of the individual slits at their observed positions along the slits are shown in light-gray. The lower and upper dashed lines in the abundance panel represent one and two times solar abundances, respectively. The right column displays the [N~II]~(top), [S~II]~(middle), and [O~I]~(bottom) BPT diagrams for the flux-weighted \hst STIS measurements (Table~\ref{ratios}; colored squares) and the APO observations for all four PAs (gray circles). The grouping of APO points in the left portion of the [O~I] diagram correspond primarily to measurements along the NLR minor axis. The demarcation lines for distinguishing ionization mechanisms are from \cite{Kewley2001, Kewley2006} and \cite{Kauffmann2003}. The left-to-right orientation matches that in Figure~\ref{structure} and all kinematic figures, with NE to the left (negative) and SW to the right (positive).}
\label{diagnostics}
\end{figure*}

We used the line ratios in Table~\ref{ratios} to create diagnostic diagrams that constrain the ionization, abundances, temperature, and density of the outflowing gas at each spatial location. In Figure~\ref{diagnostics}, we present BPT diagrams that differentiate sources of ionization by comparing lines with different ionization potentials and whose ratios vary significantly based on the spectral energy distribution (SED) of the ionizing source \citep{Baldwin1981, Veilleux1987}. At all spatial locations the results are consistent with AGN ionization, indicating the AGN's influence extends to radial distances of at least $\sim$12 kpc, in agreement with \cite{Kozlova2020}. The agreement of the \hst and APO observations indicates that AGN ionization dominates on small and large scales without localized contributions to the ionization from jets or shocks. Interestingly, the [O~III]/H$\beta$ ratios are $\sim$18 in the nucleus and steadily decrease to $\sim$6 at $\pm~3\arcsec$.

Next, we calculate the oxygen abundance using Equation (2) from \cite{Storchi-Bergmann1998} and a reference solar value ($Z_{\odot}$) of log(O/H)+12 = 8.69 \citep{Asplund2009}. As shown in the top-left panel of Figure~\ref{diagnostics}, the abundance of oxygen is $\sim$2\zsun in the nuclear regions and steadily decreases to solar values toward the west while remaining approximately constant at $\sim$1.4\zsun in the east. These radial abundance variations are in agreement with the study by \cite{Rosario2007} and the adopted average is in general agreement with \cite{Dors2020}.

The gas temperature is calculated self-consistently in photoionization models and thus observational constraints are useful for checking the validity of our modeling results. We use the \othree emission lines to derive the gas temperatures shown in the mid-left panel of Figure~\ref{diagnostics} and find typical NLR values of $\sim$10,000 -- 15,000~K \citep{Osterbrock2006}. The average gas temperature is higher toward the west; however, the weak \othree$\lambda$4363 emission line introduces large uncertainties.

Finally, the goal of our modeling process is to accurately determine the gas density and thus mass, so we derive the electron density profile from the [S~II]~$\lambda\lambda$6716/6731 line ratio. This doublet traces the low-ionization gas and is useful for that component of our photoionization models, with the \othree emitting gas typically less dense. The [S~II] densities are shown in the lower-left panel of Figure~\ref{diagnostics} and are approximately constant, except for a small decrease at the furthest west extent. The large uncertainties are driven by the S/N of the doublet in the spectra, compounded by variations in the ratio between slits that are averaged over when combining into a flux-weighted average. This is encapsulated by the dispersion of the gray points in Figure~\ref{diagnostics}, which show the densities for the individual \hst slits.

\movetabledown=2.3in
\begin{rotatetable*}
\begin{deluxetable*}{|c|c|c|c|hhhc|c|c|c|hc|c|hhc|c|c|c|c|}
\def\arraystretch{1.25}
\tabletypesize{\scriptsize}
\setlength{\tabcolsep}{0.015in} 
\tablecaption{Mrk~78 \hst STIS Flux-weighted Average Emission Line Ratios}
\tablewidth{0pt}
\tablehead{
\colhead{Position} & \colhead{[Ne~V] $\lambda$3426} & \colhead{[O~II] $\lambda$3727} & \colhead{[Ne~III] $\lambda$3869} & \nocolhead{[Ne~III] $\lambda$3969} & \nocolhead{[S~II] $\lambda$4074} & \nocolhead{H$\delta$ $\lambda$4102} & \colhead{H$\gamma$ $\lambda$4340} & \colhead{[O~III] $\lambda$4363} & \colhead{He~II $\lambda$4686} & \colhead{H$\beta$ $\lambda$4861} & \nocolhead{[O~III] $\lambda$4959} & \colhead{[O~III] $\lambda$5007} & \colhead{[O~I] $\lambda$6300} & \nocolhead{[O~I] $\lambda$6363} & \nocolhead{[N~II] $\lambda$6548} & \colhead{H$\alpha$ $\lambda$6563} & \colhead{[N~II] $\lambda$6584} & \colhead{[S~II] $\lambda$6716} & \colhead{[S~II] $\lambda$6731} & \colhead{H$\beta$ Flux}}
\startdata
-2.00	&	2.48	$\pm$	2.46	&	2.71	$\pm$	1.94	&	1.05	$\pm$	0.76	&	...	$\pm$	...	&	...	$\pm$	...	&	...	$\pm$	...	&	...	$\pm$	...	&	...	$\pm$	...	&	...	$\pm$	...	&	1.00	$\pm$	0.42	&	4.05	$\pm$	1.93	&	12.04	$\pm$	5.74	&	...	$\pm$	...	&	...	$\pm$	...	&	...	$\pm$	...	&	2.90	$\pm$	2.52	&	1.79	$\pm$	1.60	&	...	$\pm$	...	&	2.61	$\pm$	2.61	&	0.47	$\pm$	0.20	\\ \relax
-1.80	&	1.49	$\pm$	1.49	&	4.84	$\pm$	4.56	&	0.88	$\pm$	0.88	&	1.20	$\pm$	0.84	&	...	$\pm$	...	&	...	$\pm$	...	&	...	$\pm$	...	&	...	$\pm$	...	&	...	$\pm$	...	&	1.00	$\pm$	0.70	&	5.60	$\pm$	4.59	&	16.43	$\pm$	13.48	&	...	$\pm$	...	&	...	$\pm$	...	&	0.75	$\pm$	0.54	&	2.87	$\pm$	2.75	&	1.90	$\pm$	1.74	&	0.68	$\pm$	0.68	&	0.95	$\pm$	0.95	&	1.37	$\pm$	0.96	\\ \relax
-1.60	&	...	$\pm$	...	&	3.46	$\pm$	2.73	&	1.31	$\pm$	1.31	&	0.61	$\pm$	0.61	&	0.78	$\pm$	0.78	&	...	$\pm$	...	&	0.50	$\pm$	0.50	&	...	$\pm$	...	&	0.31	$\pm$	0.31	&	1.00	$\pm$	0.54	&	4.17	$\pm$	2.48	&	12.30	$\pm$	7.32	&	0.51	$\pm$	0.51	&	...	$\pm$	...	&	0.70	$\pm$	0.49	&	2.90	$\pm$	2.19	&	2.03	$\pm$	1.55	&	0.81	$\pm$	0.80	&	0.74	$\pm$	0.70	&	2.81	$\pm$	1.53	\\ \relax
-1.40	&	1.05	$\pm$	0.82	&	5.25	$\pm$	3.72	&	2.10	$\pm$	1.87	&	0.56	$\pm$	0.52	&	...	$\pm$	...	&	...	$\pm$	...	&	0.57	$\pm$	0.41	&	...	$\pm$	...	&	0.56	$\pm$	0.30	&	1.00	$\pm$	0.45	&	4.80	$\pm$	2.38	&	13.99	$\pm$	6.94	&	0.29	$\pm$	0.29	&	...	$\pm$	...	&	0.81	$\pm$	0.55	&	2.90	$\pm$	1.82	&	2.24	$\pm$	1.47	&	0.60	$\pm$	0.55	&	0.60	$\pm$	0.60	&	13.82	$\pm$	6.25	\\ \relax
-1.20	&	0.61	$\pm$	0.52	&	2.80	$\pm$	0.78	&	0.99	$\pm$	0.33	&	0.43	$\pm$	0.24	&	...	$\pm$	...	&	0.66	$\pm$	0.38	&	0.39	$\pm$	0.25	&	...	$\pm$	...	&	0.33	$\pm$	0.14	&	1.00	$\pm$	0.19	&	3.84	$\pm$	0.84	&	11.36	$\pm$	2.50	&	0.15	$\pm$	0.15	&	...	$\pm$	...	&	0.66	$\pm$	0.20	&	2.90	$\pm$	0.90	&	1.93	$\pm$	0.59	&	0.55	$\pm$	0.26	&	0.55	$\pm$	0.26	&	10.30	$\pm$	2.00	\\ \relax
-1.00	&	0.62	$\pm$	0.48	&	3.14	$\pm$	1.97	&	1.71	$\pm$	0.87	&	0.49	$\pm$	0.20	&	...	$\pm$	...	&	...	$\pm$	...	&	0.49	$\pm$	0.32	&	0.20	$\pm$	0.14	&	0.38	$\pm$	0.22	&	1.00	$\pm$	0.31	&	4.45	$\pm$	1.64	&	13.03	$\pm$	4.81	&	0.19	$\pm$	0.19	&	...	$\pm$	...	&	0.63	$\pm$	0.27	&	2.90	$\pm$	1.31	&	1.81	$\pm$	0.81	&	0.58	$\pm$	0.23	&	0.52	$\pm$	0.33	&	17.56	$\pm$	5.40	\\ \relax
-0.80	&	0.75	$\pm$	0.38	&	3.30	$\pm$	1.31	&	1.36	$\pm$	0.77	&	0.57	$\pm$	0.22	&	...	$\pm$	...	&	0.31	$\pm$	0.18	&	0.45	$\pm$	0.22	&	0.18	$\pm$	0.06	&	0.31	$\pm$	0.07	&	1.00	$\pm$	0.27	&	4.43	$\pm$	1.50	&	13.01	$\pm$	4.38	&	0.16	$\pm$	0.16	&	...	$\pm$	...	&	0.69	$\pm$	0.25	&	2.90	$\pm$	1.06	&	2.02	$\pm$	0.73	&	0.54	$\pm$	0.26	&	0.54	$\pm$	0.26	&	16.09	$\pm$	4.28	\\ \relax
-0.60	&	0.52	$\pm$	0.52	&	2.09	$\pm$	0.98	&	0.93	$\pm$	0.58	&	0.35	$\pm$	0.19	&	...	$\pm$	...	&	...	$\pm$	...	&	0.28	$\pm$	0.16	&	...	$\pm$	...	&	0.27	$\pm$	0.13	&	1.00	$\pm$	0.36	&	4.12	$\pm$	1.70	&	12.18	$\pm$	5.02	&	0.25	$\pm$	0.25	&	...	$\pm$	...	&	0.75	$\pm$	0.36	&	2.90	$\pm$	1.38	&	2.20	$\pm$	1.06	&	0.58	$\pm$	0.36	&	0.54	$\pm$	0.32	&	7.33	$\pm$	2.66	\\ \relax
-0.40	&	1.48	$\pm$	1.48	&	2.64	$\pm$	1.01	&	1.25	$\pm$	0.74	&	...	$\pm$	...	&	...	$\pm$	...	&	...	$\pm$	...	&	1.09	$\pm$	0.76	&	...	$\pm$	...	&	...	$\pm$	...	&	1.00	$\pm$	0.29	&	4.57	$\pm$	1.46	&	13.30	$\pm$	4.24	&	0.40	$\pm$	0.40	&	...	$\pm$	...	&	0.74	$\pm$	0.31	&	2.90	$\pm$	1.20	&	2.17	$\pm$	0.90	&	0.33	$\pm$	0.25	&	0.47	$\pm$	0.33	&	10.93	$\pm$	3.12	\\ \relax
-0.20	&	...	$\pm$	...	&	3.36	$\pm$	3.36	&	5.00	$\pm$	5.00	&	...	$\pm$	...	&	...	$\pm$	...	&	...	$\pm$	...	&	...	$\pm$	...	&	...	$\pm$	...	&	...	$\pm$	...	&	1.00	$\pm$	0.92	&	5.67	$\pm$	5.67	&	16.00	$\pm$	16.00	&	...	$\pm$	...	&	...	$\pm$	...	&	0.87	$\pm$	0.87	&	2.90	$\pm$	2.90	&	2.50	$\pm$	2.50	&	...	$\pm$	...	&	...	$\pm$	...	&	25.04	$\pm$	22.95	\\ \relax
$\pm$0.10	&	...	$\pm$	...	&	1.50	$\pm$	1.50	&	1.20	$\pm$	1.20	&	...	$\pm$	...	&	...	$\pm$	...	&	...	$\pm$	...	&	...	$\pm$	...	&	...	$\pm$	...	&	...	$\pm$	...	&	1.00	$\pm$	0.55	&	6.15	$\pm$	3.66	&	17.85	$\pm$	10.64	&	...	$\pm$	...	&	...	$\pm$	...	&	0.76	$\pm$	0.53	&	2.90	$\pm$	2.01	&	2.22	$\pm$	1.53	&	0.48	$\pm$	0.41	&	0.65	$\pm$	0.56	&	2.85	$\pm$	1.57	\\ \relax
0.20	&	1.05	$\pm$	0.46	&	1.62	$\pm$	0.43	&	1.49	$\pm$	0.50	&	0.31	$\pm$	0.22	&	...	$\pm$	...	&	...	$\pm$	...	&	0.38	$\pm$	0.18	&	0.25	$\pm$	0.12	&	0.40	$\pm$	0.15	&	1.00	$\pm$	0.16	&	5.36	$\pm$	1.20	&	15.64	$\pm$	3.49	&	0.26	$\pm$	0.15	&	0.09	$\pm$	0.04	&	0.55	$\pm$	0.13	&	2.90	$\pm$	0.71	&	1.59	$\pm$	0.38	&	0.35	$\pm$	0.12	&	0.38	$\pm$	0.12	&	32.90	$\pm$	5.42	\\ \relax
0.40	&	1.24	$\pm$	0.55	&	5.38	$\pm$	3.19	&	3.12	$\pm$	2.00	&	...	$\pm$	...	&	...	$\pm$	...	&	...	$\pm$	...	&	0.63	$\pm$	0.63	&	...	$\pm$	...	&	0.28	$\pm$	0.16	&	1.00	$\pm$	0.35	&	5.45	$\pm$	2.20	&	15.65	$\pm$	6.31	&	0.42	$\pm$	0.36	&	0.11	$\pm$	0.07	&	0.73	$\pm$	0.35	&	2.90	$\pm$	1.30	&	2.11	$\pm$	1.02	&	0.60	$\pm$	0.34	&	0.62	$\pm$	0.36	&	42.51	$\pm$	14.72	\\ \relax
0.60	&	3.15	$\pm$	3.15	&	3.76	$\pm$	3.76	&	4.60	$\pm$	4.33	&	...	$\pm$	...	&	...	$\pm$	...	&	...	$\pm$	...	&	0.98	$\pm$	0.98	&	...	$\pm$	...	&	...	$\pm$	...	&	1.00	$\pm$	0.66	&	4.40	$\pm$	2.96	&	12.81	$\pm$	8.65	&	1.09	$\pm$	1.09	&	...	$\pm$	...	&	...	$\pm$	...	&	2.90	$\pm$	2.58	&	1.81	$\pm$	1.65	&	1.41	$\pm$	1.41	&	0.77	$\pm$	0.77	&	3.17	$\pm$	2.10	\\ \relax
0.80	&	2.41	$\pm$	1.11	&	2.98	$\pm$	1.47	&	1.50	$\pm$	0.88	&	...	$\pm$	...	&	...	$\pm$	...	&	...	$\pm$	...	&	0.38	$\pm$	0.37	&	...	$\pm$	...	&	0.54	$\pm$	0.54	&	1.00	$\pm$	0.24	&	4.07	$\pm$	1.07	&	11.98	$\pm$	3.15	&	0.58	$\pm$	0.58	&	...	$\pm$	...	&	0.63	$\pm$	0.26	&	2.89	$\pm$	1.30	&	1.87	$\pm$	0.87	&	0.68	$\pm$	0.53	&	0.66	$\pm$	0.53	&	3.61	$\pm$	0.87	\\ \relax
1.00	&	0.81	$\pm$	0.60	&	3.90	$\pm$	1.79	&	1.73	$\pm$	0.89	&	0.55	$\pm$	0.49	&	0.70	$\pm$	0.70	&	...	$\pm$	...	&	0.43	$\pm$	0.29	&	0.47	$\pm$	0.36	&	0.32	$\pm$	0.19	&	1.00	$\pm$	0.26	&	4.05	$\pm$	1.24	&	11.91	$\pm$	3.64	&	0.15	$\pm$	0.13	&	...	$\pm$	...	&	0.59	$\pm$	0.15	&	2.90	$\pm$	1.22	&	1.65	$\pm$	0.69	&	0.54	$\pm$	0.26	&	0.55	$\pm$	0.30	&	8.31	$\pm$	2.14	\\ \relax
1.20	&	1.83	$\pm$	1.83	&	4.53	$\pm$	2.79	&	1.51	$\pm$	1.17	&	...	$\pm$	...	&	0.55	$\pm$	0.55	&	...	$\pm$	...	&	0.61	$\pm$	0.58	&	...	$\pm$	...	&	0.23	$\pm$	0.22	&	1.00	$\pm$	0.49	&	3.35	$\pm$	1.81	&	9.83	$\pm$	5.33	&	...	$\pm$	...	&	...	$\pm$	...	&	0.93	$\pm$	0.54	&	2.90	$\pm$	1.97	&	1.89	$\pm$	1.26	&	0.75	$\pm$	0.60	&	0.85	$\pm$	0.76	&	4.49	$\pm$	2.22	\\ \relax
1.40	&	...	$\pm$	...	&	4.07	$\pm$	2.08	&	1.56	$\pm$	1.12	&	0.88	$\pm$	0.83	&	...	$\pm$	...	&	0.35	$\pm$	0.35	&	0.77	$\pm$	0.77	&	0.19	$\pm$	0.11	&	0.28	$\pm$	0.19	&	1.00	$\pm$	0.33	&	3.20	$\pm$	1.19	&	9.44	$\pm$	3.52	&	0.34	$\pm$	0.34	&	...	$\pm$	...	&	0.63	$\pm$	0.33	&	2.90	$\pm$	1.65	&	1.94	$\pm$	1.16	&	0.74	$\pm$	0.50	&	0.68	$\pm$	0.47	&	3.89	$\pm$	1.28	\\ \relax
1.60	&	...	$\pm$	...	&	4.43	$\pm$	2.50	&	0.94	$\pm$	0.47	&	1.42	$\pm$	1.13	&	...	$\pm$	...	&	...	$\pm$	...	&	0.44	$\pm$	0.42	&	0.18	$\pm$	0.06	&	0.21	$\pm$	0.08	&	1.00	$\pm$	0.32	&	3.18	$\pm$	1.20	&	9.31	$\pm$	3.50	&	...	$\pm$	...	&	...	$\pm$	...	&	0.54	$\pm$	0.28	&	2.89	$\pm$	1.52	&	1.58	$\pm$	0.84	&	0.38	$\pm$	0.29	&	0.58	$\pm$	0.43	&	6.43	$\pm$	2.03	\\ \relax
1.80	&	0.65	$\pm$	0.65	&	4.88	$\pm$	3.60	&	1.11	$\pm$	0.72	&	0.65	$\pm$	0.55	&	...	$\pm$	...	&	0.79	$\pm$	0.79	&	0.60	$\pm$	0.45	&	...	$\pm$	...	&	0.35	$\pm$	0.12	&	1.00	$\pm$	0.46	&	2.98	$\pm$	1.58	&	8.76	$\pm$	4.64	&	0.46	$\pm$	0.46	&	...	$\pm$	...	&	0.61	$\pm$	0.28	&	2.90	$\pm$	1.94	&	1.75	$\pm$	1.15	&	0.70	$\pm$	0.60	&	0.68	$\pm$	0.51	&	6.23	$\pm$	2.86	\\ \relax
2.00	&	...	$\pm$	...	&	6.79	$\pm$	4.36	&	1.28	$\pm$	1.13	&	0.28	$\pm$	0.28	&	...	$\pm$	...	&	0.48	$\pm$	0.48	&	0.77	$\pm$	0.77	&	0.18	$\pm$	0.07	&	0.18	$\pm$	0.14	&	1.00	$\pm$	0.40	&	3.16	$\pm$	1.56	&	9.29	$\pm$	4.59	&	0.50	$\pm$	0.50	&	...	$\pm$	...	&	0.80	$\pm$	0.36	&	2.90	$\pm$	1.68	&	2.23	$\pm$	1.25	&	0.71	$\pm$	0.46	&	0.71	$\pm$	0.42	&	5.71	$\pm$	2.29	\\ \relax
2.20	&	...	$\pm$	...	&	5.27	$\pm$	2.03	&	0.88	$\pm$	0.32	&	0.41	$\pm$	0.21	&	0.24	$\pm$	0.13	&	0.30	$\pm$	0.22	&	0.59	$\pm$	0.28	&	0.40	$\pm$	0.40	&	0.21	$\pm$	0.12	&	1.00	$\pm$	0.21	&	2.53	$\pm$	0.68	&	7.52	$\pm$	2.02	&	0.38	$\pm$	0.38	&	...	$\pm$	...	&	0.87	$\pm$	0.23	&	2.90	$\pm$	1.07	&	2.47	$\pm$	0.88	&	0.75	$\pm$	0.33	&	0.94	$\pm$	0.41	&	3.87	$\pm$	0.81	\\ \relax
2.40	&	...	$\pm$	...	&	4.77	$\pm$	2.44	&	0.91	$\pm$	0.71	&	0.32	$\pm$	0.22	&	0.24	$\pm$	0.24	&	...	$\pm$	...	&	0.45	$\pm$	0.20	&	0.16	$\pm$	0.08	&	0.18	$\pm$	0.14	&	1.00	$\pm$	0.26	&	2.30	$\pm$	0.82	&	6.82	$\pm$	2.43	&	0.32	$\pm$	0.32	&	...	$\pm$	...	&	0.71	$\pm$	0.19	&	2.90	$\pm$	1.16	&	2.05	$\pm$	0.83	&	0.82	$\pm$	0.48	&	0.81	$\pm$	0.33	&	4.60	$\pm$	1.17	\\ \relax
2.60	&	1.07	$\pm$	1.07	&	3.82	$\pm$	1.28	&	0.74	$\pm$	0.40	&	0.40	$\pm$	0.30	&	...	$\pm$	...	&	0.21	$\pm$	0.19	&	0.47	$\pm$	0.26	&	0.17	$\pm$	0.11	&	0.14	$\pm$	0.08	&	1.00	$\pm$	0.21	&	2.15	$\pm$	0.57	&	6.41	$\pm$	1.72	&	0.26	$\pm$	0.26	&	...	$\pm$	...	&	0.63	$\pm$	0.20	&	2.90	$\pm$	0.98	&	1.87	$\pm$	0.59	&	0.80	$\pm$	0.35	&	0.71	$\pm$	0.31	&	3.76	$\pm$	0.78	\\ \relax
2.80	&	0.62	$\pm$	0.29	&	4.24	$\pm$	1.39	&	0.58	$\pm$	0.36	&	0.32	$\pm$	0.32	&	...	$\pm$	...	&	0.28	$\pm$	0.28	&	0.51	$\pm$	0.34	&	...	$\pm$	...	&	0.17	$\pm$	0.16	&	1.00	$\pm$	0.20	&	2.05	$\pm$	0.58	&	6.11	$\pm$	1.72	&	0.39	$\pm$	0.39	&	...	$\pm$	...	&	0.73	$\pm$	0.23	&	2.89	$\pm$	1.02	&	2.04	$\pm$	0.69	&	0.94	$\pm$	0.41	&	0.91	$\pm$	0.42	&	2.89	$\pm$	0.59	\\ \relax
3.00	&	0.77	$\pm$	0.77	&	3.73	$\pm$	2.02	&	1.11	$\pm$	0.77	&	0.50	$\pm$	0.47	&	...	$\pm$	...	&	...	$\pm$	...	&	0.50	$\pm$	0.42	&	...	$\pm$	...	&	0.70	$\pm$	0.70	&	1.00	$\pm$	0.34	&	2.42	$\pm$	0.98	&	7.20	$\pm$	2.93	&	0.52	$\pm$	0.52	&	...	$\pm$	...	&	0.89	$\pm$	0.41	&	2.90	$\pm$	1.57	&	2.45	$\pm$	1.27	&	0.95	$\pm$	0.56	&	0.84	$\pm$	0.47	&	1.06	$\pm$	0.36
\enddata
\vspace{-0.2em}
\begin{minipage}{9.2in}
\tablecomments{The flux-weighted average emission line ratios for slits A -- D after correcting for reddening using a galactic extinction curve \citep{Savage1979} and fixing the intrinsic H$\alpha$/H$\beta$ ratio to 2.90. The first column provides the deprojected distance from the nucleus in arcseconds, with positive towards the SW. The emission lines were fit using widths and centroids calculated from free fits to H$\alpha$ $\lambda$6563, and the listed wavelengths are approximate vacuum values. Measurements for the weak [Ne~III] $\lambda$3969, [S~II] $\lambda$4074, and H$\delta$ $\lambda$4102 lines, as well as the \othree $\lambda$4959, [O~I] $\lambda$6363, and [N~II] $\lambda$6548 lines that have relative strengths fixed to their brighter doublet counter-part (see \S3.1) have been omitted for space and are available online. The last column lists the extinction-corrected H$\beta$ flux (erg s$^{-1}$ cm$^{-2}$), which is the flux-weighted average of all measurements within each $0\farcs2$ bin, with errors propagated from the H$\beta$ flux and reddening-correction uncertainties. We required each line to have a minimum S/N of two in order to be considered a positive detection. Specifically, the height of the Gaussian fit must be at least twice the standard deviation of the flux in the continuum regions used for the fit. The uncertainty in each measurement is then the fractional uncertainty in the line flux added in quadrature with that of H$\beta$. We then add these in quadrature with the uncertainty in the reddening-correction based on the errors in the H$\alpha$/H$\beta$ ratios. In regions with low flux, this results in effective S/N ratios of less than two for the reddening-corrected measurements. Positions marked with ``...	$\pm$	..." are non-detections and those with uncertainties equal to the ratios are upper limits.}
\end{minipage}
\label{ratios}
\end{deluxetable*}
\end{rotatetable*}

\begin{figure*}[hb!]
\centering
\hspace*{0.5em}\includegraphics[width=0.46\textwidth]{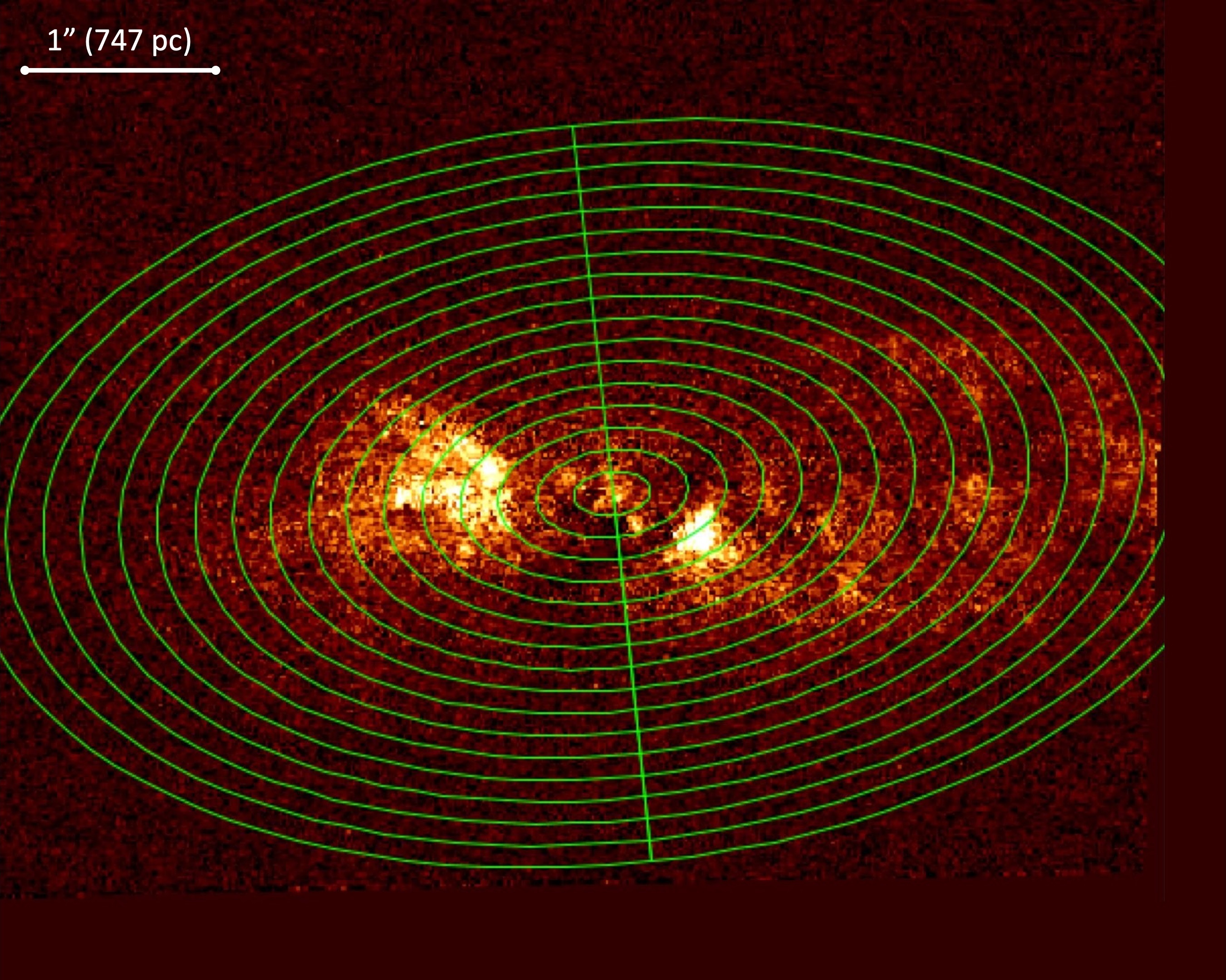}
\hspace*{0.3em}\includegraphics[width=0.52\textwidth]{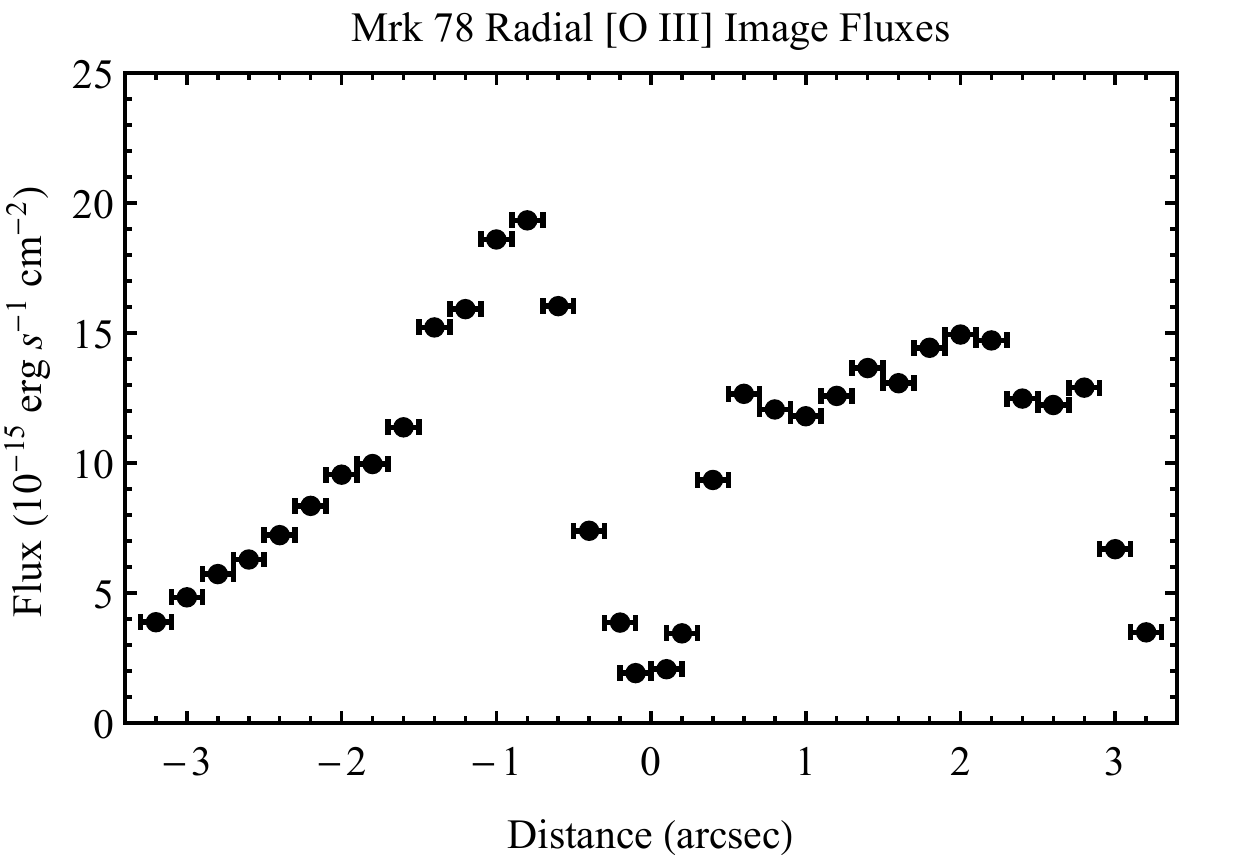}
\hspace*{0.5em}\includegraphics[width=0.46\textwidth]{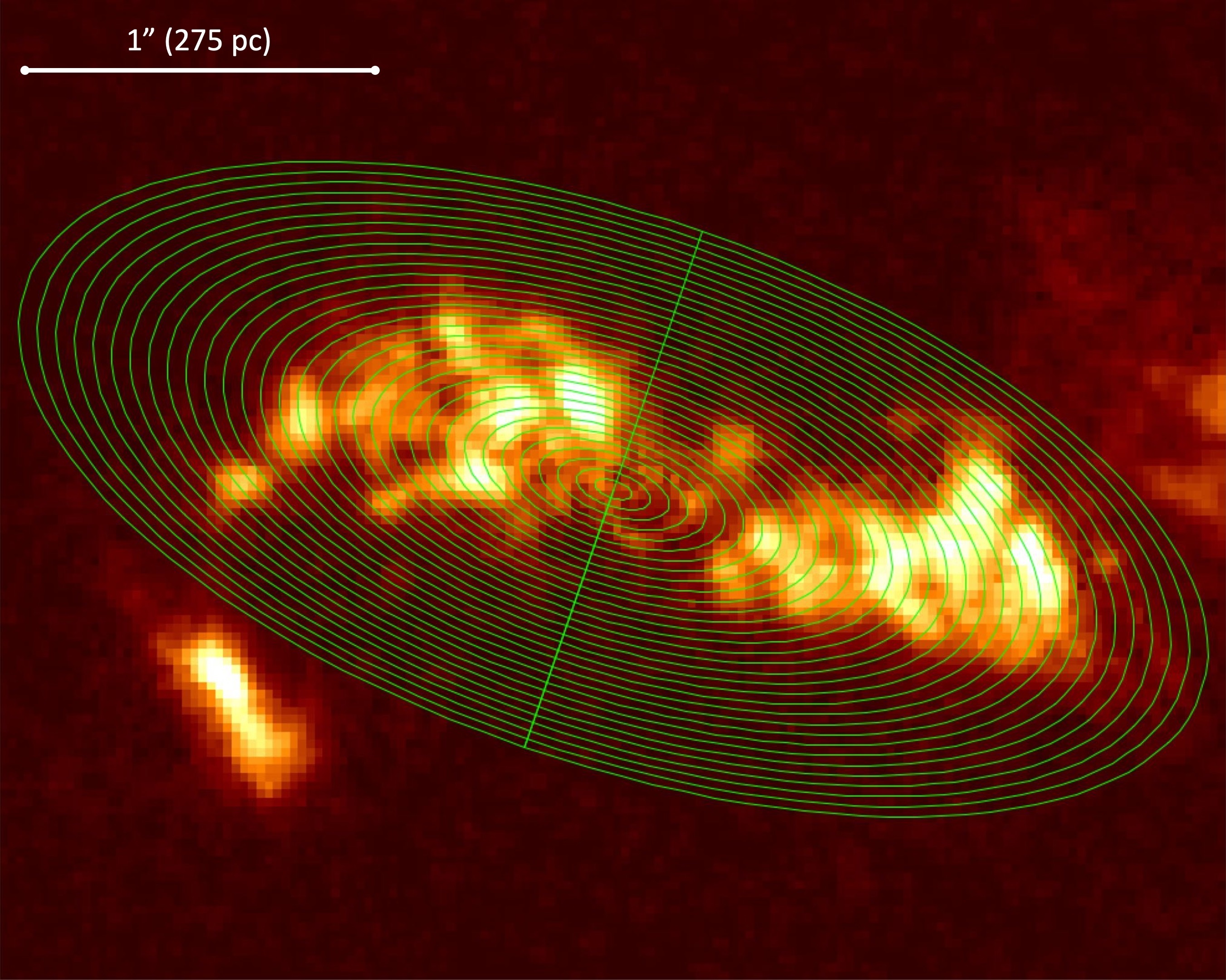}
\hspace*{0.3em}\includegraphics[width=0.52\textwidth]{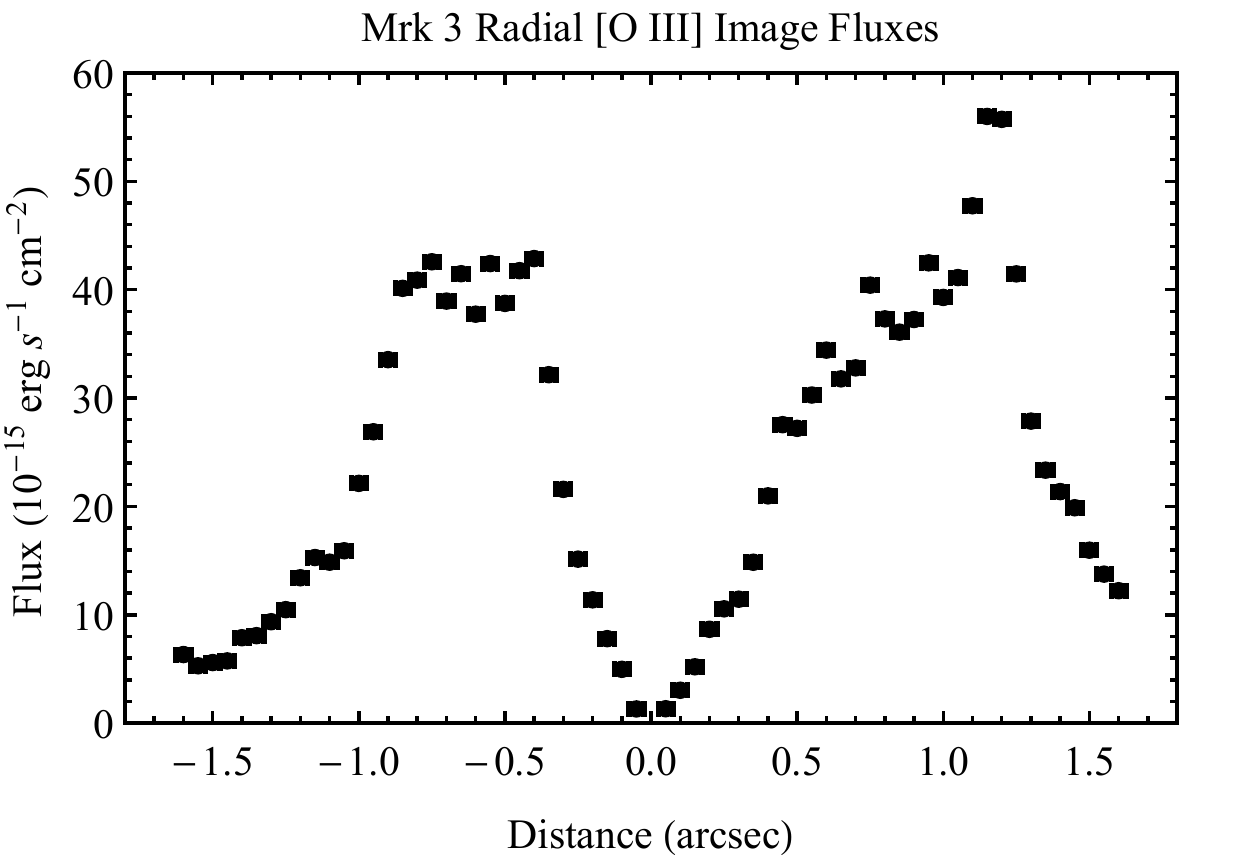}
\hspace*{0.5em}\includegraphics[width=0.46\textwidth]{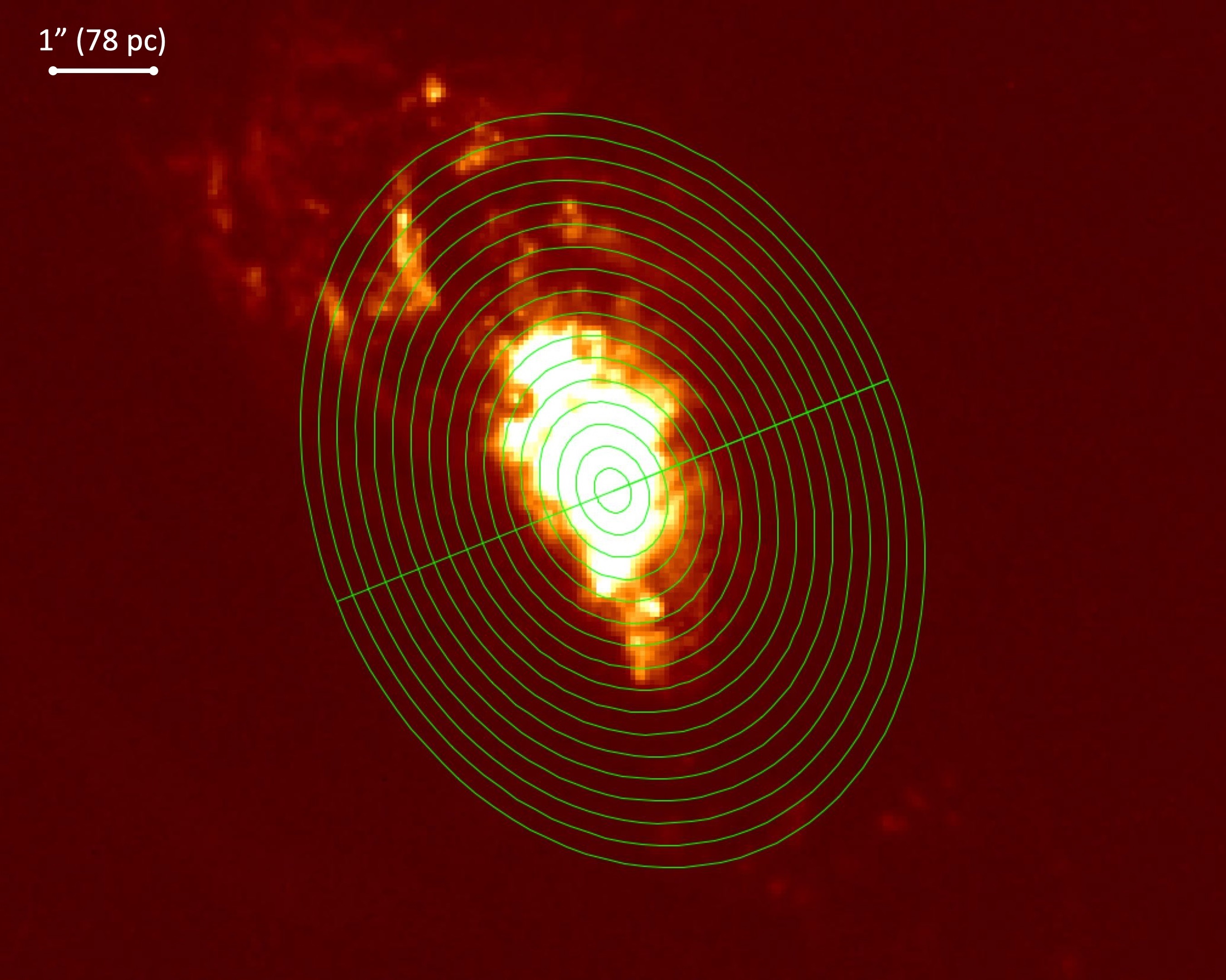}
\hspace*{0.3em}\includegraphics[width=0.52\textwidth]{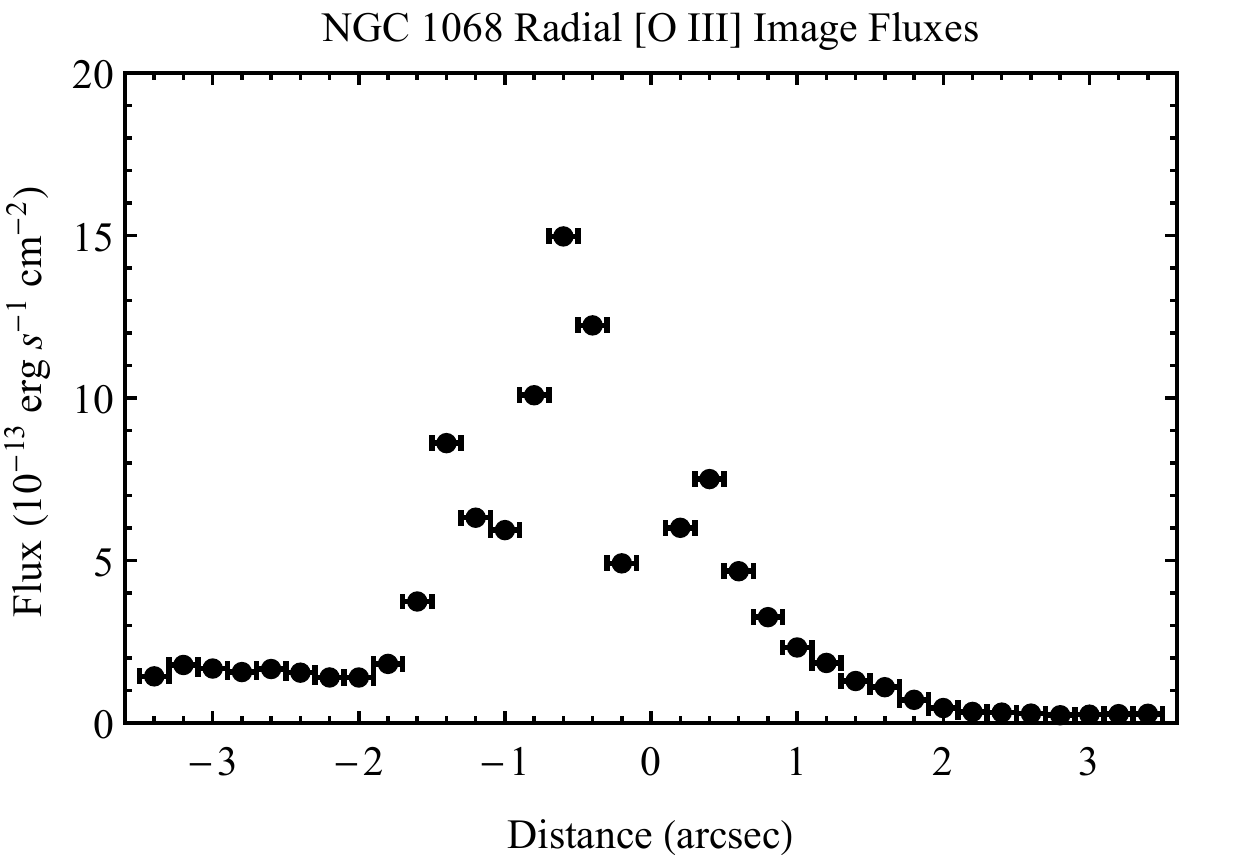}
\caption{The \othree images (left) and extracted radial flux profiles (right) for Mrk~78 (top), Mrk~3 (middle), and NGC~1068 (bottom). The annuli widths are $0\farcs2$ for Mrk~78, $0\farcs05$ for Mrk~3, which were summed to match the variable $0\farcs2$-$0\farcs3$ extractions modeled in \cite{Collins2009}, and $0\farcs2$ for NGC~1068, which were summed to match the $0\farcs4$ extractions modeled in \cite{Kraemer2000b}. The extended emission for Mrk~78 falls off the edge of the detector, resulting in a minor underestimation of the flux and mass at $r > 2\farcs80$. The annular extractions extend to larger radii than the photoionization models, and the flux profile for NGC~1068 is vertically scaled 2 dex lower than the other targets because it is significantly brighter.}
\label{imaging}
\end{figure*}

\subsection{{\normalfont [O~III]} Image Analysis}

Our photoionization modeling process accounts for the emission and gas mass within the \hst STIS slits, and we use \othree emission line images of the NLRs to calculate the total ionized gas mass as a function of distance from the nucleus for each AGN. As shown in Figure~\ref{imaging}, we determine the \othree radial flux profiles for each image by extracting fluxes within Elliptical Panda regions generated in the SAOImage DS9 software suite \citep{Joye2003}. We divide the elliptical annuli in half along the minor axis to account for asymmetries in the flux, density, and velocity profiles in each cone of the biconical outflows. The ellipses are centered on the continuum peak and are radially spaced in increments matching the spatial sampling of our line ratios and photoionization models (typically $\sim0\farcs2 - 0\farcs3$). The major axis lengths ($a$) of the ellipses are equal to the radial extent of our line ratio measurements along each slit, and the minor axis lengths ($b$) are calculated from the major axis length and the NLR inclination ($i$) via the equation $b/a = cos(i)$. The \othree images, elliptical annuli, and extracted \othree radial flux profiles are presented in Figure~\ref{imaging}. We calculate the flux errors by measuring the standard deviation ($\sigma$) of the background in line-free regions of each image. The uncertainty for each annular measurement is then equal to $\sqrt{N_{pix}} \times \sigma$, where $N_{pix}$ is the number of pixels in the annulus. The background variations ($\sigma$, erg s$^{-1}$ cm$^{-2}$ pixel$^{-1}$) for each object are: Mrk~78 $ = 1.9 \times 10^{-18}$, Mrk~3 $ = 9.9 \times 10^{-18}$, and NGC~1068 $ = 7.9 \times 10^{-18}$, resulting in typical uncertainties of $<$~1\% for the integrated \othree flux measurements.

\section{Photoionization Models}

Our analysis is based on accurately converting the \othree images fluxes to ionized gas masses at each radius. This requires accounting for local variations in the ionization state of the gas, as well as its abundances, temperature, and density. Accounting for these physical conditions is critical, because they set the gas emissivity, which directly determines the conversion between \othree luminosity and the ionized gas mass. We use our previous photoionization model results for Mrk~3 \citep{Collins2009} and NGC~1068 \citep{Kraemer2000b}, and present here our new models for Mrk~78.

\subsection{Input Parameters}

We generate photoionization models using the Cloudy spectral synthesis code (version 13.04; \citealp{Ferland2013}). A self-consistent model requires supplying the number and energy distribution of photons ionizing a gas cloud of specified composition and geometry. These conditions are encapsulated by the ionization parameter ($U$), which is the dimensionless ratio of the number of ionizing photons to atoms at the face of the gas cloud \citep[\S 13.6]{Osterbrock2006}\footnote{In some X-ray models, the ionization parameter is defined as $\xi = L_i / n_\mathrm{H} r^2$, where $L_i$ is the radiation energy density from 1 to 1000 Ry (13.6 eV -- 13.6 keV). A conversion for Seyfert power law SEDs is log($U$) $\approx$ log($\xi$) -- 1.5 \citep{Crenshaw2012}.}. This is defined as
\begin{equation}
U = \frac{Q(H)}{4 \pi r^2 n_H c},
\label{ionparam}
\end{equation}
\noindent
where $r$ is the distance from the AGN, $n_\mathrm{H}$ is the hydrogen number density (cm$^{-3}$), and $c$ is the speed of light. $Q(H)$ is the number of ionizing photons s$^{-1}$, given by $Q(H)~=~\int_{\nu_0}^{\infty} (L_{\nu}/h\nu) d\nu$, where $L_{\nu}$ is the luminosity of the AGN as a function of frequency (the SED), $h$ is Planck's constant, and $\nu_0 = 13.6$ eV/$h$ is the ionization potential of hydrogen \citep[\S 14.3]{Osterbrock2006}. We use a common power-law SED from our previous studies \citep{Kraemer2000a, Kraemer2000b} with $L_{\nu} \propto \nu^{\alpha}$. We adopt slopes of $\alpha = -0.5$ from 1 eV to 13.6 eV, $\alpha = -1.4$ from 13.6 eV to 0.5 keV, $\alpha = -1$ from 0.5 keV to 10 keV, and $\alpha = -0.5$ from 10 keV to 100 keV, with cutoffs below 1 eV and above 100 keV.

We determine the bolometric luminosity of the AGN by summing the observed \othree fluxes in our APO long-slit observations and applying the correction factor from \cite{Heckman2004}, namely, \lbol~=~3500 $\times$ $L_{\mathrm{[O~III]}}$, yielding \lbol~=~7.9$\times$10$^{45}$ erg s$^{-1}$ (log \lbol~$\approx$ 45.9). This estimate is in excellent agreement with the values found by \cite{Whittle1988} and \cite{GonzalezDelgado2001} when rescaled to our adopted distance. We then numerically compute the above integral, normalized to the bolometric luminosity, and find $Q(H) = 3.8 \times 10^{54}$ photons s$^{-1}$ (log $Q(H)$ $\approx$ 54.58), in general agreement with \cite{Wilson1988}, \cite{Whittle2004}, and \cite{Rosario2007}.

The gas composition is set by the elemental abundances, dust content, and depletion fractions of elements onto dust grains. We found that the abundances vary with radial distance from the nucleus (Figure~\ref{diagnostics}), but are consistent with an average value of $\sim$1.3~$Z_{\odot}$ within the uncertainties across the NLR, and we adopt this average value for our models.  The exact logarithmic values relative to hydrogen by number for dust free models are: He = -0.96, C = -3.46, N = -3.94, O = -3.20, Ne = -3.96, Na = -5.65, Mg = -4.29, Al = -5.44, Si = -4.38, P = -6.48, S = -4.77, Ar = -5.49, Ca = -5.55, Fe = -4.39, Ni = -5.67. The strong low-ionization lines are better reproduced when including a dusty component, and for models with a dust level of 50\% relative to the interstellar medium we accounted for depletion of elements in graphite and silicate grains \citep{Seab1983, Snow1996, Collins2009}. The logarithmic abundances relative to hydrogen by number for the dusty models are: He = -0.96, C = -3.63, N = -3.94, O = -3.32, Ne = -3.96, Na = -5.65, Mg = -4.57, Al = -5.70, Si = -4.66, P = -6.48, S = -4.77, Ar = -5.49, Ca = -5.81, Fe = -4.67, Ni = -5.93.

\subsection{Model Selection}

We account for ionization stratification within the gas at each location by using up to three model components with different densities. These are denoted according to the value of their ionization parameter ($U$) as HIGH, MED, and LOW ION. At each radius, the only unknown quantities in Equation~\ref{ionparam} are $U$ and $n_\mathrm{H}$, so we choose a range of $U$ values to produce the observed emission and solve for the corresponding density to maintain physical consistency. We then generate a grid of models and add fractional combinations of the HIGH, MED, and LOW components to create a composite model that matches the emission line ratios and is normalized to the H$\beta$ luminosity at each location along the slit.

We determine the best-fitting model for each radius using a simple optimization scheme that compares all fractional permutations of the HIGH, MED, and LOW model components in 5\% intervals across our range of ionization parameters. The best-fitting model may be composed of one, two, or three components, and the simplest model matching all of the emission line criteria is selected. The criteria for a satisfactory fit vary by emission line, with sensitive diagnostic lines that constrain the gas density and mass having the strictest limits.

We initially require the predicted model line ratios to match the data (Table~\ref{ratios}) within the following tolerances: The He~II $\lambda$4686 ratio that is sensitive to the column density and the \othree $\lambda$5007 ratio that determines our flux-to-mass scaling must match within 20\%, which is smaller than the measurement uncertainties at each location along the slit (Table~\ref{ratios}). The \othree $\lambda$4363, [O~I] $\lambda$6300, [N~II] $\lambda$6584, and [S~II] $\lambda\lambda$6716,6731 ratios must match within a factor of two. The remaining emission lines must match the observations to within a factor of four. We successfully created models satisfying all of these starting criteria for 14/26 spatial positions. When these criteria resulted in multiple solutions, we incrementally tightened the criteria for the key diagnostic lines until a single best match was found. Similarly, when the initial criteria resulted in no solutions, we incrementally relaxed the criteria for all lines until a match was found. In cases with similar competing models, we selected the composite model that best matched the He~II $\lambda$4686 and \othree $\lambda$5007 lines, which are most sensitive to the gas column density and flux-to-mass scale factor, making them critical for determining accurate gas masses. The input and output parameters for our best-fitting Cloudy models are provided in Table~\ref{cloudy}.

\begin{figure*}
\centering
\includegraphics[width=0.495\textwidth]{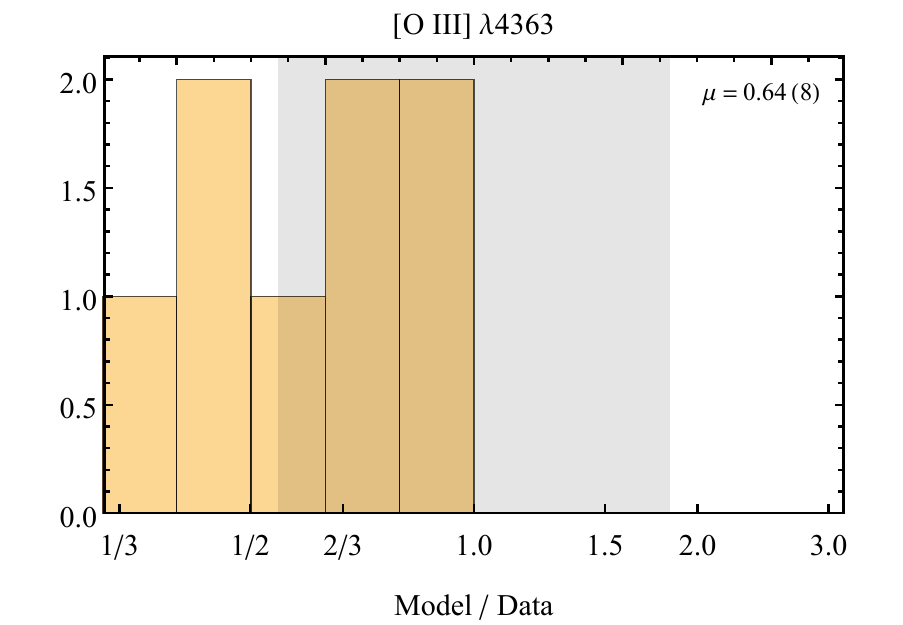}
\includegraphics[width=0.495\textwidth]{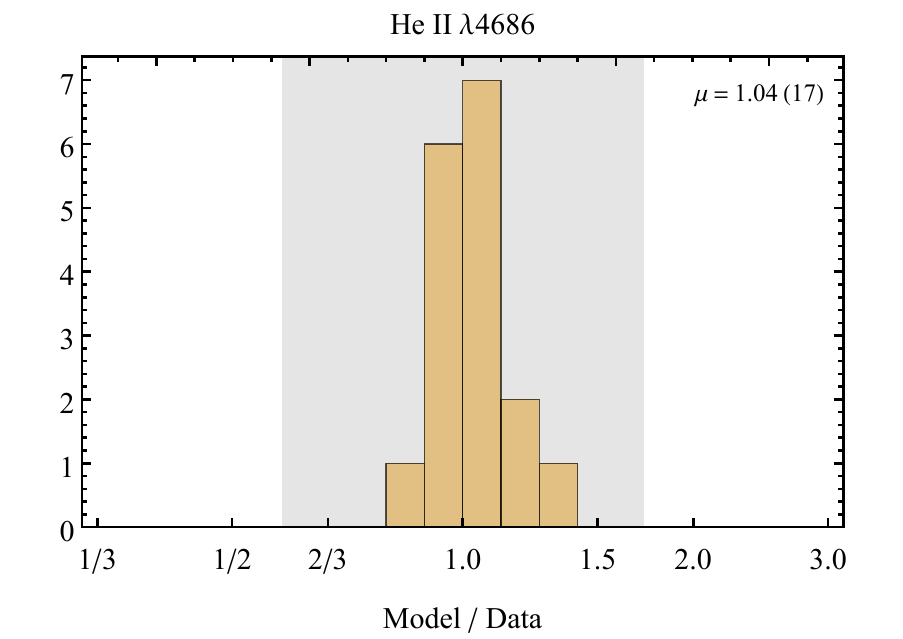}
\includegraphics[width=0.495\textwidth]{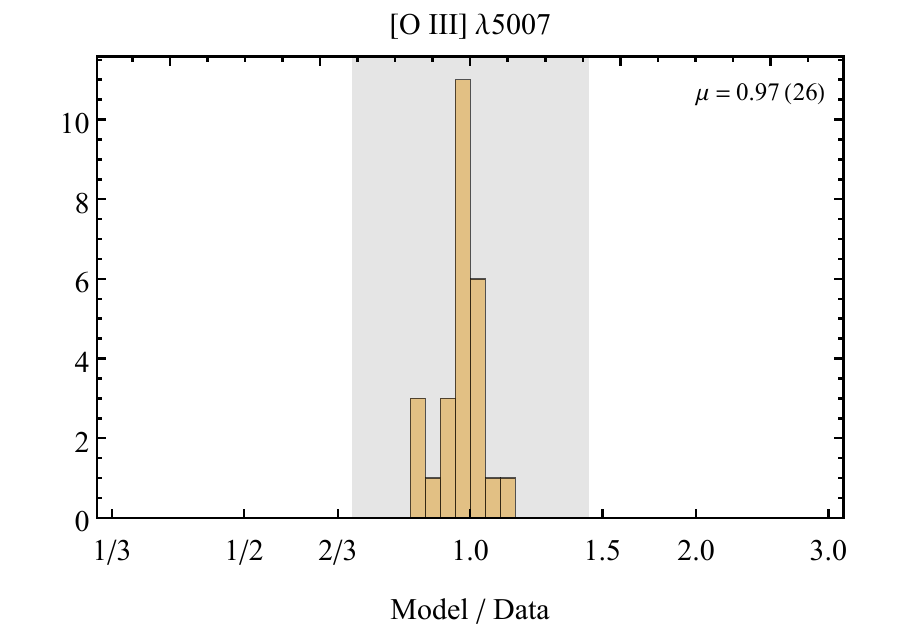}
\includegraphics[width=0.495\textwidth]{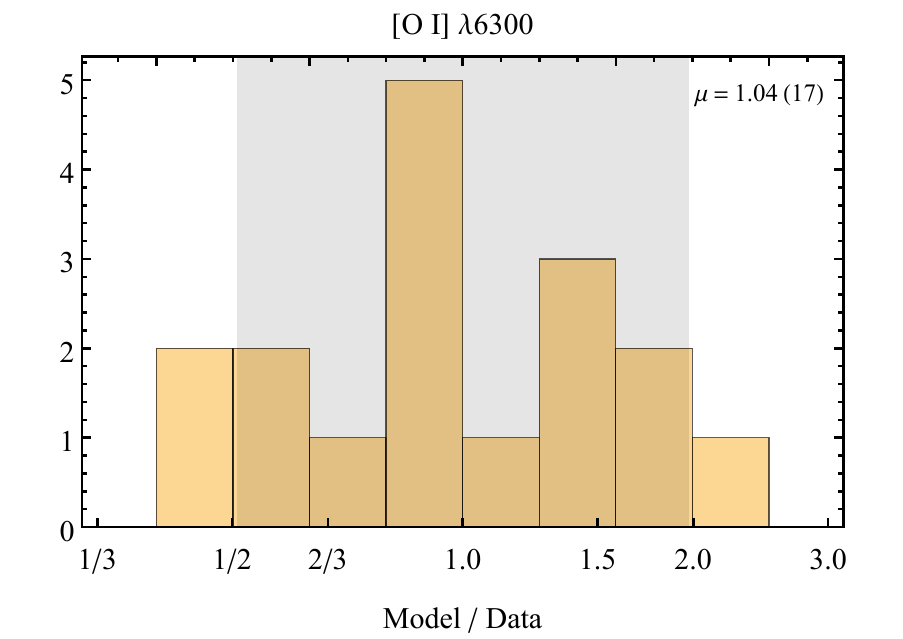}
\includegraphics[width=0.495\textwidth]{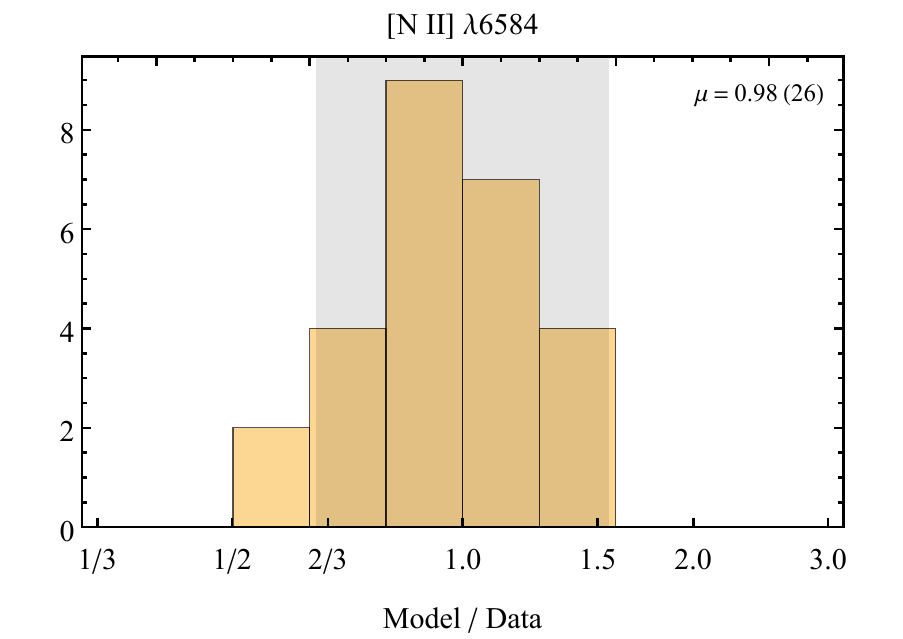}
\includegraphics[width=0.495\textwidth]{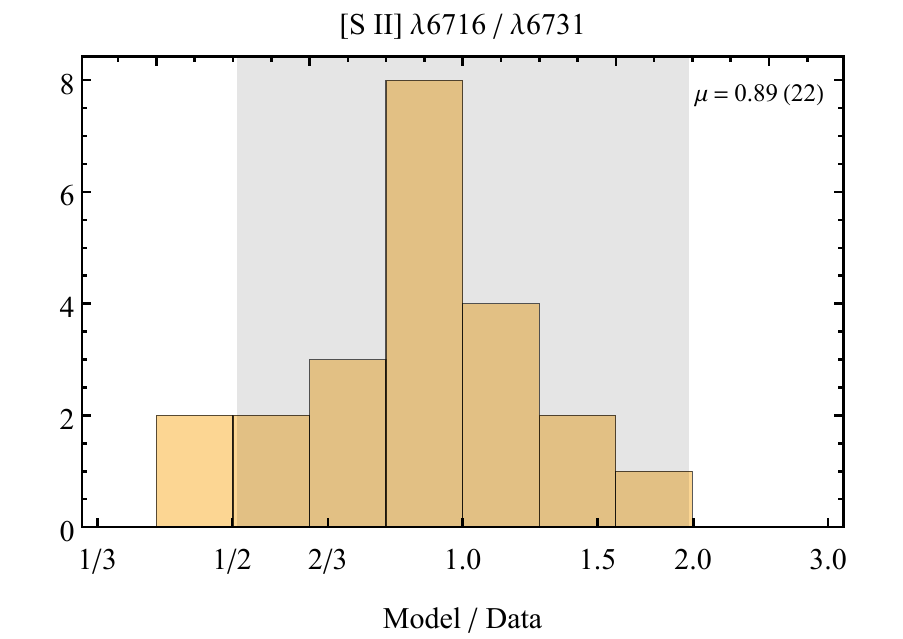}
\caption{Histograms of the Mrk~78 model line ratios divided by the dereddened values (Table~\ref{ratios}) for select emission lines that are most important for constraining the gas masses. These are: [O~III] $\lambda$4363 (temperature), He II $\lambda$4686 (column density), [O~III] $\lambda$5007 (flux-to-mass), [O~I] $\lambda$6300 (column density and SED), [N~II] $\lambda$6584 (abundances), and [S~II] $\lambda$6716/$\lambda$6731 (density). A value of unity indicates an exact match between the model and data, while the shaded regions are the mean uncertainties for each emission line over all positions. The mean of each distribution ($\mu$) is shown in the upper right corner, along with the number of measurements in parentheses. The horizontal tick marks are logarithmically spaced for even distribution around unity. Points above unity are over-predicted, while points below unity are under-predicted by the models. Figures comparing the data and models for all emission lines at all positions are available in the Appendix. In general, all models match the data within the uncertainties, except for several under-predictions of [O~III] $\lambda$4363, indicating a mild under-prediction of the gas temperature at some locations. The dispersion in [O~I] $\lambda$6300 may be due to its combined sensitivity to the SED, gas turbulence, temperature, and column density.}
\label{models}
\end{figure*}

\subsection{Comparison to the Observations}

A comparison of the model and data line ratios for the most important lines are shown in Figure~\ref{models}, with the results for all lines and positions provided in Appendix~A. Several factors contribute to the observed deviations, such as a poor Gaussian fit, S/N of the measurements, quality of atomic data for each element, and the accuracy of our multi-component models.

There is a mild under-prediction for the overall strength of the [S~II] $\lambda \lambda$6716, 6731 lines at most locations, which may indicate the presence of more dust, that some of the [S~II] emission arises from the edges of the ionized NLR bicone, or that the low-ionization gas is exposed to a partially-absorbed SED from a closer-in absorber (e.g. \citealp{Maksym2016, Maksym2017, Mingozzi2019}) as we found for Mrk 573 \citep{Revalski2018a} and Mrk 3 \citep{Collins2009}. There is also a small but systematic offset between the observed and predicted [S~II] doublet ratios ($\sim$11\%), indicating the low-ionization model densities are slightly over-predicted. The model values are within the measurement uncertainties, but could result in a minor under-prediction of the mass in the low-ionization component of the gas.

The general under-prediction of [O~III] $\lambda$4363 is a minor concern as it indicates an under-prediction of the temperature in more highly ionized zones; however, the most discrepant points have the largest uncertainties (see Figure~\ref{diagnostics}) and may be partially attributed to blending with H$\gamma$. The [Ne~V] and [O~II] lines are generally under-predicted, which may indicate an overzealous reddening correction due to using a Galactic extinction curve, in agreement with the slight under-prediction of H$\gamma$ at some positions. These lines were allowed less-stringent limits to properly match the key diagnostic lines that constrain the gas number and column densities that are used to calculate the gas masses.

Generating successful models for Mrk 78 is more difficult than for the majority of the targets in our sample because we are using a flux-weighted average of all measurements at each radial distance within four slits. These multiple extractions encompass emission with a larger range of physical conditions than are observed in a single slit. In general, there are insufficient high-ionization lines to tightly constrain the HIGH model component. While adding a third model component can improve the fit, it must be physically consistent with detected and non-detected emission lines. The locations with the strongest [Ne~V] detections, such as -$2\farcs0$, -$0\farcs4$, +$0\farcs6$, and +$0\farcs8$, all have contributions from a HIGH model component. Similarly, regions with weak or non-detections of [Ne~V] (e.g. -$1\farcs0$, -$1\farcs6$) either have no HIGH component, or it has a smaller contribution to the luminosity. The S/N was insufficient to measure emission line ratios at $-3\farcs0$, $-2\farcs8$, $-2\farcs6$, $-2\farcs4$, and $-2\farcs2$, and for these positions we adopt the physically-consistent models from their positive counterparts. Overall, our models are able to successfully match all of the key diagnostic emission lines to within a factor of two or better at most locations across the NLR.

Finally, to confirm that our models are physically plausible, we also derive the surface areas ($A = L_{\mathrm{H}\beta}/F_{\mathrm{H}\beta}$) and thicknesses ($N_{\mathrm{H}}/n_{\mathrm{H}}$) of the emitting clouds, which must fit within the four \hst spectral slits. This criterion is satisfied for all but two positions, which are at large radii and are attributed to the significant line ratio uncertainties at those locations. We also calculate the depths of the clouds into the plane of the sky by dividing the cloud area by the projected slit width ($\sim$150 pc for Mrk~78) to verify that they are less than or equal to the line-of-sight distance across the bicone at each location. It is important to note that each ionized component may not be co-located within the slit, as the emission is spread across $0\farcs2$ in the spatial and dispersion directions. These physical quantities are presented in Table~\ref{cloudy}.

\setlength{\tabcolsep}{0.1in}
\tabletypesize{\footnotesize}
\startlongtable
\begin{deluxetable*}{|c|c|c|c|c|c||c|c|c|c|c|}
\def\arraystretch{0.98}
\tablecaption{Cloudy Model Parameters for Mrk~78}
\tablehead{
\colhead{Distance} & \colhead{Comp} & \colhead{Ionization} & \colhead{Column} & \colhead{Number} & \colhead{Dust} & \colhead{Fraction} & \colhead{log($F_{\mathrm{H}\beta}$)} & \colhead{Cloud} & \colhead{Cloud} & \colhead{Cloud\vspace{-0.75em}}\\
\colhead{from} & \colhead{ION} & \colhead{Parameter} & \colhead{Density} & \colhead{Density} & \colhead{Content} & \colhead{of} & \colhead{Model} & \colhead{Surface} & \colhead{Model} & \colhead{Model\vspace{-0.75em}}\\
\colhead{Nucleus} & \colhead{Name} & \colhead{log($U$)} & \colhead{log($N_\mathrm{H}$)} & \colhead{log($n_\mathrm{H}$)} & \colhead{Relative} & \colhead{Total} & \colhead{Flux (erg} & \colhead{Area} & \colhead{Thickness} & \colhead{Depth\vspace{-0.75em}}\\
\colhead{(arcsec)}& \colhead{} & \colhead{(unitless)} & \colhead{(cm$^{-2}$)} & \colhead{(cm$^{-3}$)} & \colhead{to ISM} & \colhead{Model} & \colhead{s$^{-1}$ cm$^{-2}$)} & \colhead{($10^3$ pc$^2$)} & \colhead{(pc)} & \colhead{(pc)\vspace{-0.75em}}\\
\colhead{(1)} & \colhead{(2)} & \colhead{(3)} & \colhead{(4)} & \colhead{(5)} & \colhead{(6)} & \colhead{(7)} & \colhead{(8)} & \colhead{(9)} & \colhead{(10)} & \colhead{(11)}
}
\startdata
-2.00	 & 	High		 & 	-1.20	 & 	21.40	 & 	0.89	 & 	0.0	 & 	0.15	 & 	-2.76	 & 	4.0	 & 	104.9	 & 	26.8	 \\ 
-2.00	 & 	Med		 & 	-1.80	 & 	21.40	 & 	1.49	 & 	0.5	 & 	0.50	 & 	-2.37	 & 	5.4	 & 	26.3	 & 	36.1	 \\ 
-2.00	 & 	Low		 & 	-3.60	 & 	19.80	 & 	3.29	 & 	0.5	 & 	0.35	 & 	-2.21	 & 	2.7	 & 	$<$0.1	 & 	17.8	 \\ 
-1.80	 & 	High		 & 	...	 & 	...	 & 	...	 & 	...	 & 	0.00	 & 	...	 & 	...	 & 	...	 & 	...	 \\ 
-1.80	 & 	Med		 & 	-1.80	 & 	21.40	 & 	1.58	 & 	0.5	 & 	0.85	 & 	-2.28	 & 	16.4	 & 	21.4	 & 	109.7	 \\ 
-1.80	 & 	Low		 & 	-3.20	 & 	20.20	 & 	2.98	 & 	0.5	 & 	0.15	 & 	-2.12	 & 	2.0	 & 	0.1	 & 	13.5	 \\ 
-1.60	 & 	High		 & 	...	 & 	...	 & 	...	 & 	...	 & 	0.00	 & 	...	 & 	...	 & 	...	 & 	...	 \\ 
-1.60	 & 	Med		 & 	-1.80	 & 	21.20	 & 	1.68	 & 	0.5	 & 	0.65	 & 	-2.19	 & 	14.0	 & 	10.7	 & 	94.0	 \\ 
-1.60	 & 	Low		 & 	-3.80	 & 	19.80	 & 	3.68	 & 	0.5	 & 	0.35	 & 	-2.02	 & 	5.1	 & 	$<$0.1	 & 	34.0	 \\ 
-1.40	 & 	High		 & 	...	 & 	...	 & 	...	 & 	...	 & 	0.00	 & 	...	 & 	...	 & 	...	 & 	...	 \\ 
-1.40	 & 	Med		 & 	-1.80	 & 	20.80	 & 	1.80	 & 	0.5	 & 	0.50	 & 	-2.39	 & 	56.3	 & 	3.2	 & 	376.6	 \\ 
-1.40	 & 	Low		 & 	-3.20	 & 	20.20	 & 	3.20	 & 	0.5	 & 	0.50	 & 	-1.90	 & 	18.1	 & 	$<$0.1	 & 	121.0	 \\ 
-1.20	 & 	High		 & 	-1.40	 & 	21.20	 & 	1.53	 & 	0.0	 & 	0.10	 & 	-2.28	 & 	8.3	 & 	15.2	 & 	55.4	 \\ 
-1.20	 & 	Med		 & 	-2.80	 & 	20.20	 & 	2.93	 & 	0.5	 & 	0.85	 & 	-1.80	 & 	23.5	 & 	0.1	 & 	157.3	 \\ 
-1.20	 & 	Low		 & 	-3.00	 & 	20.40	 & 	3.13	 & 	0.5	 & 	0.05	 & 	-1.77	 & 	1.3	 & 	0.1	 & 	8.7	 \\ 
-1.00	 & 	High		 & 	-1.80	 & 	20.60	 & 	2.09	 & 	0.0	 & 	0.10	 & 	-2.27	 & 	10.8	 & 	1.0	 & 	72.3	 \\ 
-1.00	 & 	Med		 & 	-1.80	 & 	21.60	 & 	2.09	 & 	0.5	 & 	0.40	 & 	-1.75	 & 	13.2	 & 	10.5	 & 	88.5	 \\ 
-1.00	 & 	Low		 & 	-3.00	 & 	20.00	 & 	3.29	 & 	0.5	 & 	0.50	 & 	-1.64	 & 	12.6	 & 	$<$0.1	 & 	84.1	 \\ 
-0.80	 & 	High		 & 	-1.20	 & 	21.40	 & 	1.68	 & 	0.0	 & 	0.05	 & 	-1.97	 & 	4.5	 & 	17.0	 & 	29.9	 \\ 
-0.80	 & 	Med		 & 	-1.80	 & 	21.60	 & 	2.28	 & 	0.5	 & 	0.40	 & 	-1.56	 & 	14.1	 & 	6.8	 & 	94.3	 \\ 
-0.80	 & 	Low		 & 	-3.00	 & 	20.00	 & 	3.48	 & 	0.5	 & 	0.55	 & 	-1.45	 & 	14.7	 & 	$<$0.1	 & 	98.6	 \\ 
-0.60	 & 	High		 & 	-1.40	 & 	21.80	 & 	2.13	 & 	0.0	 & 	0.15	 & 	-1.15	 & 	1.2	 & 	15.2	 & 	7.7	 \\ 
-0.60	 & 	Med		 & 	-1.80	 & 	21.60	 & 	2.53	 & 	0.5	 & 	0.30	 & 	-1.32	 & 	3.4	 & 	3.8	 & 	22.7	 \\ 
-0.60	 & 	Low		 & 	-3.00	 & 	20.60	 & 	3.73	 & 	0.5	 & 	0.55	 & 	-1.16	 & 	4.3	 & 	$<$0.1	 & 	29.0	 \\ 
-0.40	 & 	High		 & 	-1.20	 & 	21.40	 & 	2.28	 & 	0.0	 & 	0.20	 & 	-1.37	 & 	5.1	 & 	4.3	 & 	34.2	 \\ 
-0.40	 & 	Med		 & 	-2.60	 & 	21.00	 & 	3.68	 & 	0.5	 & 	0.70	 & 	-0.83	 & 	5.1	 & 	0.1	 & 	34.1	 \\ 
-0.40	 & 	Low		 & 	-3.00	 & 	20.20	 & 	4.08	 & 	0.5	 & 	0.10	 & 	-0.83	 & 	0.7	 & 	$<$0.1	 & 	4.9	 \\ 
-0.20	 & 	High		 & 	...	 & 	...	 & 	...	 & 	...	 & 	0.00	 & 	...	 & 	...	 & 	...	 & 	...	 \\ 
-0.20	 & 	Med		 & 	-2.40	 & 	21.00	 & 	4.09	 & 	0.5	 & 	1.00	 & 	-0.24	 & 	6.5	 & 	$<$0.1	 & 	43.8	 \\ 
-0.20	 & 	Low		 & 	...	 & 	...	 & 	...	 & 	...	 & 	0.00	 & 	...	 & 	...	 & 	...	 & 	...	 \\ 
$\pm$0.10	 & 	High		 & 	...	 & 	...	 & 	...	 & 	...	 & 	0.00	 & 	...	 & 	...	 & 	...	 & 	...	 \\ 
$\pm$0.10	 & 	Med		 & 	-1.80	 & 	21.80	 & 	4.09	 & 	0.5	 & 	0.95	 & 	0.25	 & 	0.5	 & 	0.2	 & 	3.0	 \\ 
$\pm$0.10	 & 	Low		 & 	-3.60	 & 	19.40	 & 	5.89	 & 	0.5	 & 	0.05	 & 	0.33	 & 	$<$0.1	 & 	0.0	 & 	0.1	 \\ 
0.20	 & 	High		 & 	-1.40	 & 	20.60	 & 	3.09	 & 	0.0	 & 	0.10	 & 	-1.35	 & 	5.5	 & 	0.1	 & 	37.1	 \\ 
0.20	 & 	Med		 & 	-2.00	 & 	21.60	 & 	3.69	 & 	0.5	 & 	0.85	 & 	-0.30	 & 	4.1	 & 	0.3	 & 	27.6	 \\ 
0.20	 & 	Low		 & 	-3.20	 & 	19.60	 & 	4.89	 & 	0.5	 & 	0.05	 & 	-0.41	 & 	0.3	 & 	$<$0.1	 & 	2.1	 \\ 
0.40	 & 	High		 & 	...	 & 	...	 & 	...	 & 	...	 & 	0.00	 & 	...	 & 	...	 & 	...	 & 	...	 \\ 
0.40	 & 	Med		 & 	-1.80	 & 	21.40	 & 	2.88	 & 	0.5	 & 	0.45	 & 	-0.98	 & 	10.8	 & 	1.1	 & 	72.5	 \\ 
0.40	 & 	Low		 & 	-3.00	 & 	20.20	 & 	4.08	 & 	0.5	 & 	0.55	 & 	-0.83	 & 	9.4	 & 	$<$0.1	 & 	63.2	 \\ 
0.60	 & 	High		 & 	-1.00	 & 	21.20	 & 	1.73	 & 	0.0	 & 	0.10	 & 	-2.21	 & 	5.1	 & 	9.6	 & 	34.0	 \\ 
0.60	 & 	Med		 & 	-1.80	 & 	21.60	 & 	2.53	 & 	0.5	 & 	0.55	 & 	-1.32	 & 	3.6	 & 	3.8	 & 	24.0	 \\ 
0.60	 & 	Low		 & 	-3.40	 & 	20.20	 & 	4.13	 & 	0.5	 & 	0.35	 & 	-1.16	 & 	1.6	 & 	$<$0.1	 & 	10.6	 \\ 
0.80	 & 	High		 & 	-1.40	 & 	21.00	 & 	1.88	 & 	0.0	 & 	0.25	 & 	-2.15	 & 	7.7	 & 	4.3	 & 	51.4	 \\ 
0.80	 & 	Med		 & 	-1.80	 & 	21.60	 & 	2.28	 & 	0.5	 & 	0.50	 & 	-1.56	 & 	4.0	 & 	6.8	 & 	26.4	 \\ 
0.80	 & 	Low		 & 	-3.20	 & 	20.40	 & 	3.68	 & 	0.5	 & 	0.25	 & 	-1.41	 & 	1.4	 & 	$<$0.1	 & 	9.2	 \\ 
1.00	 & 	High		 & 	-1.20	 & 	21.40	 & 	1.49	 & 	0.0	 & 	0.05	 & 	-2.16	 & 	3.0	 & 	26.3	 & 	19.9	 \\ 
1.00	 & 	Med		 & 	-1.80	 & 	21.40	 & 	2.09	 & 	0.5	 & 	0.65	 & 	-1.77	 & 	15.7	 & 	6.6	 & 	104.9	 \\ 
1.00	 & 	Low		 & 	-3.80	 & 	19.20	 & 	4.09	 & 	0.5	 & 	0.30	 & 	-1.67	 & 	5.8	 & 	$<$0.1	 & 	38.6	 \\ 
1.20	 & 	High		 & 	...	 & 	...	 & 	...	 & 	...	 & 	0.00	 & 	...	 & 	...	 & 	...	 & 	...	 \\ 
1.20	 & 	Med		 & 	-1.20	 & 	21.80	 & 	1.33	 & 	0.0	 & 	0.50	 & 	-1.80	 & 	7.0	 & 	95.6	 & 	46.6	 \\ 
1.20	 & 	Low		 & 	-3.40	 & 	20.20	 & 	3.53	 & 	0.5	 & 	0.50	 & 	-1.76	 & 	6.4	 & 	$<$0.1	 & 	42.7	 \\ 
1.40	 & 	High		 & 	-1.40	 & 	21.40	 & 	1.40	 & 	0.0	 & 	0.10	 & 	-2.14	 & 	2.0	 & 	32.4	 & 	13.5	 \\ 
1.40	 & 	Med		 & 	-1.80	 & 	21.20	 & 	1.80	 & 	0.5	 & 	0.40	 & 	-2.07	 & 	6.8	 & 	8.1	 & 	45.6	 \\ 
1.40	 & 	Low		 & 	-3.60	 & 	19.80	 & 	3.60	 & 	0.5	 & 	0.50	 & 	-1.90	 & 	5.8	 & 	$<$0.1	 & 	38.8	 \\ 
1.60	 & 	High		 & 	-1.80	 & 	21.20	 & 	1.68	 & 	0.0	 & 	0.15	 & 	-2.03	 & 	4.4	 & 	10.7	 & 	29.2	 \\ 
1.60	 & 	Med		 & 	-1.80	 & 	21.40	 & 	1.68	 & 	0.5	 & 	0.35	 & 	-2.18	 & 	14.4	 & 	17.0	 & 	96.3	 \\ 
1.60	 & 	Low		 & 	-3.80	 & 	19.20	 & 	3.68	 & 	0.5	 & 	0.50	 & 	-2.08	 & 	16.3	 & 	$<$0.1	 & 	109.0	 \\ 
1.80	 & 	High		 & 	-0.80	 & 	21.80	 & 	0.58	 & 	0.0	 & 	0.10	 & 	-2.77	 & 	12.3	 & 	537.8	 & 	82.2	 \\ 
1.80	 & 	Med		 & 	-2.00	 & 	21.40	 & 	1.78	 & 	0.5	 & 	0.15	 & 	-2.21	 & 	5.1	 & 	13.5	 & 	33.9	 \\ 
1.80	 & 	Low		 & 	-3.00	 & 	20.40	 & 	2.78	 & 	0.5	 & 	0.75	 & 	-2.12	 & 	20.4	 & 	0.1	 & 	136.8	 \\ 
2.00	 & 	High		 & 	-1.60	 & 	21.40	 & 	1.29	 & 	0.0	 & 	0.45	 & 	-2.22	 & 	15.8	 & 	41.7	 & 	105.5	 \\ 
2.00	 & 	Med		 & 	-3.00	 & 	20.00	 & 	2.69	 & 	0.5	 & 	0.05	 & 	-2.23	 & 	1.8	 & 	0.1	 & 	12.2	 \\ 
2.00	 & 	Low		 & 	-3.80	 & 	19.20	 & 	3.49	 & 	0.5	 & 	0.50	 & 	-2.26	 & 	19.6	 & 	$<$0.1	 & 	130.9	 \\ 
2.20	 & 	High		 & 	...	 & 	...	 & 	...	 & 	...	 & 	0.00	 & 	...	 & 	...	 & 	...	 & 	...	 \\ 
2.20	 & 	Med		 & 	-1.80	 & 	21.20	 & 	1.40	 & 	0.5	 & 	0.50	 & 	-2.47	 & 	24.2	 & 	20.4	 & 	162.1	 \\ 
2.20	 & 	Low		 & 	-3.60	 & 	19.40	 & 	3.20	 & 	0.5	 & 	0.50	 & 	-2.34	 & 	18.0	 & 	$<$0.1	 & 	120.4	 \\ 
2.40	 & 	High		 & 	-1.40	 & 	21.60	 & 	0.93	 & 	0.0	 & 	0.15	 & 	-2.38	 & 	7.1	 & 	151.6	 & 	47.4	 \\ 
2.40	 & 	Med		 & 	-1.80	 & 	21.60	 & 	1.33	 & 	0.5	 & 	0.25	 & 	-2.51	 & 	16.0	 & 	60.3	 & 	107.2	 \\ 
2.40	 & 	Low		 & 	-4.00	 & 	18.80	 & 	3.53	 & 	0.5	 & 	0.60	 & 	-2.56	 & 	42.7	 & 	$<$0.1	 & 	285.4	 \\ 
2.60	 & 	High		 & 	-1.20	 & 	21.80	 & 	0.66	 & 	0.0	 & 	0.35	 & 	-2.46	 & 	19.0	 & 	447.4	 & 	127.2	 \\ 
2.60	 & 	Med		 & 	-1.80	 & 	21.60	 & 	1.26	 & 	0.5	 & 	0.10	 & 	-2.58	 & 	7.2	 & 	70.9	 & 	48.0	 \\ 
2.60	 & 	Low		 & 	-4.00	 & 	18.80	 & 	3.46	 & 	0.5	 & 	0.55	 & 	-2.63	 & 	43.7	 & 	$<$0.1	 & 	292.2	 \\ 
2.80	 & 	High		 & 	-1.40	 & 	21.60	 & 	0.79	 & 	0.0	 & 	0.35	 & 	-2.52	 & 	16.7	 & 	209.2	 & 	111.7	 \\ 
2.80	 & 	Med		 & 	-3.60	 & 	19.80	 & 	2.99	 & 	0.5	 & 	0.15	 & 	-2.51	 & 	7.0	 & 	$<$0.1	 & 	46.9	 \\ 
2.80	 & 	Low		 & 	-4.00	 & 	19.00	 & 	3.39	 & 	0.5	 & 	0.50	 & 	-2.58	 & 	27.6	 & 	$<$0.1	 & 	184.5	 \\ 
3.00	 & 	High		 & 	-1.60	 & 	20.80	 & 	0.93	 & 	0.0	 & 	0.20	 & 	-3.26	 & 	28.9	 & 	24.0	 & 	193.2	 \\ 
3.00	 & 	Med		 & 	-2.60	 & 	18.60	 & 	1.93	 & 	0.5	 & 	0.40	 & 	-4.36	 & 	730.6	 & 	$<$0.1	 & 	4887.7	 \\ 
3.00	 & 	Low		 & 	-3.80	 & 	19.60	 & 	3.13	 & 	0.5	 & 	0.40	 & 	-2.58	 & 	12.0	 & 	$<$0.1	 & 	80.2	 \\ 
\enddata
\tablecomments{The best fit Cloudy model input (columns 1-6) and output (columns 7-11) parameters. The columns are: (1) position with positive values toward the SW, (2) component name, (3) log ionization parameter, (4) log column density, (5) log number density, (6) dust fraction relative to the ISM, (7) fraction of model contributing to the H$\beta$ luminosity, (8) log H$\beta$ model flux (erg s$^{-1}$ cm$^{-2}$), (9) surface area of the gas divided by $10^3$, (10) gas cloud thickness ($N_\mathrm{H}$/$n_\mathrm{H}$), and (11) depth into the plane of the sky.}
\label{cloudy}
\end{deluxetable*}

\textcolor{white}{.}\\
\textcolor{white}{.}\\
\textcolor{white}{.}\\

\section{Calculations}

\subsection{Mass of the Ionized Gas}

We use the parameters from our photoionization models and the H$\beta$ luminosities to calculate the gas mass as a function of radius for the emission encompassed by the \hst STIS slits. The mass at each location is given by
\begin{equation}
M_{slit} = N_\mathrm{H} \mu m_p \left(\frac{L_{\mathrm{H}\beta}}{F_{\mathrm{H}\beta_m}}\right)
\label{masseqn}
\end{equation}
\noindent
\citep{Crenshaw2015}. In Equation~\ref{masseqn}, $N_\mathrm{H}$ is the model hydrogen column density, $\mu$ is the mean mass per proton ($\sim$ 1.4 for our abundances), $m_p$ is the proton mass, $F_{\mathrm{H}\beta_m}$ is the H$\beta$ model flux, and $L_{\mathrm{H}\beta}$ is the luminosity of H$\beta$ calculated from the extinction-corrected flux and distance. We calculate the masses for each of our HIGH, MED, and LOW ionization components separately by dividing up the H$\beta$ luminosity by the model fractional contributions and then sum the mass in each component. Conceptually, this process finds the area of the emitting clouds through the ratio of the luminosities and fluxes, multiplies by the column density (projected particles per unit area) to yield the total number of particles, which is multiplied by the mean mass per particle to give the total ionized mass. The results of Equation~\ref{masseqn} describe the number of H$\beta$ photons per unit mass that can be scaled to \othree based on the [O~III]/H$\beta$ ratios. With the mass per unit \othree flux at each location, we determine the total ionized gas mass outside of the slit by multiplying the mass in the slit by the flux ratio $F_{image} / F_{slit}$. This formalism eliminates the scale factor used in our previous investigations and simplifies the calculations. For Mrk~78, we retain the scale factor because the flux-weighted average line ratio luminosities do not allow for a direct comparison between the slit and annuli fluxes. The errors in the masses are dominated by the uncertainties in the H$\beta$ luminosities, which are determined from the emission line fit residuals and the reddening uncertainty. In addition, there is an uncertainty of $\pm0.05$ dex ($\sim$12\%) in the model column densities, and thus the final masses, due to the Cloudy model grid step sizes.

\subsection{Outflow Parameters}

Our goal is to determine the impact of the outflows on their host galaxies. The energy carried in the outflows can be quantified using six primary metrics: the mass, kinetic energy, momentum, and their respective outflow rates. The mass outflow rate ($\dot{M}_{out}$) at each distance is
\begin{equation}
\dot{M}_{out} = \left(\frac{M_{out}v}{\delta r} \right),
\end{equation}
\noindent
where $M_{out}$ is the outflowing mass in each annulus (which can be less than the total mass when a portion of the gas is in rotation, as we found for Mrk 34 in \citealp{Revalski2018b}), $v$ is the deprojected velocity corrected for inclination and position angle on the sky (\S 3.2 of \citealp{Revalski2018a}), and $\delta r$ is the deprojected width of each annulus. The kinetic energy ($E$), kinetic energy flow rate ($\dot E$), momentum ($p$), and momentum flow rate ($\dot p$) at each radius are given by
\begin{equation}
E = \frac{1}{2}{M}_{out} v^2,
\end{equation}
\begin{equation}
\dot E = \frac{1}{2} \dot{M}_{out} v^2,
\end{equation}
\begin{equation}
p = M_{out}v,
\end{equation}
\begin{equation}
\dot p = \dot M_{out}v.
\end{equation}
We do not include contributions to the energetics from velocity dispersion, such as turbulence, which would add a $\sigma_v$ term to the expressions. We obtain a single radial profile for each of these quantities by azimuthally summing the values derived for each of the semi-annuli.

\section{Results}

\begin{figure*}
\centering
\includegraphics[width=0.89\textwidth]{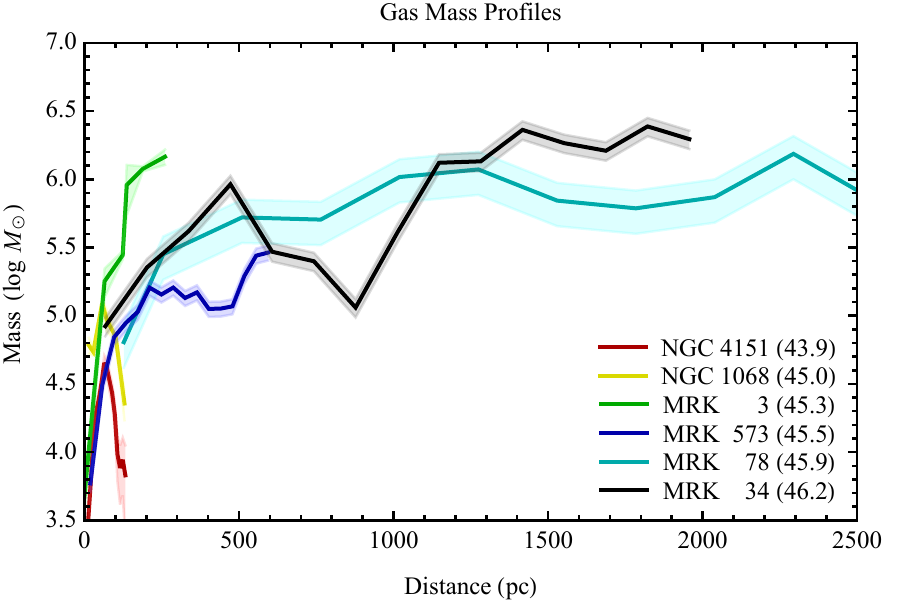}\\
\vspace{0.5em}
\includegraphics[width=0.89\textwidth]{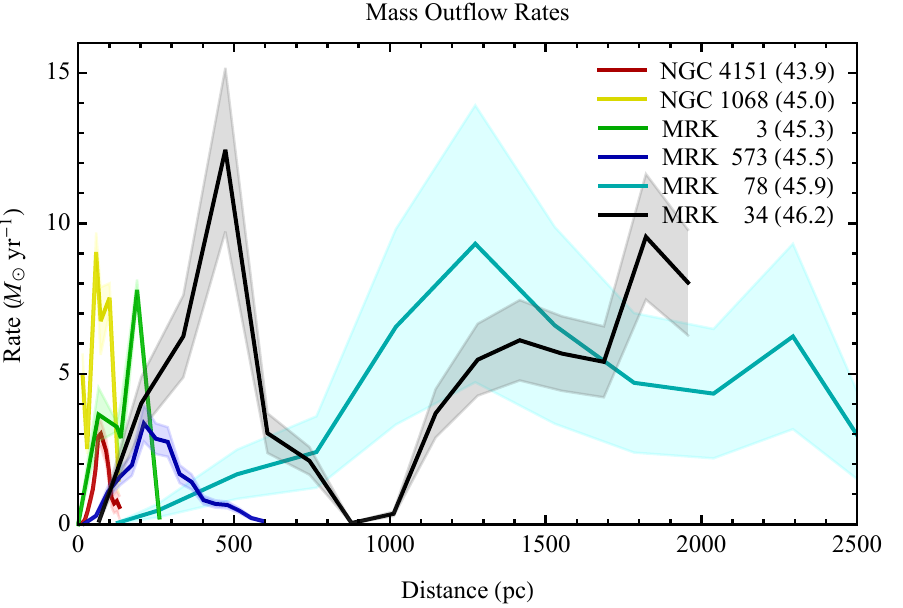}
\vspace{-0.5em}
\caption{The spatially-resolved gas mass profiles (top) and mass outflow rates (bottom) for NGC~4151 \citep{Crenshaw2015}, NGC~1068 (this work), Mrk~3 (this work), Mrk~573 \citep{Revalski2018a}, Mrk~78 (this work), and Mrk~34 \citep{Revalski2018b}. The logarithmic bolometric luminosity for each AGN is provided in parentheses. The masses and outflow rates are the quantity in each radial bin, not enclosed totals, and each target has a different bin size. The uncertainties are propagated from the errors in the reddening-corrected H$\beta$ luminosities and the resolution of the parameters in the photoionization model grids. The integrated masses and peak rates are compared in Figure~\ref{results}, and the results are available in tabular form in the Appendix.}
\label{mdots}
\end{figure*}

\begin{figure}
\centering
\includegraphics[width=0.48\textwidth]{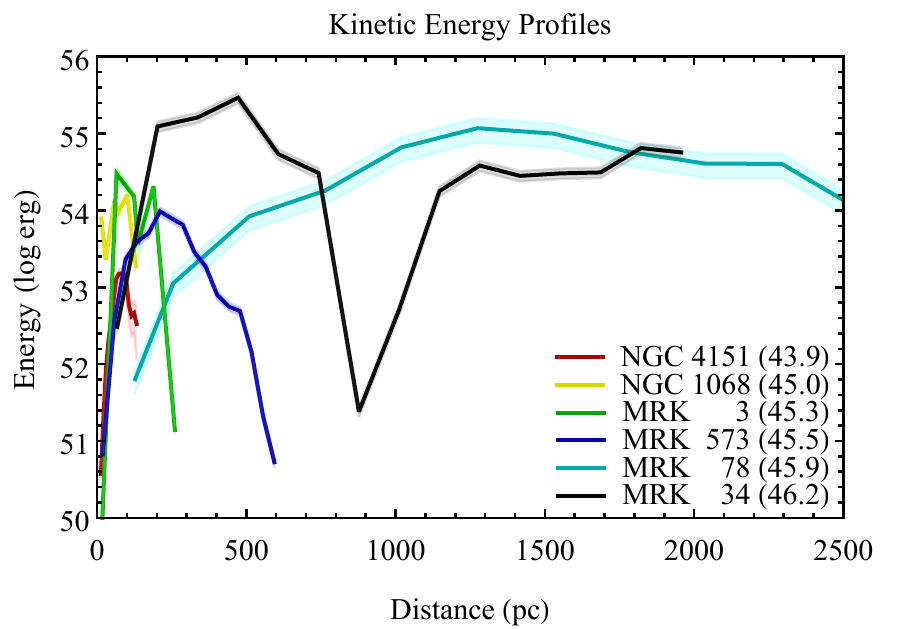}\\
\vspace{0.25em}
\includegraphics[width=0.48\textwidth]{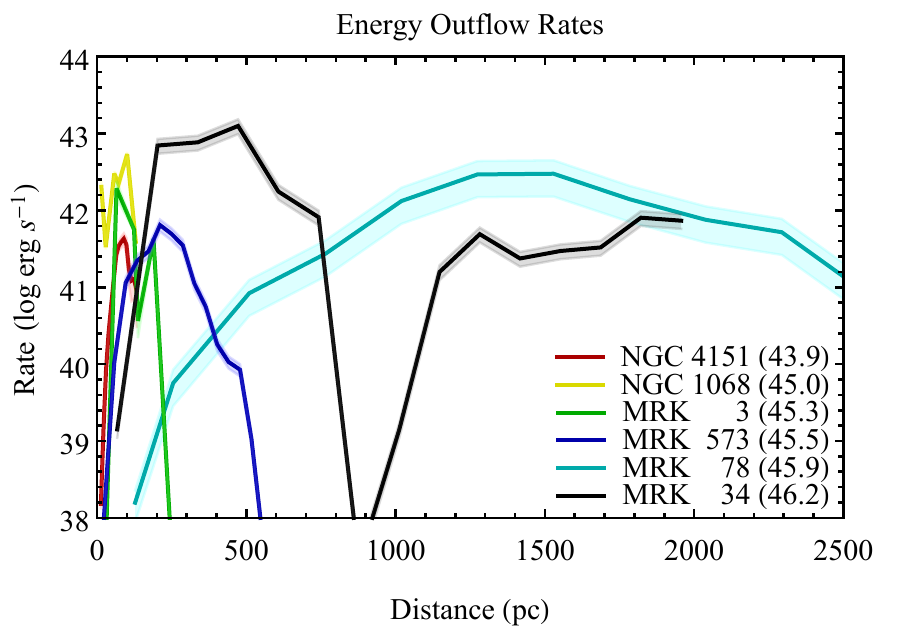}
\caption{The same as Figure~\ref{mdots} for the spatially-resolved kinetic energy and kinetic energy outflow rates.}
\label{edots}
\end{figure}

\begin{figure}
\centering
\includegraphics[width=0.48\textwidth]{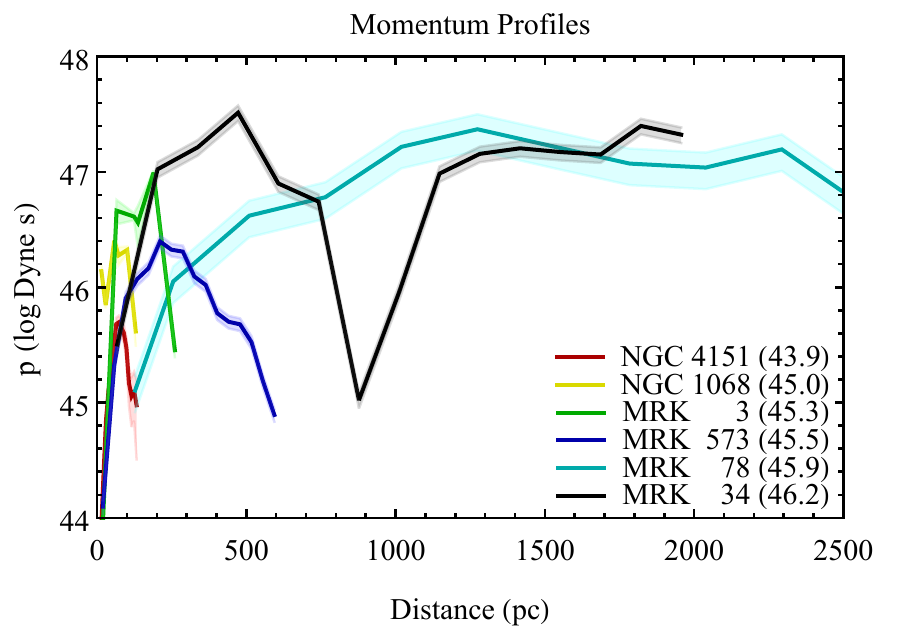}\\
\vspace{0.25em}
\includegraphics[width=0.48\textwidth]{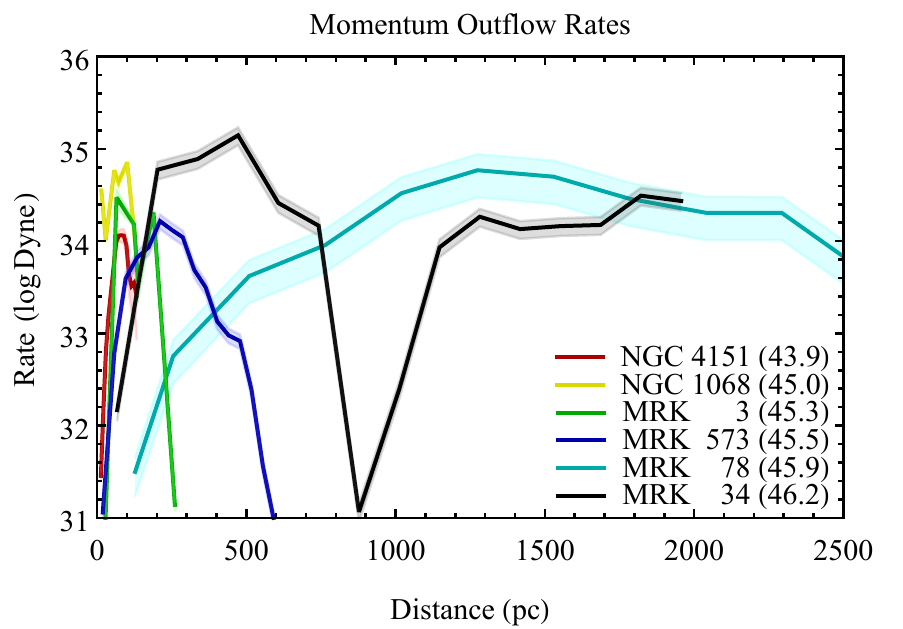}
\caption{The same as Figure~\ref{mdots} for the spatially-resolved momenta and momenta outflow rates.}
\label{pdots}
\end{figure}

We present the spatially-resolved gas mass profiles and mass outflow rates for each AGN in Figure~\ref{mdots}, as well as the kinetic energies, momenta, and their outflow rates in Figures~\ref{edots}~and~\ref{pdots}. In general, the rates rise from zero at the nucleus, where the outflow velocities are zero, to maximum values at hundreds of pcs from the nuclei in the lower luminosity targets. The higher luminosity AGN, Mrk~34 and Mrk~78, reach their peak mass outflow rates at $\sim$0.5 and $\sim$1.2~kpc, respectively, and display the largest peak mass outflow rates of $\dot M_{out} \approx 10$ $M_{\odot}$ yr$^{-1}$. The ionized outflows extend to radial distances of $r \approx 0.1 - 3$~kpc from the nucleus, with signatures of disturbed and/or rotational kinematics at larger radii.

\begin{figure*}
\centering
\includegraphics[width=0.497\textwidth]{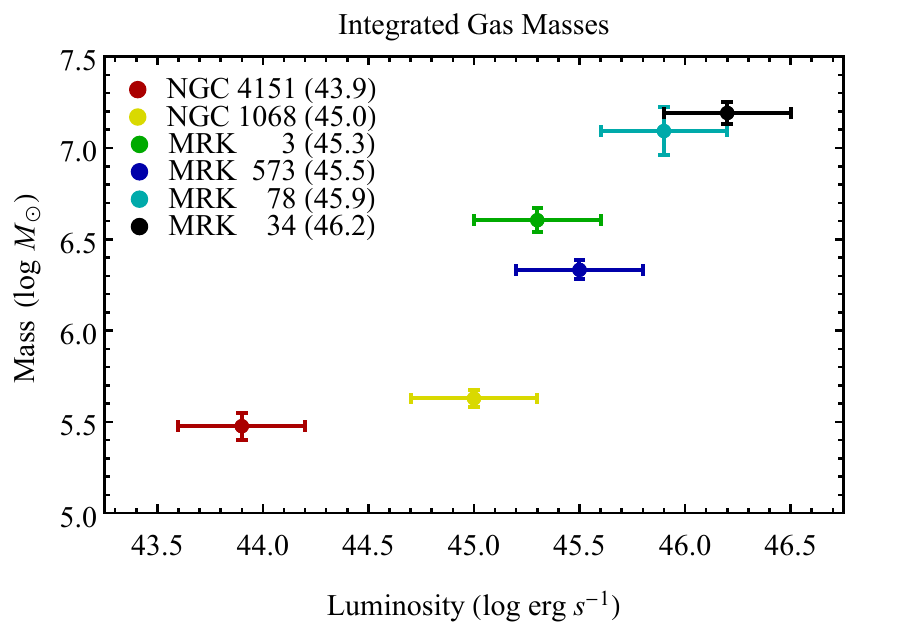}
\includegraphics[width=0.497\textwidth]{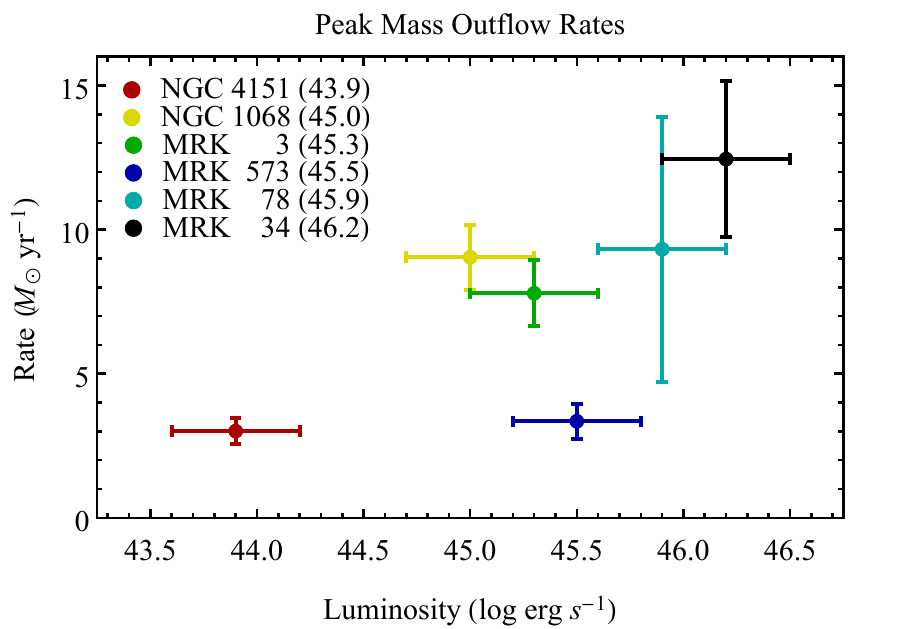}\\
\vspace{0.5em}
\includegraphics[width=0.497\textwidth]{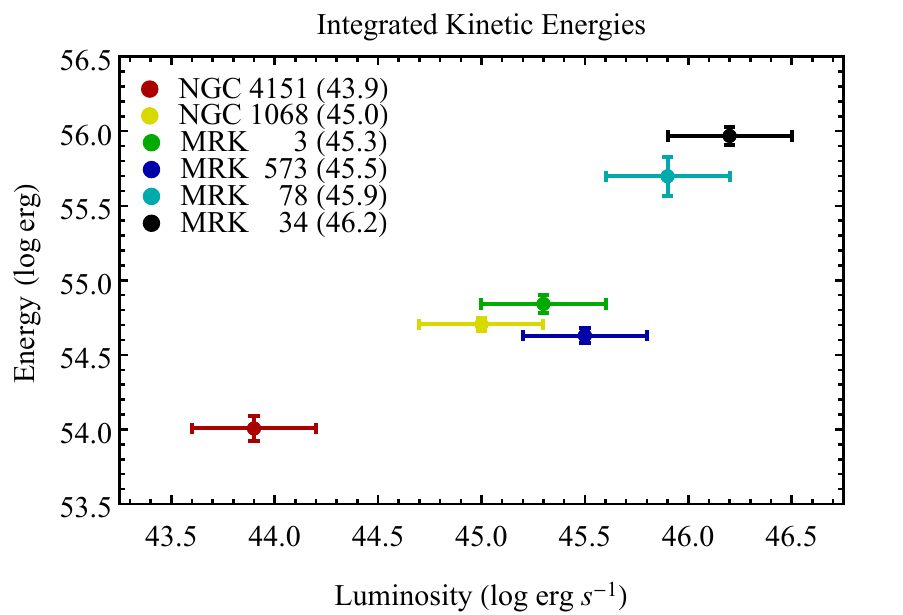}
\includegraphics[width=0.497\textwidth]{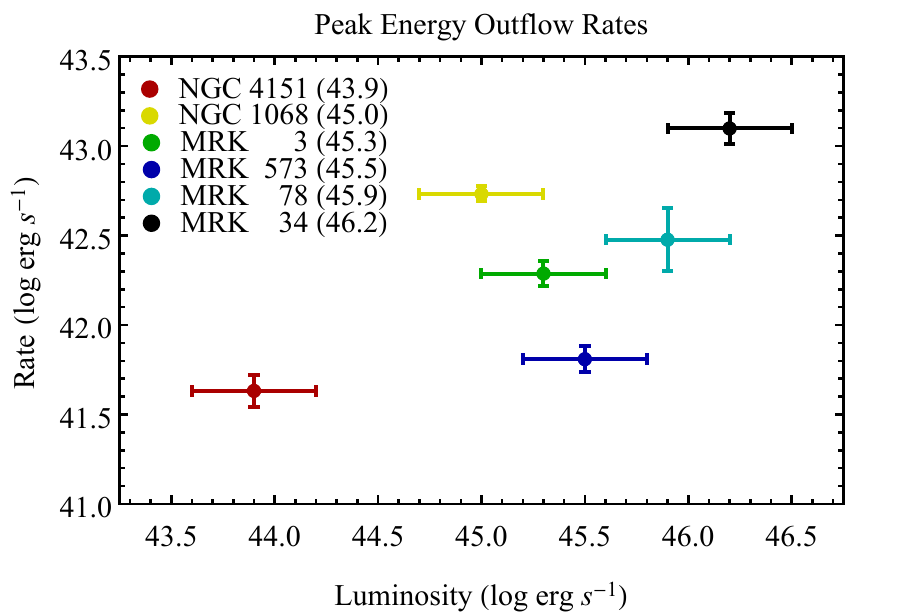}\\
\vspace{0.5em}
\includegraphics[width=0.497\textwidth]{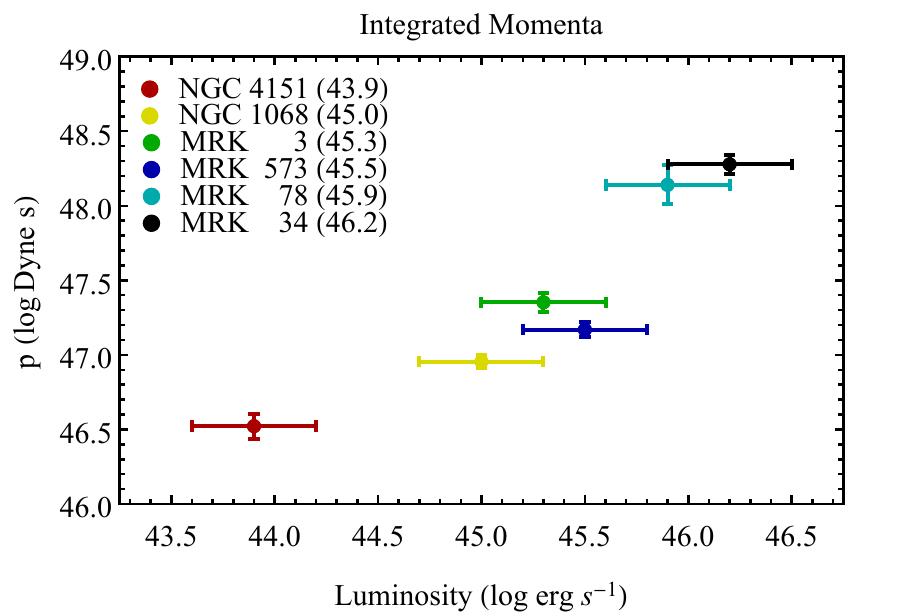}
\includegraphics[width=0.497\textwidth]{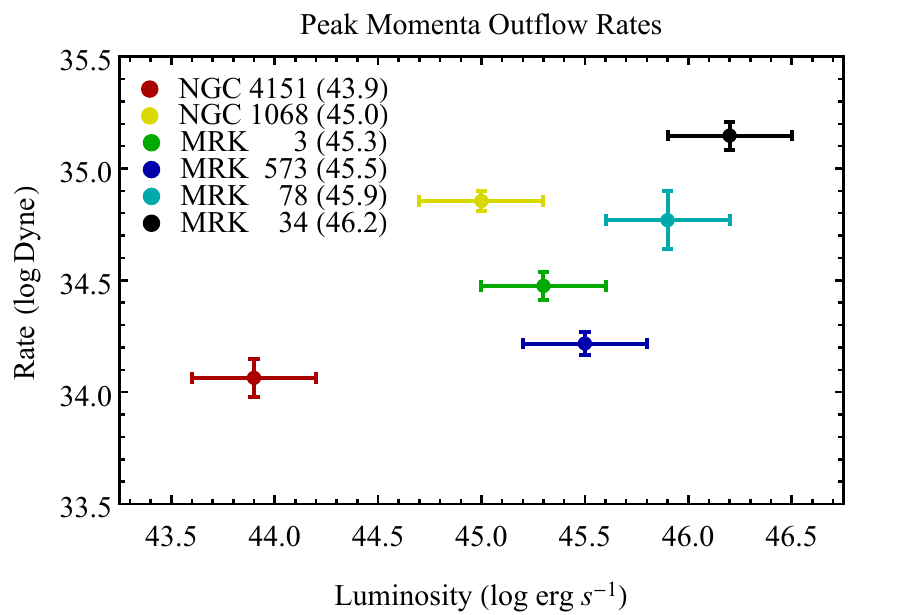}
\caption{The total mass, kinetic energy, and momentum of the outflows (left column), as well as the maximum mass, kinetic energy, and momentum outflow rates (right column) for each AGN as a function of the bolometric luminosity, with exact values provided in parentheses. The uncertainties in luminosity are adopted from the typical $\pm$0.3 dex dispersion in the \othree to bolometric luminosity scaling relationship, while the vertical errors are driven by the emission line fit residuals, reddening-correction uncertainty, and the resolution of the parameters in the photoionization model grids (see text in \S5.1). In general, all of the total and peak energetic quantities are positively correlated with increasing AGN luminosity (see text in \S6.1).}
\label{results}
\end{figure*}

\subsection{Correlations with Luminosity}

Overall, several trends are visible in the shapes of the profiles and magnitudes of the rates. The masses of the ionized outflows and their radial extents from the nuclei increase almost monotonically with AGN luminosity. The kinetic energies, momenta, and their outflow rates also scale with bolometric luminosity; however, the relative difference between each AGN varies depending on the adopted metric. To explore these correlations, we present the integrated gas masses, kinetic energies, and momenta, as well as their maximal outflow rates, in Figure~\ref{results} and Table~\ref{results2}. In general, the dispersions in these relationships are smallest for the integrated properties (left column of Figure~\ref{results}), which are less sensitive to differences in the specific model parameters that are averaged over as compared to the peak quantities (right column of Figure~\ref{results}). While the integrated quantities may provide a better measure of the total outflow impact, it is important to note that accurately determining these values required spatially-resolved spectroscopy and photoionization models to account for variations in the gas density and outflow velocity at each location in the NLR.

The peak quantities also trend with increasing luminosity, with two notable outliers. The peak outflow rate for Mrk~573 is somewhat lower than targets of comparable luminosity. This may be due to true physical dispersion or a minor under-estimation in mass at the location of the peak outflow rate from using a single model component at large radial distances. This was required because there were insufficient emission lines to create multi-component photoionization models over the full spatial extent of the outflow in Mrk~573. Oppositely, the peak energy and momentum outflow rates for NGC~1068 are higher than those for targets of comparable luminosity, due to a combination of higher than average mass and outflow velocity at small radii.

In addition to the trends explored in Figure~\ref{results}, it is well-known that the extents of outflows and photoionized gas are correlated with AGN luminosity (e.g., \citealp{Schmitt2003, Greene2011, Liu2013, Hainline2014, Bae2017, Fischer2018, Kang2018, Storchi-Bergmann2018, Sun2018, Luo2020}). However, we defer an analysis of the outflow radii, as these measurements are strongly affected by the depth and sensitivity of the observations, which as discussed in \S3.2, are not uniform for our sample.

These correlations are in good qualitative agreement with recent studies. \cite{Fiore2017} presents an extensive review of molecular, ionized, and X-ray outflow scaling relationships, primarily for AGN with \lbol~$>$ 10$^{44}$ erg s$^{-1}$. The relationship between mass outflow rate and \lbol~is shallower than we find in Figure~\ref{results}, which is in agreement with the conclusions of \cite{Shimizu2019} and \cite{Baron2019} that there is a steepening of these relationships at lower luminosities. Our results sit above the relationship in Figure 11 of \cite{Kakkad2020}, but both results are consistent when we consider that we have a significant mass contribution from lower density, higher ionization gas that is not included in a single-density medium that primarily traces the \othree emission line gas. We present further comparisons with recent studies in the discussion section.

\subsection{Spatially-Resolved Gas Properties}

Our multi-component photoionization models allow us to explore the ionization stratification and range of densities in the outflows. The total ionized gas mass within the outflows ($r <$ $3\arcsec$) of Mrk~78 as derived from the \othree imaging is $M_{ion} = 3.5\times 10^7$ $M_{\odot}$. The four \hst STIS slits used to construct the models encompass $\sim$35\% of the \othree image flux, and thus mass, of the NLR. The mass is divided among the HIGH, MED, and LOW ION model components. Their contributions to the outflow gas mass (with model uncertainties) are:
\begin{displaymath}
\mathrm{HIGH} = 1.5(\pm 0.2) \times 10^7 M_{\odot}\rightarrow 42(\pm6)\%
\end{displaymath}
\begin{displaymath}
\mathrm{MED} = 1.9(\pm 0.2) \times 10^7  M_{\odot}\rightarrow 56(\pm5)\%
\end{displaymath}
\begin{displaymath}
\mathrm{LOW} = 7.0(\pm 0.7) \times 10^5 M_{\odot}\rightarrow 2(\pm1)\%
\end{displaymath}
\noindent
The densities of these components across all spatial locations vary by several orders of magnitude, from $\log(n_{\mathrm{H}}) \approx 0.6 - 6$~cm$^{-3}$. The LOW ION component generally has the highest density and  contributes $\sim$20 -- 50\% of the H$\beta$ luminosity (Table~\ref{cloudy}), but contains only $\sim$2\% of the ionized gas mass. This is a stark demonstration of the selection effects present when calculating gas masses. The luminosity is weighted towards higher density gas that emits efficiently because the free electrons can recombine quickly, but this component contains only a small fraction of the mass. The result is that even modest underestimates of the density for this luminous component significantly over-estimates the gas mass.

The gas masses derived for Mrk~3 and Mrk~78 are consistent with the results of \cite{Collins2009} and \cite{Rosario2007}, respectively, with our estimates moderately larger due to the inclusion of emission outside of the \hst STIS slits. The rapid rise in the gas mass profile for Mrk~3 may be due to a more compact reservoir that is externally fueled through its interaction with the nearby spiral galaxy UGC 3422 \citep{Noordermeer2005, Collins2009, Gnilka2020}. NGC~1068 displayed an enhanced mass outflow rate of $\sim$17 $M_{\odot}$ yr$^{-1}$ at $0\farcs8$ ($\sim$62 pc), which may be the result of a shock that results in additional radiation without additional mass, so we replaced it with the mean of itself and the adjacent two values.

The mass outflow rates for Mrk~78 are consistent with our previous studies, where gas is accelerated in situ at all radii. If the outflow were a steady nuclear flow, the peak outflow rate calculated from the mass in the central bin ($M_{ion} = 6.4 \times 10^4$ $M_{\odot}$) at the location of the peak velocity ($r = 1.2$ kpc) would be $\sim$0.4 $M_{\odot}$ yr$^{-1}$. This is $\sim$24 times smaller than the observed peak outflow rate, indicating that material is entrained in the outflow and/or material is accelerated from reservoirs of gas at each radius. The outflow energetics are consistent with being radiatively driven for at least four of the AGN (Table~\ref{results2}), which have peak momentum outflow rates that are less than the AGN photon momentum (\lbol/c). The energetics are also comparable to recent simulations of radiative driving \citep{Mosallanezhad2019}, but require additional mass modeling of the bulge and galaxy potentials for a proper comparison. Alternatives such as the disk-wind driving model of \cite{Menci2019} can also reproduce some of the observed correlations.

Finally, it is important to note that the masses, kinetic energies, and momenta can be summed over all radii to obtain enclosed totals; however, the outflow rates cannot. Integrating the outflow rates ($\dot M = M v / \delta r$) over radius ($\delta r$) simply returns the momentum ($p = Mv$) and calculating a ``total" mass outflow rate is incorrect. As noted by \cite{Shimizu2019}, this is because the rate needs to be calculated within a common radius. This differs from summing the rates azimuthally to obtain the radial profiles, as these measurements represent the mass flux through common boundaries.

\begin{deluxetable*}{lccccccccc}
\tabletypesize{\small}
\setlength{\tabcolsep}{0.03in} 
\tablecaption{Integrated and Peak Outflow Properties}
\tablehead{
\colhead{Catalog} & \colhead{$\log$(\lbol)} & \colhead{Total $M$} &\colhead{$\dot M_{max}$} & \colhead{Total $E$} & \colhead{$\dot E_{max}$} & \colhead{Total $p$} & \colhead{$\dot p_{max}$} & \colhead{$\dot E_{max}$/\lbol} & \colhead{$\dot pc$/\lbol \vspace{-0.5em}}\\
\colhead{Name} & \colhead{(erg s$^{-1}$)} & \colhead{($\log M_{\odot}$)} &\colhead{($M_{\odot}$ yr$^{-1}$)} & \colhead{($\log$ erg)} & \colhead{($\log$ erg s$^{-1}$)} & \colhead{($\log$ dyne s)} & \colhead{($\log$ dyne)} & \colhead{(\%)} & \colhead{(\%) \vspace{-0.5em}}\\
\colhead{(1)} & \colhead{(2)} & \colhead{(3)} &\colhead{(4)} & \colhead{(5)} & \colhead{(6)} & \colhead{(7)} & \colhead{(8)} & \colhead{(9)} & \colhead{(10)}
}
\startdata
NGC 4151	 & 	43.9	$\pm$	0.3	 & 	5.48	$\pm$	0.07	 & 	3.01	$\pm$	0.45	 & 	54.01	$\pm$	0.08	 & 	41.63	$\pm$	0.09	 & 	46.52	$\pm$	0.08	 & 	34.06	$\pm$	0.08	 & 	0.54	$\pm$	0.11	 & 	437	$\pm$	85	 \\ 
NGC 1068	 & 	45.0	$\pm$	0.3	 & 	5.63	$\pm$	0.05	 & 	9.04	$\pm$	1.13	 & 	54.70	$\pm$	0.05	 & 	42.73	$\pm$	0.04	 & 	46.96	$\pm$	0.05	 & 	34.86	$\pm$	0.05	 & 	0.54	$\pm$	0.05	 & 	214	$\pm$	22	 \\ 
Mrk 3	 & 	45.3	$\pm$	0.3	 & 	6.60	$\pm$	0.07	 & 	7.79	$\pm$	1.15	 & 	54.84	$\pm$	0.06	 & 	42.29	$\pm$	0.07	 & 	47.35	$\pm$	0.06	 & 	34.47	$\pm$	0.06	 & 	0.10	$\pm$	0.02	 & 	44	$\pm$	6	 \\ 
Mrk 573	 & 	45.5	$\pm$	0.3	 & 	6.33	$\pm$	0.05	 & 	3.35	$\pm$	0.60	 & 	54.63	$\pm$	0.05	 & 	41.81	$\pm$	0.07	 & 	47.17	$\pm$	0.05	 & 	34.22	$\pm$	0.05	 & 	0.02	$\pm$	0.01	 & 	15	$\pm$	1	 \\ 
Mrk 78	 & 	45.9	$\pm$	0.3	 & 	7.09	$\pm$	0.13	 & 	9.32	$\pm$	4.60	 & 	55.70	$\pm$	0.13	 & 	42.48	$\pm$	0.17	 & 	48.14	$\pm$	0.13	 & 	34.77	$\pm$	0.13	 & 	0.04	$\pm$	0.02	 & 	22	$\pm$	6	 \\ 
Mrk 34	 & 	46.2	$\pm$	0.3	 & 	7.19	$\pm$	0.06	 & 	12.45	$\pm$	2.72	 & 	55.97	$\pm$	0.06	 & 	43.10	$\pm$	0.09	 & 	48.28	$\pm$	0.06	 & 	35.15	$\pm$	0.06	 & 	0.08	$\pm$	0.02	 & 	26	$\pm$	3
\enddata
\tablecomments{The tabulated results from Figure~\ref{results}. The columns are (1) target name, (2) bolometric luminosity, (3) total mass of the ionized gas, (4) peak mass outflow rate, (5) total kinetic energy, (6) peak kinetic energy outflow rate, (7) total momentum, (8) peak momentum outflow rate, (9) peak kinetic energy outflow rate divided by the bolometric luminosity (percentage), and (10) the peak momentum outflow rate divided by the photon momentum (percentage).}
\label{results2}
\vspace{-2em}
\end{deluxetable*}

\section{Discussion}

We discuss the assumptions of our analysis (\S~\ref{assumptions}), compare our results with recent outflow studies (\S~\ref{recent}), highlight connections with X-ray outflows (\S~\ref{multiphase}), and explore the implications of our results in the context of AGN feedback (\S~\ref{implications})\footnote{A portion of this discussion has been adapted from \cite{Revalski2019}: ``Quantifying Feedback from Narrow Line Region Outflows in Nearby Active Galaxies", Dissertation, Georgia State University, 2019. \url{https://scholarworks.gsu.edu/phy_astr_diss/114}.}. The result-driven reader may choose to forgo \S~\ref{assumptions} without a loss of continuity.

\subsection{Assumptions\label{assumptions}}

Our goal is to produce high-precision measurements of the radial gas mass profiles, outflow rates, and energetics. We discuss here the assumptions underlying our techniques that were not explicitly addressed elsewhere and attempt to characterize their effects on our results.

\paragraph{Galaxy Distances}

Uncertainties in the adopted distance to each AGN have the potential to affect our results. We have calculated these distances using Hubble's Law, which determines the distance to each galaxy based on its measured recessional velocity ($D = v/H_0$). This process assumes that a galaxy's motion is dominated by Hubble flow due to the expansion of the universe. However, galaxies also have their own peculiar velocities that may be as large as $\sim$600 km s$^{-1}$. This effect is negligible for galaxies at distances of $>$ 85 Mpc ($z > 0.02$), introducing an uncertainty of $<$ 10\%. The closest galaxy in our sample is NGC~4151, for which \cite{Crenshaw2015} adopted a distance of $D = 13.3$ Mpc. Subsequently, the distance to NGC 4151 has been independently measured to be $16.1 \pm 0.5$ Mpc using Cepheid variables \citep{Yuan2020}. Similarly, a Tully-Fisher estimate of the distance for NGC~1068 using a $B$-band calibration \citep{Tully2008} and a $k$-corrected apparent magnitude of $9.23 \pm 0.01$  yields a distance of $D = 13.0 \pm 3.8$ Mpc, which agrees with the redshift estimate (Table~\ref{sample}) within the uncertainties (J. Robinson et al., private communication).

Changing the adopted distance for a galaxy corresponds to a shift in our adopted spatial scale, which would alter the outflow radii and bin sizes used to calculate the gas masses and outflow rates. If a new distance of $D_1$ is adopted, then compared to the previous $D_2$ the spatial scale would change by a factor of $(D_1/D_2)$. In addition, the luminosities used to calculate the gas mass profiles would change by a factor of $(D_1/D_2)^2$. While the effects of distance uncertainties are small for our sample based on the accuracy of our adopted distances, the kinematic and modeling results can be scaled to calibrate the outflow results for different distance estimates.

\paragraph{Outflow Geometries}

Our results rely on adopting a geometric model of the host galaxy and outflow orientations to deproject the observed radial distances and line-of-sight velocities to their intrinsic values. In the cases of Mrk~573 and Mrk~34, the data support outflow along the galactic disk due to alignment between the ionizing bicone and the disk, based on the structure observed in \hst imaging. Specifically, the [O~III] emission corresponds to arcs of emission that can be traced to inner spiral dust lanes, as well as fueling flows of warm molecular gas in the case of Mrk~573 \citep{Fischer2017}, implying driving of the outflows off of spiral dust lanes within the host galaxy disk. The orientations of the disks were constrained with either kinematic models of the stellar velocity fields or isophotal ellipse fitting to continuum images \citep{Fischer2017, Fischer2018}. In the cases of NGC~4151, NGC~1068, Mrk~3, and Mrk~78, the kinematics are well fit by biconical outflow models with material flowing along the axes of the bicone \citep{Ruiz2001, Das2005, Crenshaw2010, Fischer2011}. These models are nevertheless consistent with in situ acceleration because the galactic disks have finite thicknesses that allow for ionization and acceleration of disk material by the ionizing bicone regardless of the orientation of the bicone with respect to the disk \citep{Crenshaw2010b, Fischer2011}. Interestingly, \cite{Takeo2020} suggest that the biconical morphology may play a key role in fueling AGN. If these systems were interpreted in the framework of the disk-flow model, then the deprojected outflow velocities would be higher, leading to larger outflow rates by up to factors of a few.

\paragraph{Azimuthal Symmetry}

The measured gas kinematics and quantities derived from our photoionization models are based on the emission that occupies the narrow \hst STIS long-slits. We then use \hst \othree images of the NLRs to calculate the total ionized gas mass and outflow rate at each radius. This process requires us to assume that the quantities derived within the long-slit are symmetric over all azimuthal angles at each distance, which may not be the case considering the biconical morphology of the NLR. Specifically, the outflow velocity, density, and reddening laws that are derived within the spectral slits are assumed to hold elsewhere in the NLR.

As a first measure, we quantify the \othree flux outside of the nominal bicone in each AGN by dividing the elliptical annuli into smaller azimuthal segments. We find that the large solid angle of weaker emission along the bicone minor axes correspond to a small fraction of the total \othree luminosity. Specifically, for NGC~4151, Mrk~573, and Mrk~34, adopting bicone half-opening angles of $\theta = 33\degr$, $38\degr$, and $40\degr$, respectively, results in only 31\%, 19\%, and 20\% of the [O~III] flux outside of the nominal bicones \citep{Revalski2019}.

Supplementary studies of these AGN (\citealp{Das2005, Storchi-Bergmann2010, Fischer2017}) indicate that to first order the outflow velocity is not a strong function of azimuthal angle within the ionizing bicones. In addition, while the derived density and reddening laws used to calculate the ionized gas masses are unlikely to hold precisely along the NLR minor axis, our APO DIS observations indicate that this gas is AGN ionized; however, see the recent \hst study by \cite{Ma2020}. While the material may be susceptible to more foreground reddening, it would result in only a minor underestimation of the gas mass along the NLR minor axis, which already corresponds to $\lesssim 30\%$ of the total \othree luminosity. We further reduced the impact of assuming azimuthal symmetry by dividing the elliptical annuli into two sections, modeling each half of the bicone independently. This refinement was important for Mrk~34 and Mrk~78, where the density and related physical conditions were bimodal across the NLR bicone.

\paragraph{Ionizing Continuum}

In our Cloudy photoionization models we adopted a standard Seyfert power-law SED that has been derived primarily from studies of Type 1 AGN (e.g., \citealp{Schmitt1997, Alonso-Herrero2003, Jin2012}). Due to the obscured nature of Type 2 AGN, their ionizing source cannot be directly detected. We normalized the SEDs to calibrated measures such as the $2-10$ keV or \othree luminosities. In the case of Mrk~34, the AGN is Compton-thick and even higher energy X-rays were used to model the SED \citep{Gandhi2014}. These quantities are believed to be isotropic and thus are used to calculate the bolometric luminosities based on calibrations derived from Type~1 AGN.

The scatter in these relationships are typically factors of $\sim$3 -- 4 and ultimately affect our estimate of the number of ionizing photons per second ($Q_{ion}$) emitted by the AGN that are used in our Cloudy models. $Q_{ion}$ and the deprojected distances were used to constrain the gas densities as functions of the ionization parameter at each distance. Based on changes in the predicted emission line ratios over small ranges of ionization parameter and density, it is likely that changes in $Q_{ion}$ by factors of $\sim$2 -- 4 would be indistinguishable within the uncertainties.

As each of the Cloudy model components may have a different ionization parameter and density, and thus a different mass, quantifying the resulting change in the total gas mass is not straightforward. However, larger changes in $Q_{ion}$ that may occur if the AGN is changing in luminosity on timescales comparable to the light-crossing time of the NLR would be captured, as the best-fit models would be unable to reproduce the density-sensitive emission line ratios at the ionization parameters required to match their overall strengths relative to the H$\beta$ emission line. In the case of Mrk~78, the initial estimate of the bolometric luminosity from \cite{Woo2002} was too low by a factor of 40, resulting in low model densities that could not match all of the key diagnostic lines and overestimated the gas masses.

\paragraph{Extinction Curves}

We adopted a Milky Way Galactic extinction curve \citep{Savage1979, Cardelli1989} to correct the observed emission line ratios that were used to model the ionized gas mass. This assumes that the standard Galactic reddening law applies within both our Galaxy and within the AGN host galaxy. This is unlikely to be the case for all nearby AGN, and in the case of Mrk~3 \cite{Collins2005} found the attenuation to be consistent with an LMC-type extinction curve. It is difficult to distinguish between models without extensive UV spectroscopy, as the differences between various extinction curves are largest in the UV and they have a smaller impact in the optical. For the majority of our galaxies, using a Milky Way Galactic extinction curve results in reddening-corrected emission line ratios that agree well with the model predictions, particularly for the H and He recombination lines that are robust and largely unaffected by the specific Cloudy model parameters.

\subsection{Comparison with Recent Studies\label{recent}}

We find peak mass outflow rates of $\dot M_{out} \approx 3 - 12$ $M_{\odot}$ yr$^{-1}$ that are correlated with AGN luminosity. These outflow rates are comparable to those found by several recent studies. At low redshifts, \cite{Storchi-Bergmann2010} and \cite{Barbosa2014} found global mass outflow rates of $\sim$2 $M_{\odot}$ yr$^{-1}$ for NGC~4151 and NGC~1068, respectively, in agreement with our average values for these targets. Interestingly, our mass outflow rates are also comparable to those of local Ultraluminous Infrared Galaxies (ULIRGs) that are fueled by major mergers, with \cite{Rose2018} reporting global mass outflow rates of $\dot M_{out} \approx 0.1 - 11$ $M_{\odot}$ yr$^{-1}$ for a sample of nine local ($z < 0.15$) galaxies. In a spatially-resolved study of NGC~5728, \cite{Shimizu2019} found a peak outflow rate of $\sim$0.1 $M_{\odot}$ yr$^{-1}$ at $\sim$250 pc from the central AGN, with the outflows reaching radial extents of $\sim$600 pc. This target has a comparable luminosity to NGC~4151 and displays strong signatures of inflow fueling the AGN.

At higher redshifts ($z \approx 0.6 - 2$), \cite{ForsterSchreiber2019} found mass outflow rates of $\sim$0.2 -- 20 $M_{\odot}$ yr$^{-1}$, with AGN-driven outflows having significantly higher densities than those driven by star-formation (see also \citealp{Swinbank2019} and \citealp{Fluetsch2020}). Similarly, \cite{Leung2019} found a mean mass outflow rate of $\dot M_{out} \approx 13$ $M_{\odot}$ yr$^{-1}$ and comparable energy outflow rates to our AGN at even higher redshifts ($z \approx 1.4 - 3.8$), with a small number of targets displaying more energetic outflows. Finally, \cite{Kakkad2020} recently completed an investigation of AGN-driven outflows in a sample of more luminous AGN at $z \approx 2$, and when combined with the results of other recent investigations, find a similar trend of increasing mass outflow rates across five orders of magnitude in luminosity.

The commonality in the majority of these studies is that they adopt gas densities of $n_{\mathrm{H}}\sim$1000 cm$^{-3}$, which is significantly higher than the values of $\sim$100 cm$^{-3}$ used in earlier investigations. These higher densities correspond to lower mass and outflow rate measurements as $M_{ion} \propto L/n_{\mathrm{H}}$. These higher densities are consistent with our MED and LOW ION model components that produce the majority of the observed luminosity, and suggest that at least some previously reported outflow rates that adopt densities of $n_{\mathrm{H}}\sim$100 cm$^{-3}$ are overestimated. This trend toward adopting higher densities in mass outflow rate calculations is well-supported by our results and recent studies. Previous estimates of the gas electron density have been obtained using the [S~II] emission line doublet. As noted by \cite{Kraemer2000c}, \cite{Kakkad2018}, \cite{Baron2019}, \cite{Kewley2019}, \cite{Revalski2019}, \cite{Shimizu2019}, \cite{Davies2020ric}, \cite{Comeron2021}, and others, these lines only probe a single, low-ionization component of the outflows. The density derived from the doublet can be lower than the majority of the gas producing the observed luminosity, which significantly overestimates the mass of the ionized outflows. In a future study we will present several methods for determining spatially-resolved gas densities and compare the resulting gas masses with those derived from multi-component photoionization models.

\subsection{Connecting Multiphase Outflows \label{multiphase}}

Our results account for the optical and UV emission line gas; however, powerful outflows are also observed in the hot X-ray, neutral, and cold molecular gas phases (e.g. \citealp{Perna2017, Bischetti2019, Fluetsch2019, Roberts-Borsani2019, Catalan-Torrecilla2020, Lutz2020, Veilleux2020}). Understanding the multi-phase and multi-scale properties of these outflows requires multiwavelength datasets at the highest possible spatial resolution \citep{Gaspari2020}. These data are being obtained with the current generation of observatories, including the {\it Chandra} X-ray Observatory, the Atacama Large Millimeter Array (ALMA), and will be enhanced with next generation instruments on the \textit{James Webb Space Telescope}. The connection to molecular outflows is discussed later, and in the X-rays all of the targets in our sample have been observed by \textit{Chandra}. \cite{Wang2011a, Wang2011b, Wang2011c} conducted a detailed analysis of the X-ray gas in NGC~4151 and found a global mass outflow rate of $\dot M_{out} \approx 2$ $M_{\odot}$ yr$^{-1}$. This was expanded to a spatially-resolved analysis by \cite{Kraemer2020}, which confirmed a peak outflow rate of $\dot M_{out} \approx 2$ $M_{\odot}$ yr$^{-1}$ at a distance of $\sim$150 pc from the nucleus. In NGC~1068, \cite{Kraemer2015} found the X-ray gas mass is an order of magnitude larger than the optical emission line gas, while in Mrk~3 \cite{Bogdan2017} found an X-ray gas mass of $M_{ion} \approx 1.1 \times 10^7 M_{\odot}$, which is comparable to the mass of the optical outflows that we and \cite{Collins2009} calculated. \textit{Chandra} observations of Mrk~573 were modeled by \cite{Gonzalez-Martin2010} and \cite{Bianchi2010}, with the latter finding two photoionized components required to reproduce the observed emission and ionization parameters that are natural extensions of our optical model parameters. This supports the idea that the optical, UV, and soft X-ray emission arises from a single photoionized region \citep{Bianchi2006, Mazzalay2010}, with localized contributions from shocks detected in the X-rays in similar AGN (e.g. \citealp{Maksym2019}). A similar multiwavelength study of NGC~1365 by \cite{Venturi2018} using MUSE and \textit{Chandra} find comparable mass outflow rates for the optical and X-ray emission line gas, supporting the increased impact of multi-phase outflows. These results indicate that outflows in the X-ray emission line gas may have a common origin with the optical outflows and can have comparable outflow energetics.

\subsection{Implications for Feedback\label{implications}}

A critical open question is: Do outflows provide effective feedback to their host galaxies? The answer to this question depends on the criteria adopted for the definition of \textit{effective} feedback. Commonly used criteria are theoretical thresholds for the peak kinetic luminosity (peak energy outflow rate), which needs to reach $\sim$0.5 -- 5\% of the AGN bolometric luminosity to significantly impact the host galaxy \citep{DiMatteo2005, Hopkins2010}. Interestingly, the peak energy outflow rates of NGC~4151 and NGC~1068, our two lowest luminosity targets, both exceed $\sim$0.5\% of their bolometric luminosities, while the higher luminosity targets average $\sim$0.05\% (see Table~\ref{results2}). However, these criteria are primarily intended for high redshift galaxies that are establishing their bulge structure through ongoing evolution with the central AGN radiating near the Eddington limit. In the local universe, AGN host galaxies have fully-established bulges, and defining effective feedback in the context of evacuating reservoirs of potential star-forming gas and the disruption of star-formation may be more relevant (e.g. \citealp{Smith2020, Garcia-Bernete2021}).

The gas masses and outflow rates for our sample indicate that the ionized gas can be evacuated from the inner bulges on timescales of $\tau \approx M/\dot M \approx 10^6$ years. Recent observations of Mrk~573 with ALMA reveal that the nuclear region ($r<1.4$ kpc) contains $\sim$10$^8$ $M_{\odot}$ of cold molecular gas (Wiklind et al., private communication), which is consistent with the cold-to-ionized gas ratio of similar AGN. If the ionized gas reservoir is continually replenished through the ionization of the cold molecular gas as seen in NGC~4151 \citep{May2020}, then the evacuation timescale for the cold gas in the bulge of Mrk~573 is $\tau \approx M/\dot M \approx 10^8$ years, which is comparable to the duty cycle of an AGN. However, if the molecular gas is also outflowing with comparable velocities, then the enhanced outflow rate would lead to more rapid depletion (see, e.g. \citealp{Baron2020}, \citealp{Fluetsch2020}, and \citealp{Garcia-Bernete2021}). Observations of the molecular gas with ALMA and bulge-mass decomposition modeling \citep{Fischer2017}  of each AGN are required to calculate more precise evacuation timescales and determine if the outflows reach escape velocities.

We also observe that AGN ionized gas in the ENLR extends to kpc scales and can encompass $>$ 10$^{8}$ $M_{\odot}$ of gas. This may inhibit star formation by heating the gas in the galaxy disk and inducing turbulence that prevents collapse into star-forming regions, as evidenced by the high FWHM lines that often extend beyond the outflow regions \citep{Fischer2018}. This heating and turbulence induced feedback mechanism has been suggested by several recent studies (e.g. \citealp{Cheung2016, Morganti2017, Chen2019, Lacerda2020, Wylezalek2020, Zhuang2020, Zinger2020}), and there is evidence that at least some nearby AGN sit below the Kennicutt-Schmidt law \citep{Kennicutt1998, Wang2007} that describes the star-formation rate per unit area as a function of gas surface density. This is worthy of future investigation and will require determining the ENLR gas mass of each AGN from deep \othree imaging, measuring their star formation rates, and comparing them with a matched sample of quiescent galaxies with similar host galaxy types, colors, and gas masses (e.g. \citealp{Rosario2018, Rosario2019}).

\section{Conclusions}

We provide the largest sample to date of spatially-resolved NLR mass outflow energetics that are based on multi-component photoionization models. These results provide important constraints for determining accurate gas masses and outflow rates using spatially-resolved spectroscopy. Our main conclusions are the following:

\begin{enumerate}
\item Spatially-resolved observations are required to properly constrain the properties of ionized outflows. These include the outflow velocity profile, luminosity distribution, radial extent, and separation of kinematic components. Global techniques that utilize spatially integrated spectra and provide a single estimate of the outflow mass and energetics are susceptible to strong selection effects.

\item The adopted gas densities have a profound effect on the derived gas masses, and modeling the optical emission line gas is a multiphase problem that requires multiple density components. When using a single density, estimates from the [S~II] doublet or assuming a constant density of $n_\mathrm{H} = 100$ cm$^{-3}$ can overestimate the gas mass for NLR outflows by more than an order of magnitude in some instances.

\item The outflows are photoionized by the central AGN and are likely driven by radiation pressure, as evidenced by the correlation of mass outflow properties with bolometric luminosity. Shocks are not required for ionizing or driving the optical emission line gas, with a localized contribution from shocks detected in only one object, NGC~1068.

\item The outflows contain total ionized gas masses of $M \approx 10^{5.5} - 10^{7.5}$~$M_{\odot}$ and reach peak velocities of $v \approx 800 - 2000$ km s$^{-1}$. They extend to radial distances of $r \approx 0.1 - 3$ kpc from the nucleus, reaching maximum mass outflow rates of $\dot M_{out} \approx 3 - 12$ $M_{\odot}$ yr$^{-1}$ and encompassing total kinetic energies of $E \approx 10^{54} - 10^{56}$ erg.

\item The ionized gas masses, outflow rates, energetics, and radial extents of the outflows are all positively correlated with the AGN bolometric luminosity. The dispersion in these trends is smaller for the integrated quantities as compared to the peak rates.

\item The outflow rates are consistent with in situ acceleration, where gas is accelerated at multiple radii by the radiation field, rather than a steady nuclear flow. The mass, kinetic energy, and momentum profiles may be summed radially to obtain enclosed totals; however, the radial rates cannot, because they must be sampled within a common radius.
\end{enumerate}

As interest in spatially-resolved outflow studies grows, it has become clear that the choice of gas densities plays a critical role in calculating accurate outflow gas masses. In a forthcoming paper, we will present the results of several density estimate techniques and compare the spatially-resolved gas masses and outflow rates to those obtained using multi-component photoionization models.

In addition, we will expand our sample to properly quantify the slopes and dispersions in the luminosity scaling relations through an approved Cycle 28 \hst program (\hst PID 16246, PI: M.~Revalski) that is obtaining \othree and continuum images of additional Seyfert galaxies with archival \hst STIS spectroscopy. When combined with archival targets, our expanded sample of 22 AGN will span $\sim$4 dex in bolometric luminosity.

\acknowledgements

M.R. thanks Jenny Novacescu and the STScI Library Staff for ADS reference support and Justin H. Robinson for helpful discussions. The authors thank the anonymous referee for helpful comments that improved the clarity of this paper.

This work was supported by the STScI Directors Discretionary Research Fund (DDRF) Proposal 82490. W.P.M. acknowledges support by Chandra grants GO8-19096X, GO5-16101X, GO7-18112X, and GO8-19099X. Support for this work was provided by NASA through grant number HST-GO-15350.001-A from the Space Telescope Science Institute, which is operated by AURA, Inc., under NASA contract NAS 5-26555. 

This paper used the photoionization code Cloudy, which can be obtained from \url{http://www.nublado.org} and the Atomic Line List available at \url{http://www.pa.uky.edu/~peter/atomic/}. This research has made use of NASA's Astrophysics Data System. This research has made use of the NASA/IPAC Extragalactic Database (NED), which is operated by the Jet Propulsion Laboratory, California Institute of Technology, under contract with the National Aeronautics and Space Administration. IRAF is distributed by the National Optical Astronomy Observatories, which are operated by the Association of Universities for Research in Astronomy, Inc., under cooperative agreement with the National Science Foundation.

\facilities{HST(STIS, WFPC2, FOC), ARC(DIS)}

\software{IRAF \citep{Tody1986, Tody1993},
MultiNest \citep{Feroz2019},
Cloudy \citep{Ferland2013},
Mathematica \citep{Mathematica}, Python (\citealp{VanRossum2009}, \url{https://www.python.org}), Interactive Data Language (IDL, \url{https://www.harrisgeospatial.com/Software-Technology/IDL}).
}

\bibliography{references}{}
\bibliographystyle{aasjournal}

\appendix
\restartappendixnumbering
\section{Photoionization Modeling Results for Mrk~78}

\begin{figure*}[h!]
\centering
\includegraphics[width=0.49\textwidth]{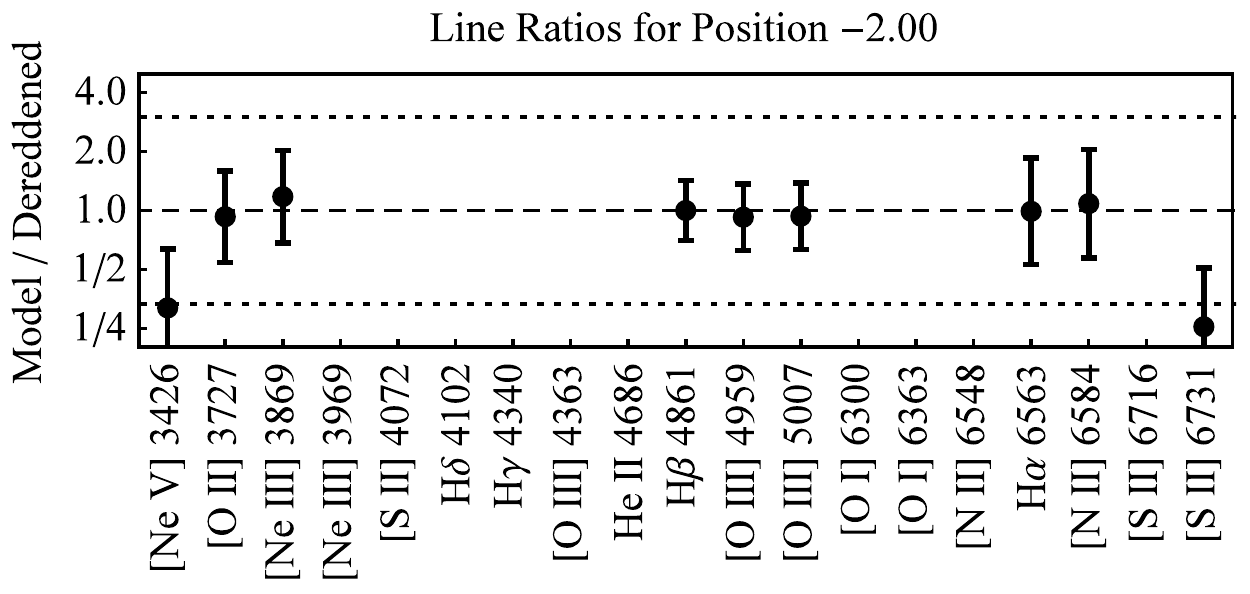}
\includegraphics[width=0.49\textwidth]{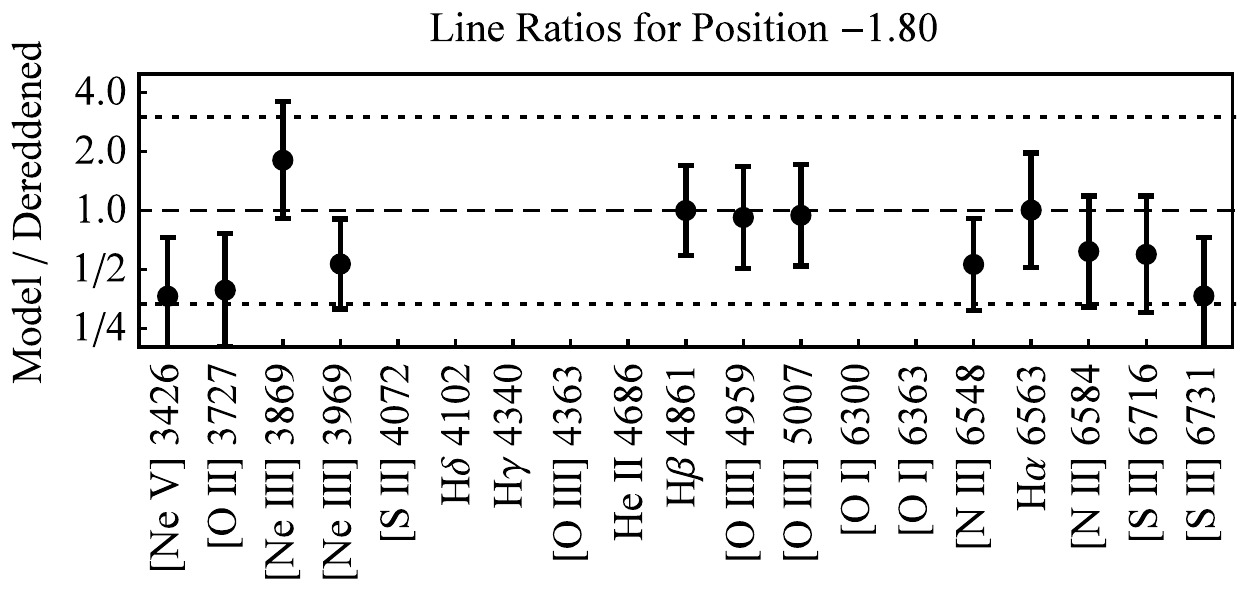}
\includegraphics[width=0.49\textwidth]{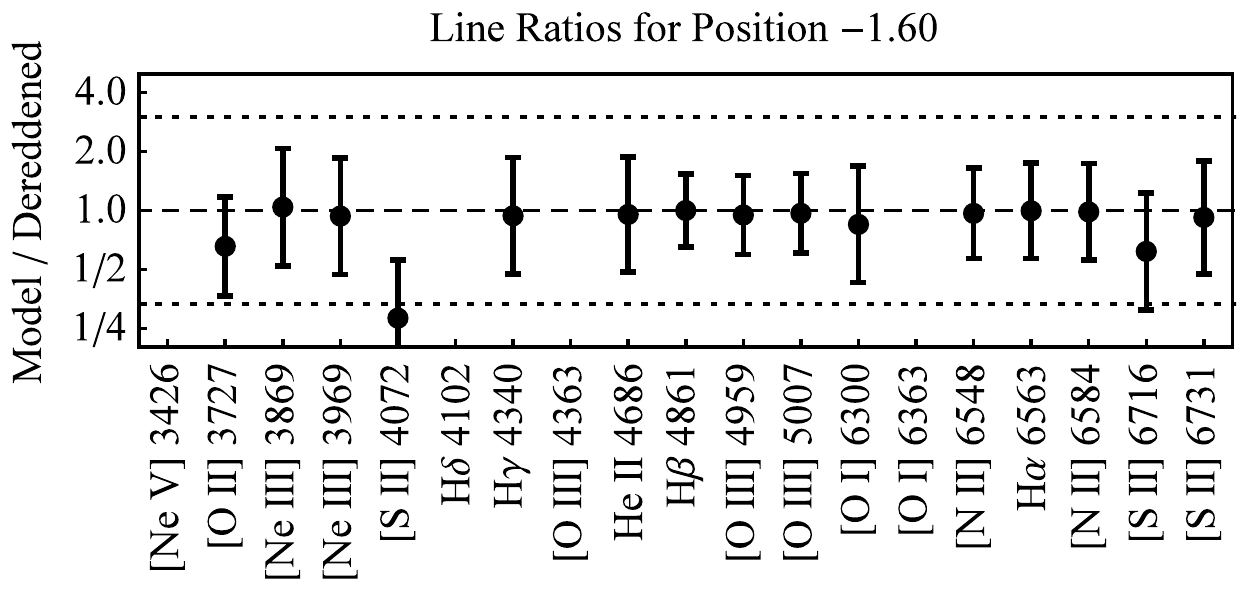}
\includegraphics[width=0.49\textwidth]{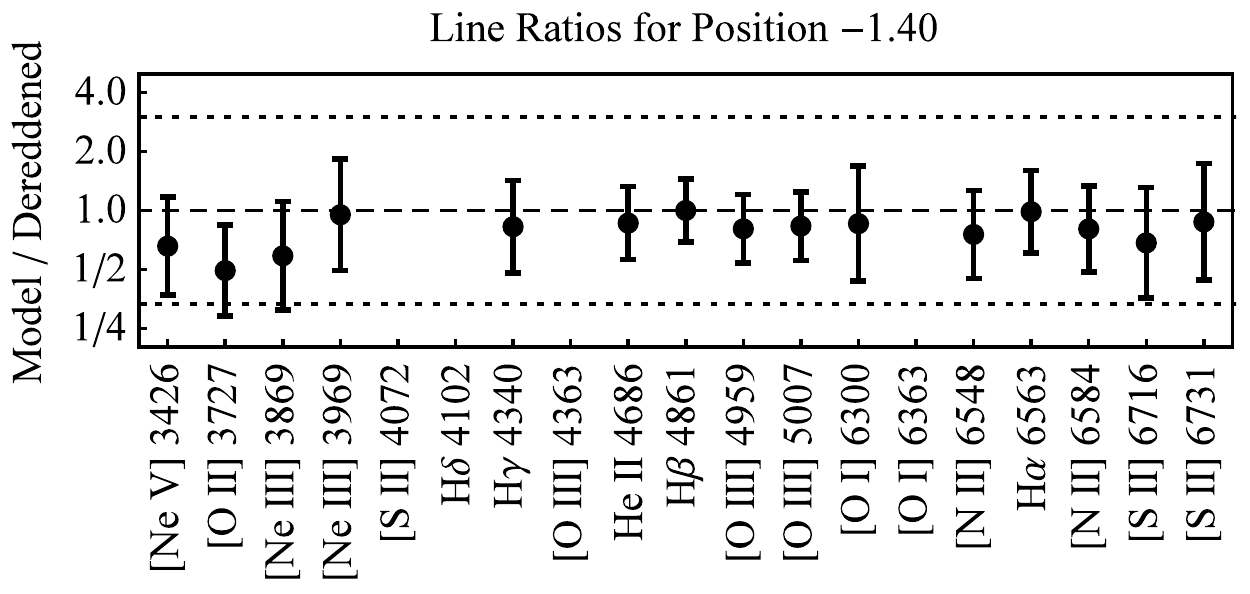}
\includegraphics[width=0.49\textwidth]{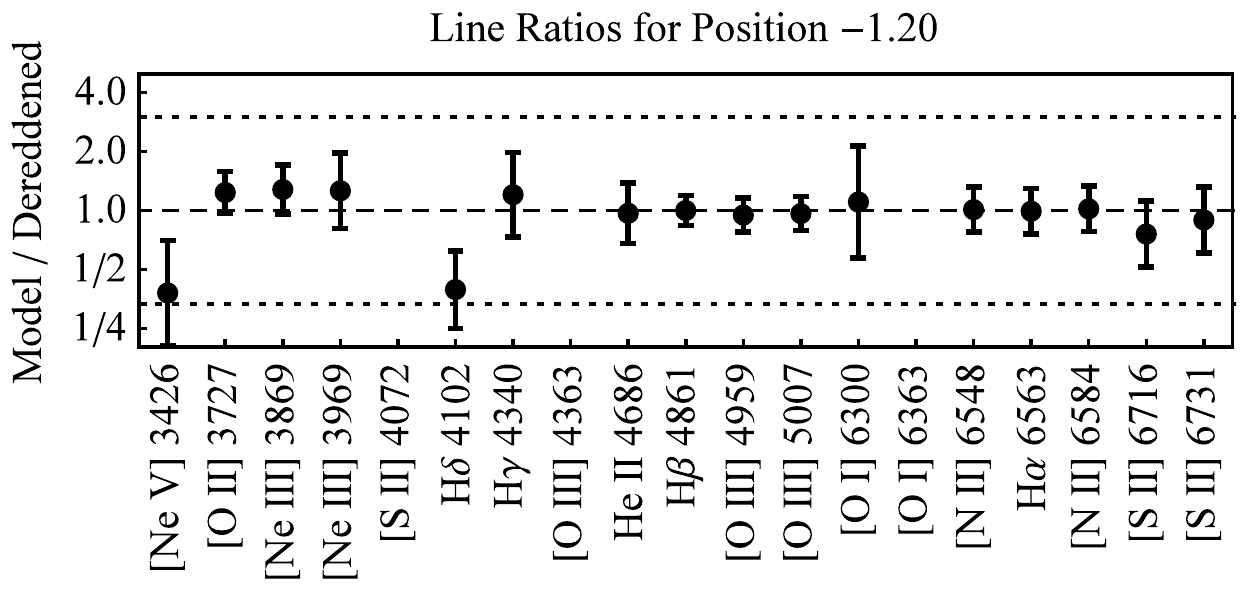}
\includegraphics[width=0.49\textwidth]{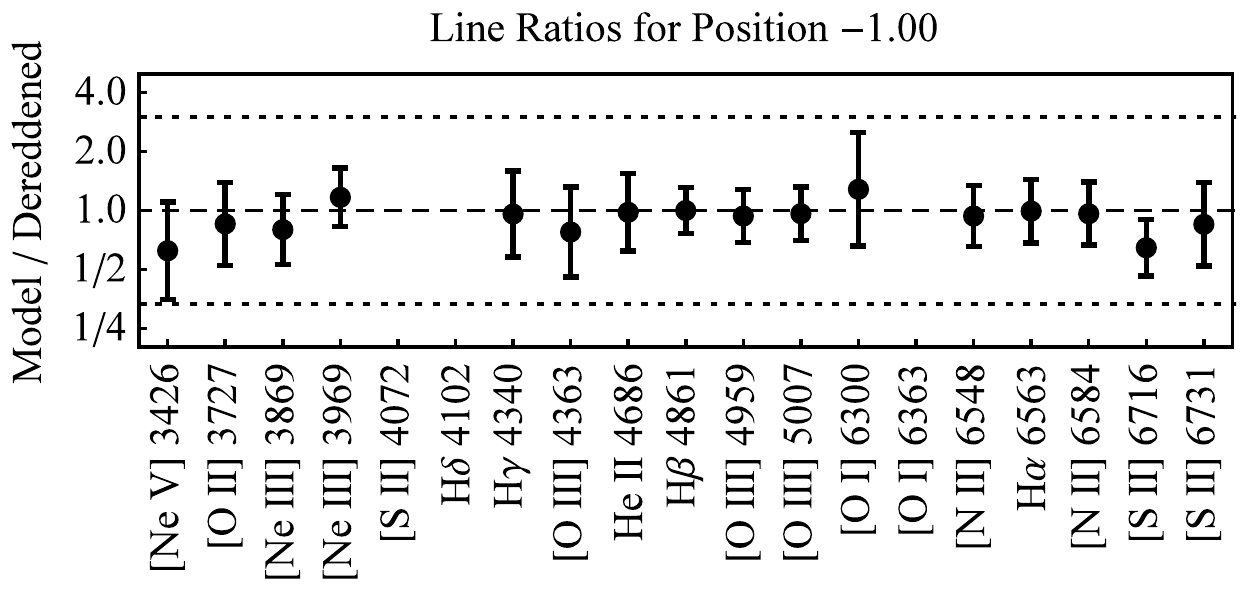}
\includegraphics[width=0.49\textwidth]{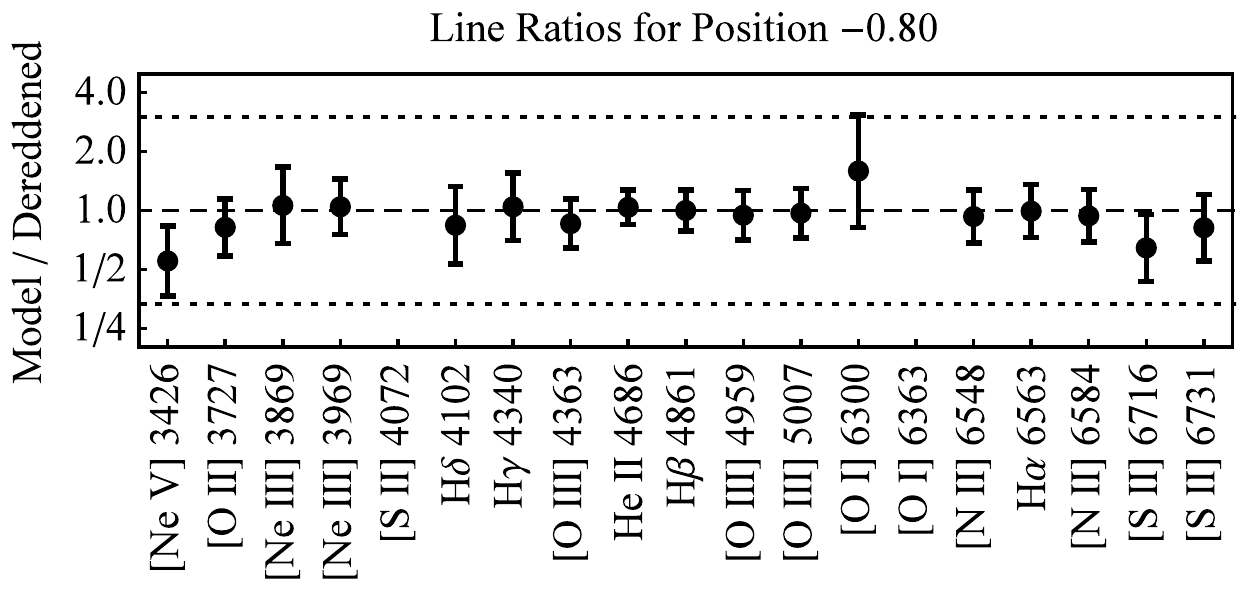}
\includegraphics[width=0.49\textwidth]{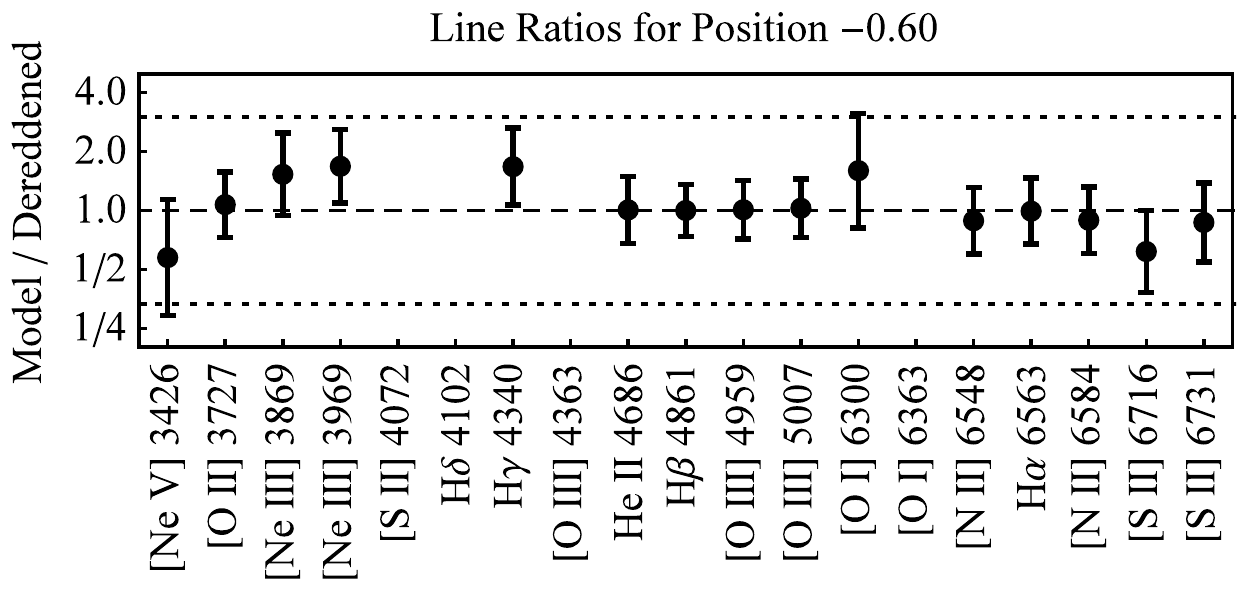}
\caption{The composite model line ratios divided by the dereddened values (Table~\ref{ratios}) for each position (see \S4). The dashed unity lines indicate an exact match between the model and data, while the dotted lines are factor of three boundaries. The vertical tick marks are logarithmically spaced for even distribution around the unity line. Points above this line are over-predicted, while points below are under-predicted by the model.}
\end{figure*}

\addtocounter{figure}{-1}

\begin{figure*}
\centering
\includegraphics[width=0.49\textwidth]{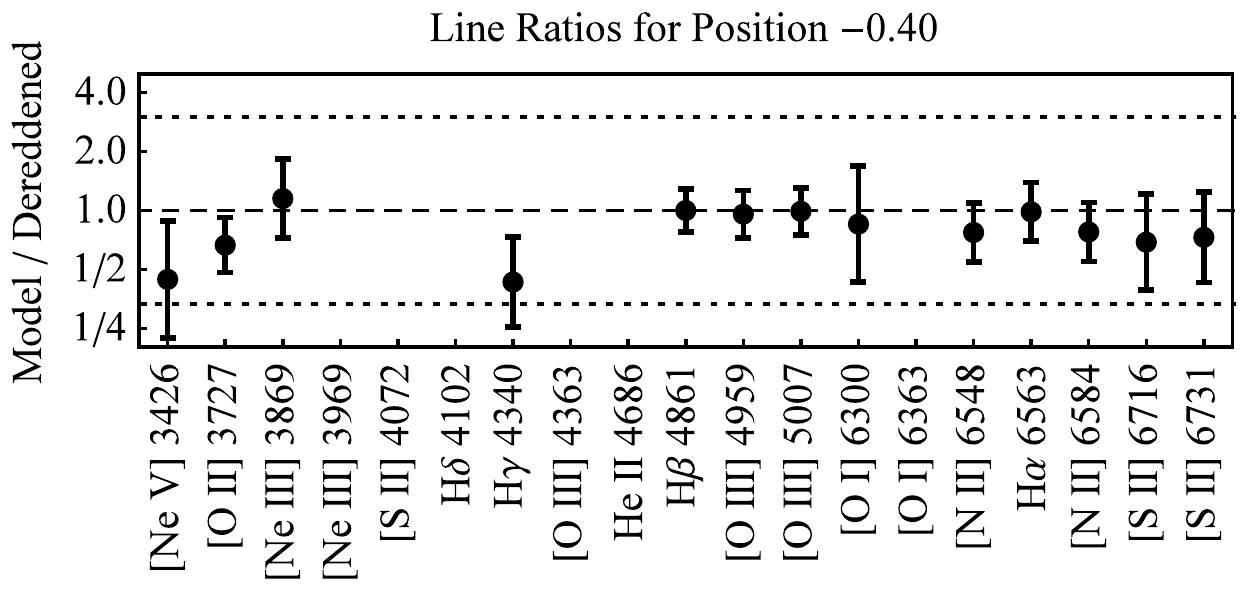}
\includegraphics[width=0.49\textwidth]{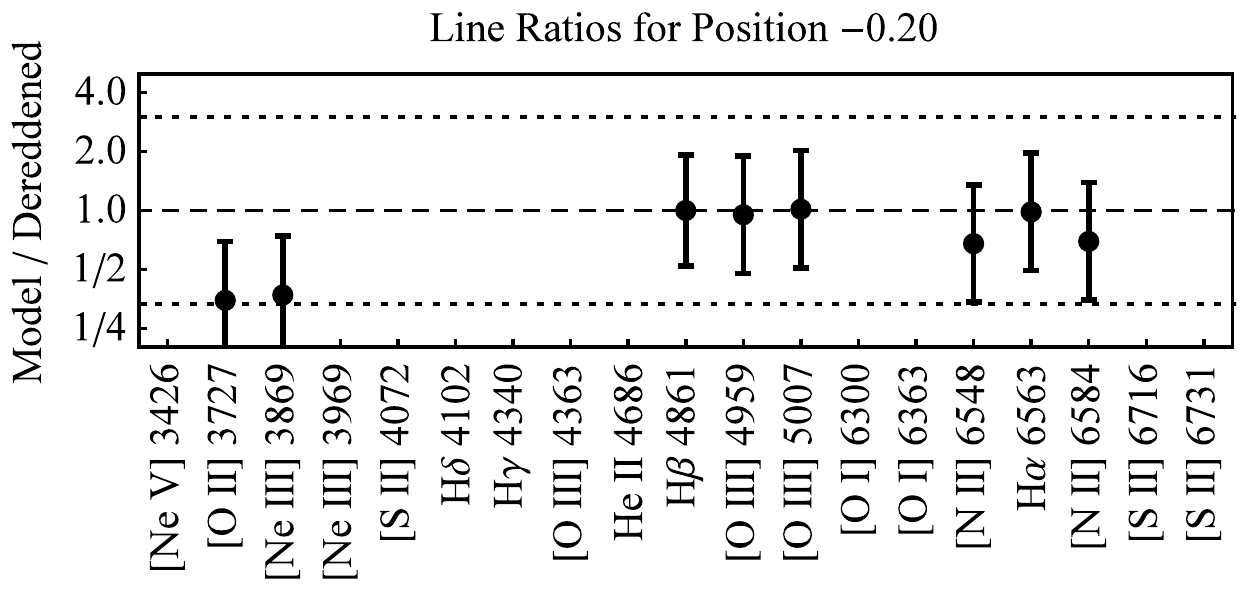}
\includegraphics[width=0.49\textwidth]{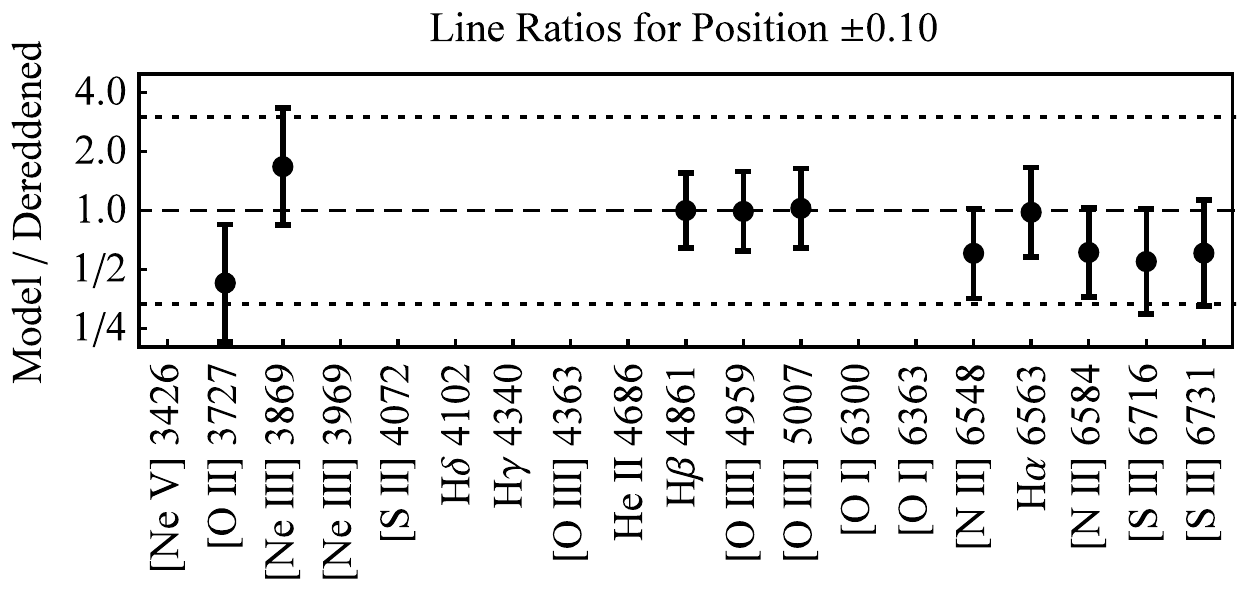}
\includegraphics[width=0.49\textwidth]{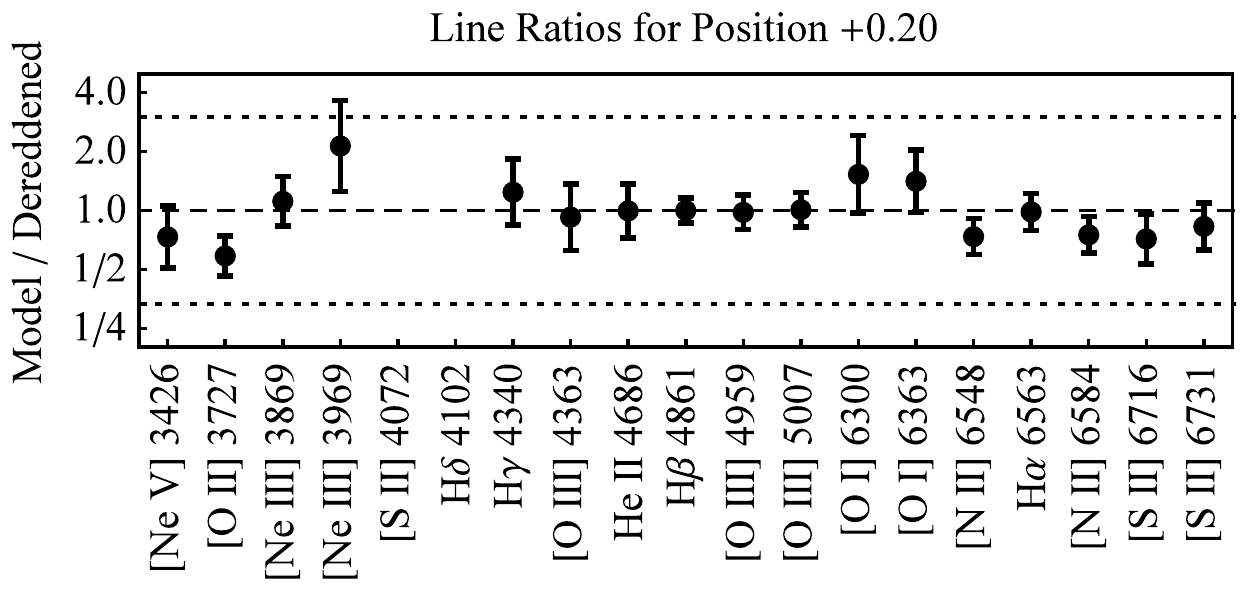}
\includegraphics[width=0.49\textwidth]{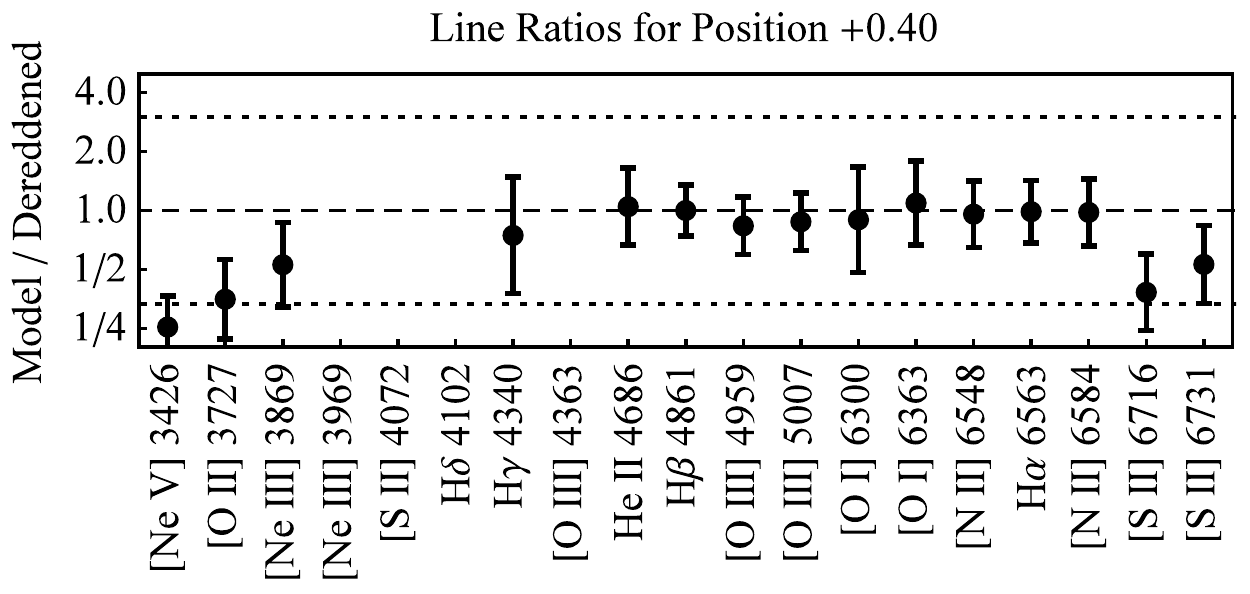}
\includegraphics[width=0.49\textwidth]{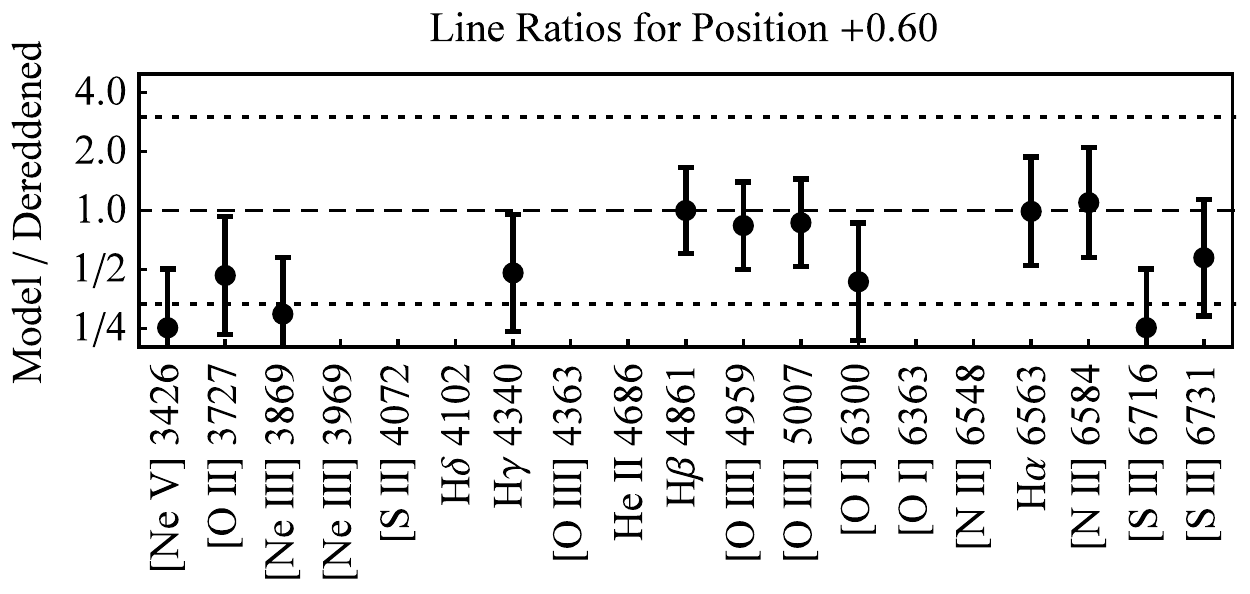}
\includegraphics[width=0.49\textwidth]{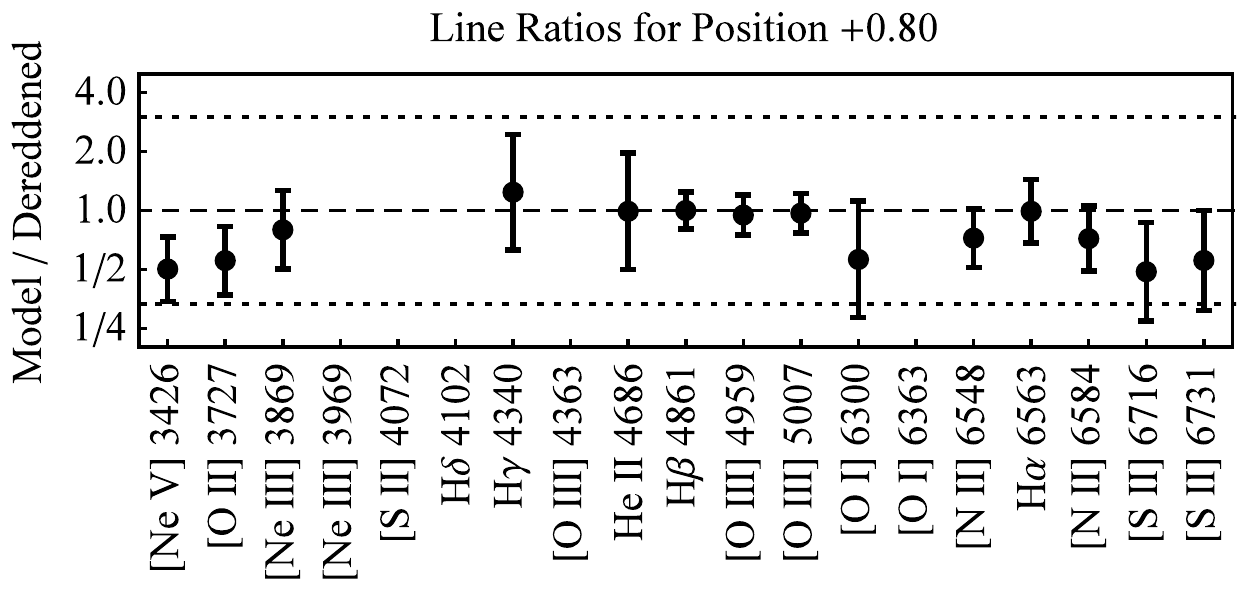}
\includegraphics[width=0.49\textwidth]{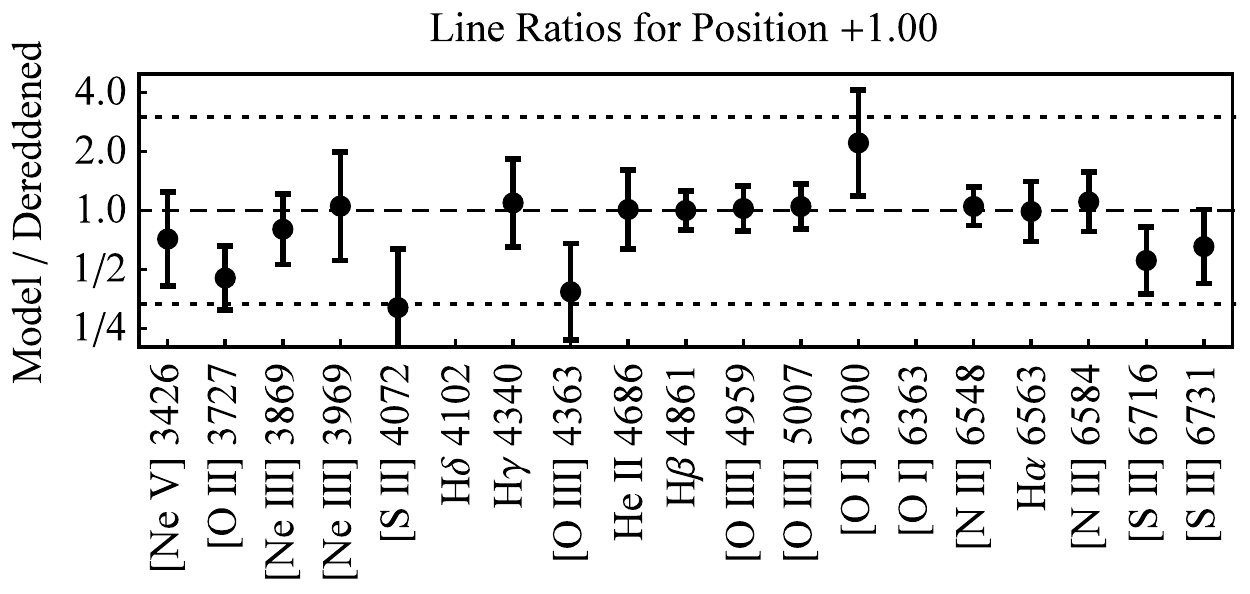}
\includegraphics[width=0.49\textwidth]{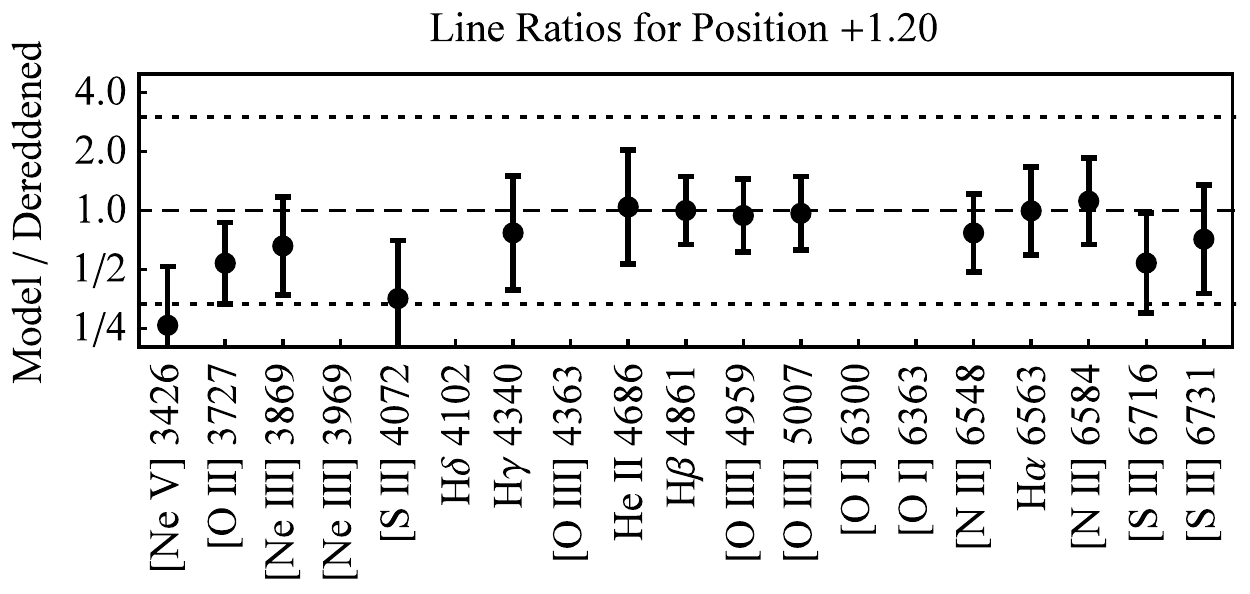}
\includegraphics[width=0.49\textwidth]{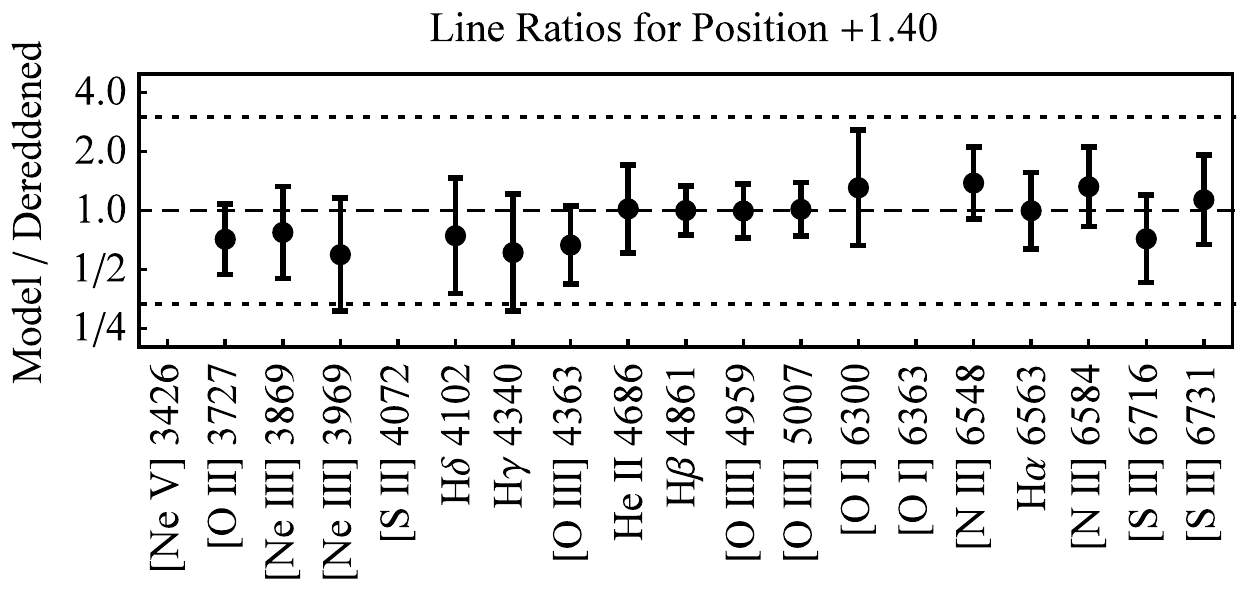}
\caption{{\it continued.}}
\end{figure*}

\addtocounter{figure}{-1}

\begin{figure*}
\centering
\includegraphics[width=0.49\textwidth]{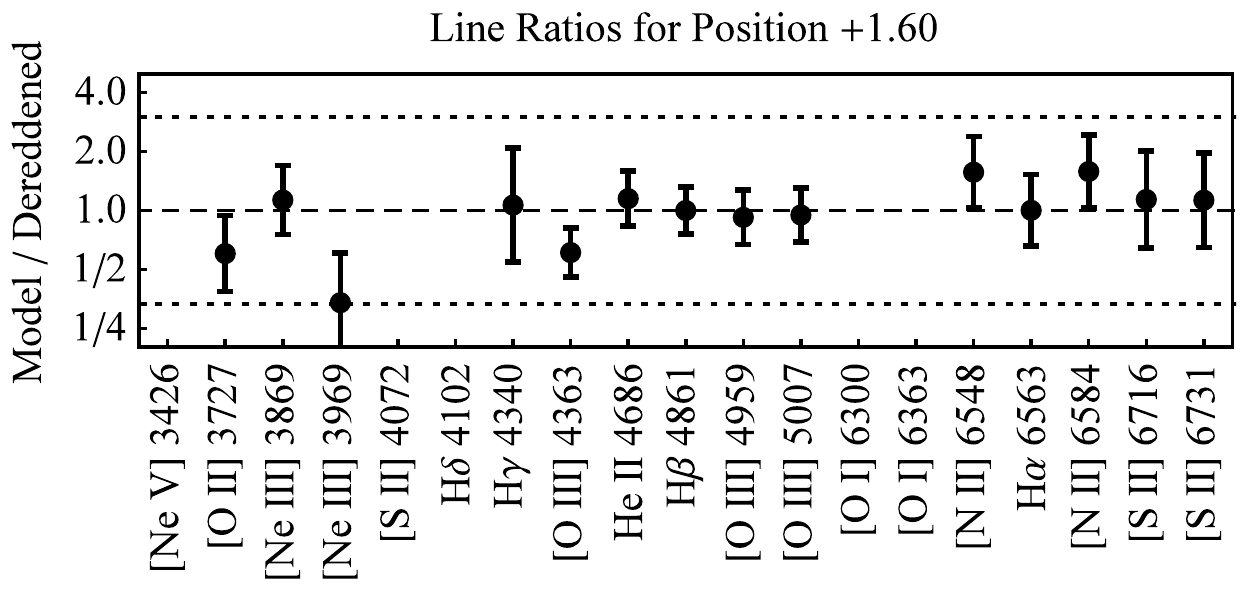}
\includegraphics[width=0.49\textwidth]{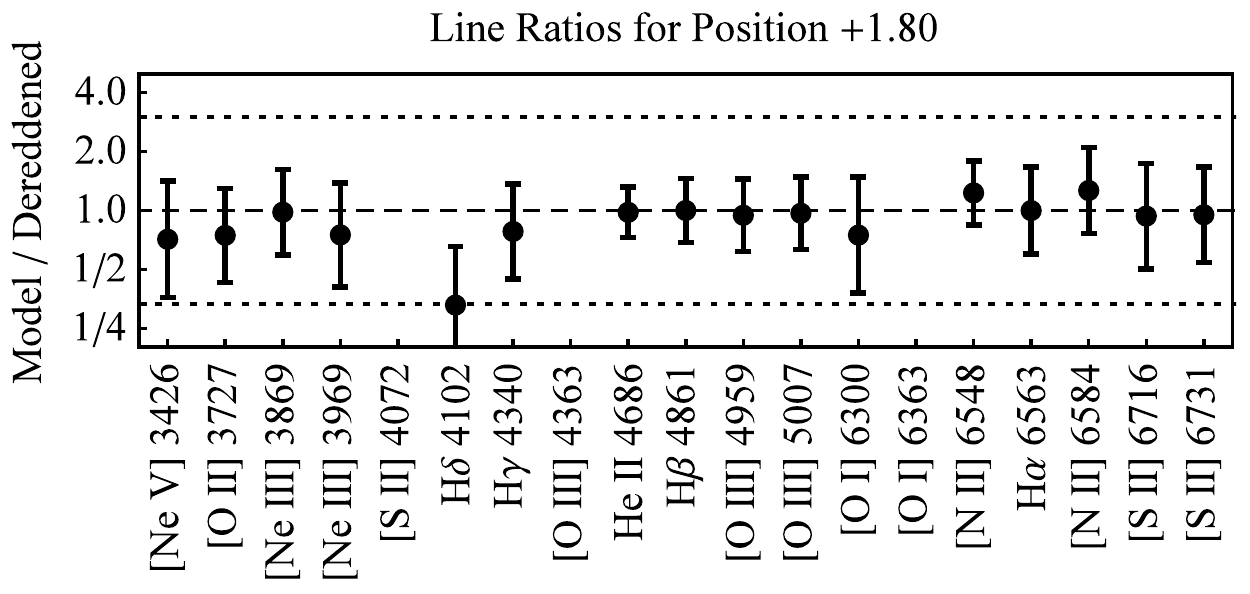}
\includegraphics[width=0.49\textwidth]{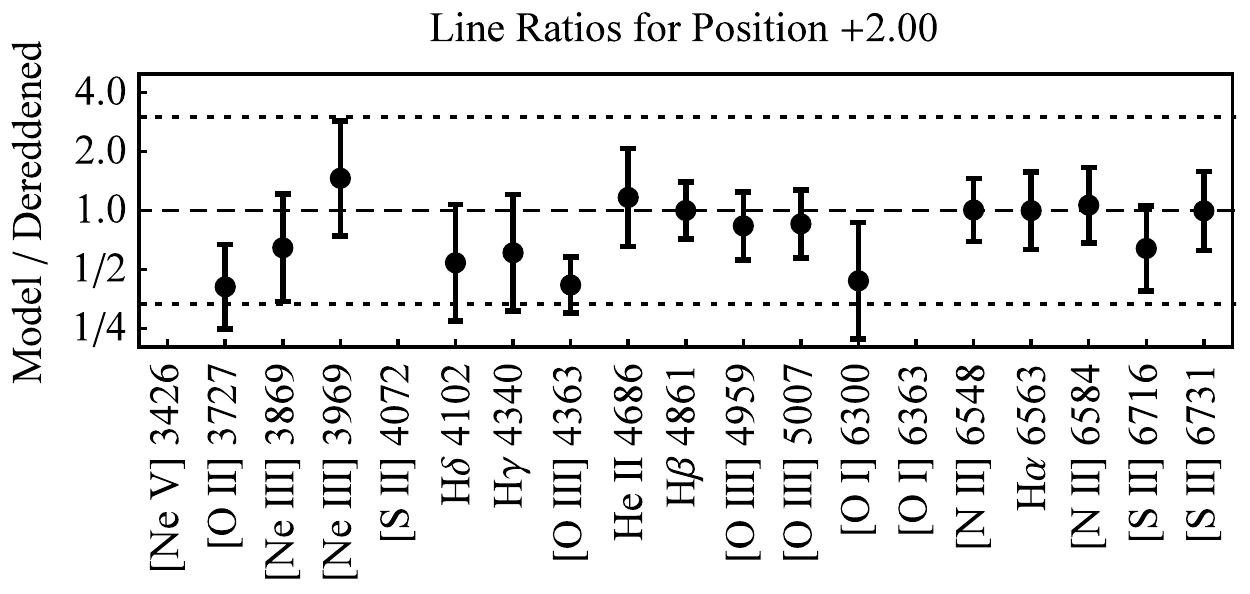}
\includegraphics[width=0.49\textwidth]{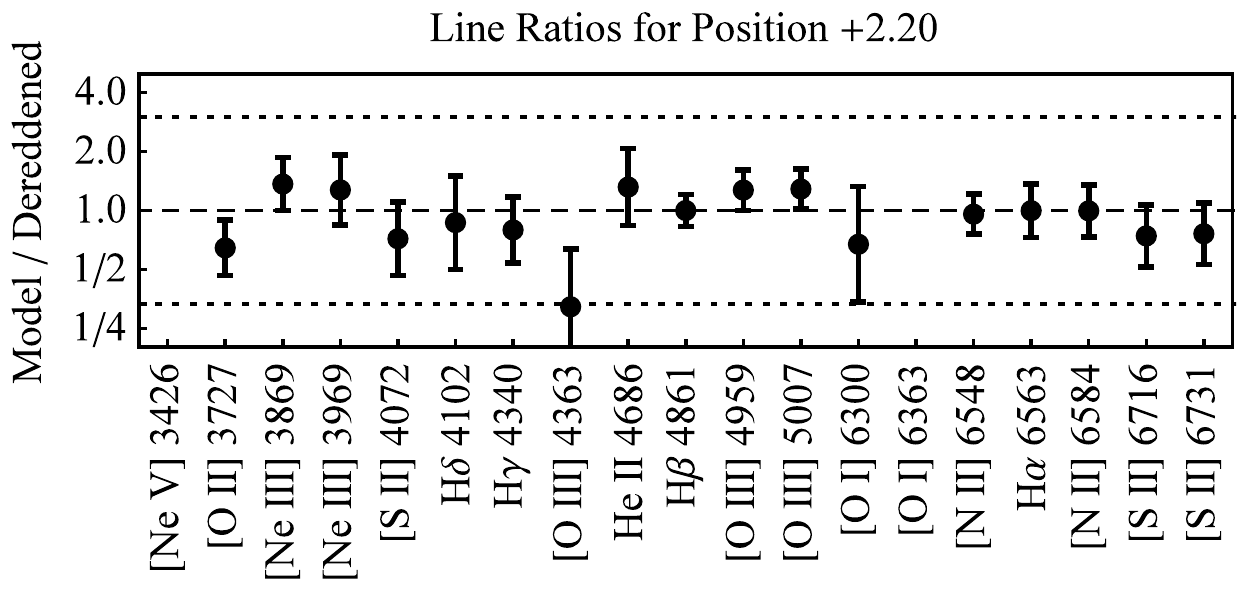}
\includegraphics[width=0.49\textwidth]{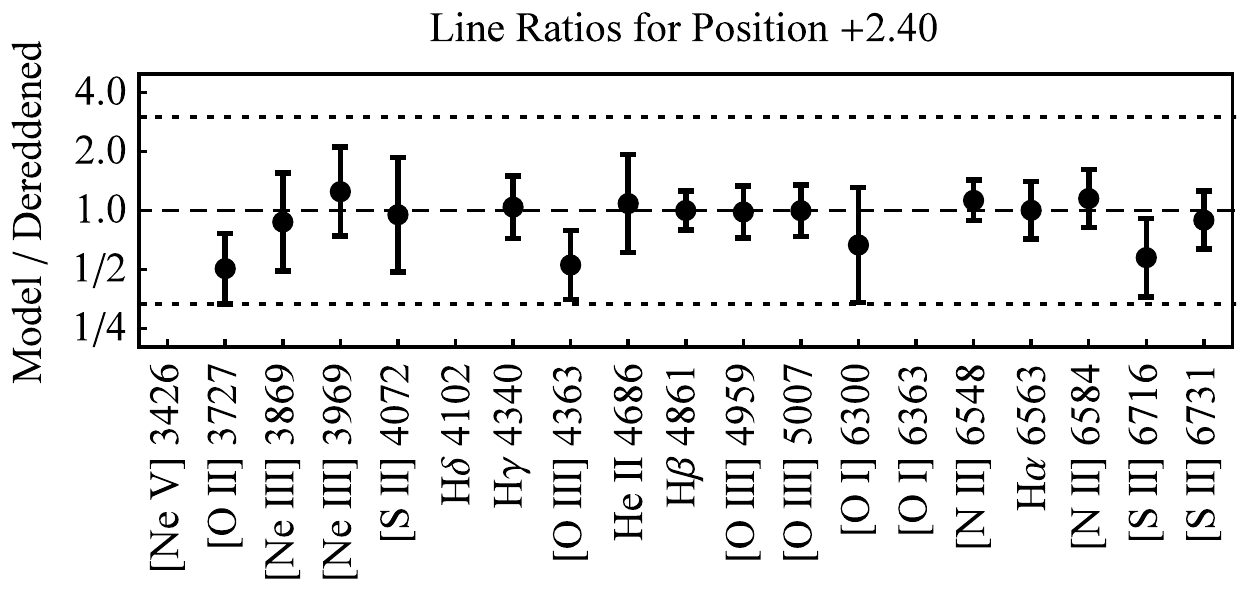}
\includegraphics[width=0.49\textwidth]{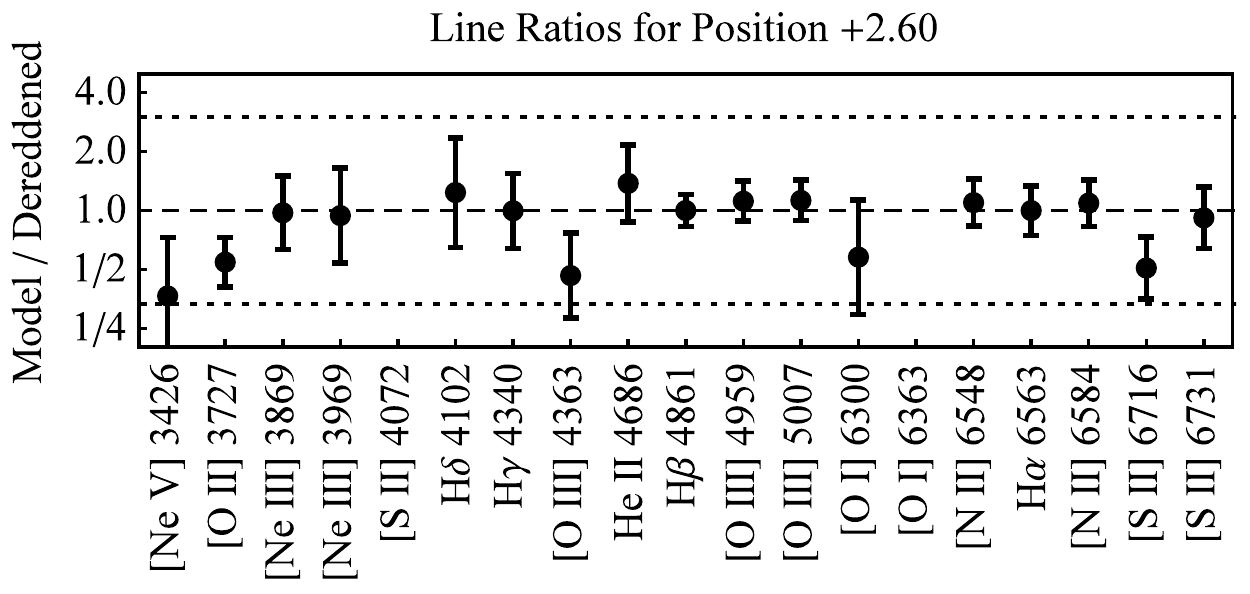}
\includegraphics[width=0.49\textwidth]{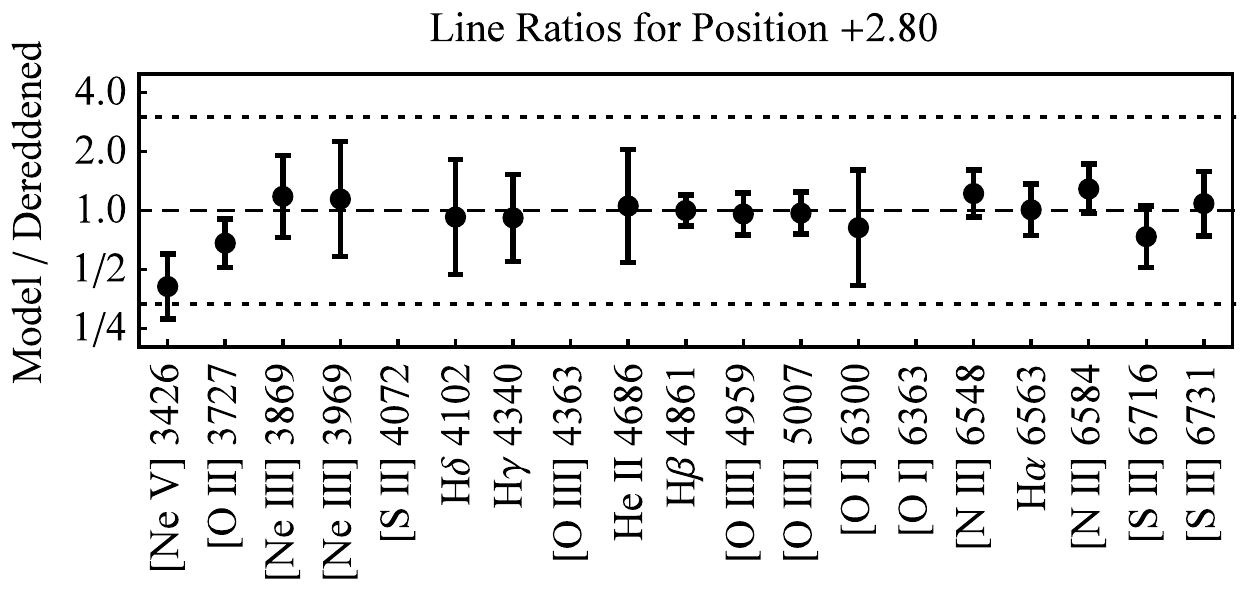}
\includegraphics[width=0.49\textwidth]{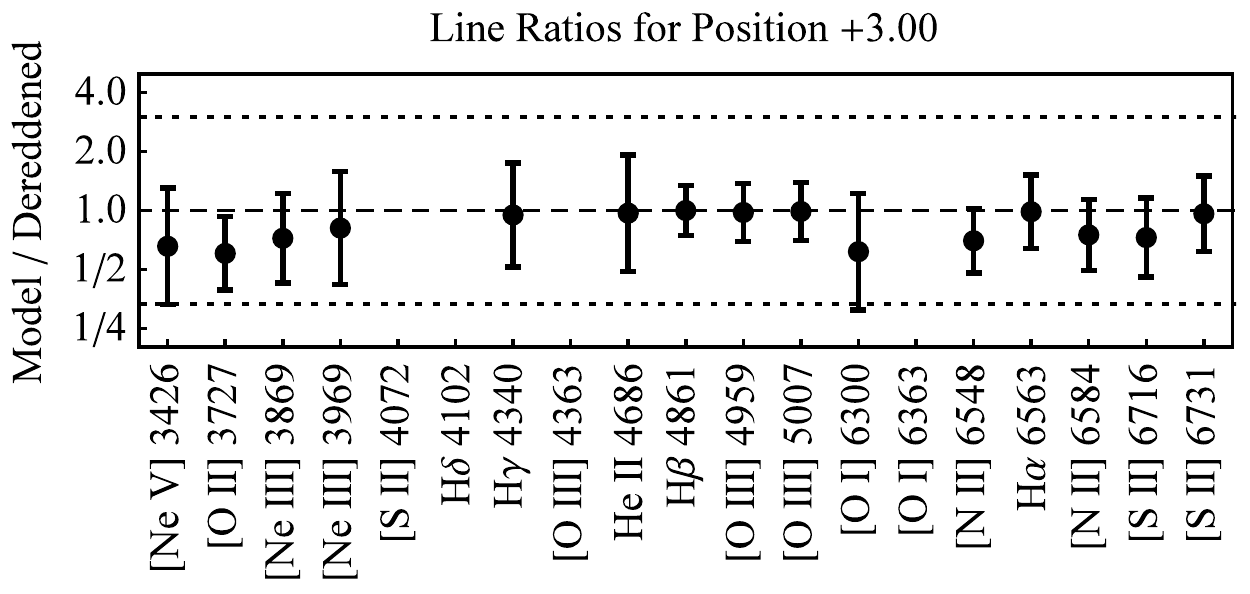}
\caption{{\it continued.}}
\end{figure*}

\setlength{\tabcolsep}{0.1in}
\tabletypesize{\scriptsize}
\begin{deluxetable*}{|c|c|c|c|c|c|c|c|}
\vspace{-0.75em}
\def\arraystretch{0.75}
\tablecaption{Radial Mass Outflow \& Energetic Results}
\tablehead{
\colhead{Distance} & \colhead{Velocity} & \colhead{Mass} & \colhead{$\dot{M}$} & \colhead{Energy} & \colhead{$\dot{E}$} & \colhead{Momentum} & \colhead{$\dot{p}$ \vspace{-1.0em}}\\
\colhead{(pc)} & \colhead{(km s$^{-1}$)} & \colhead{($\log M_{\odot}$)} & \colhead{($M_{\odot}$ yr$^{-1}$)} & \colhead{($\log$ erg)} & \colhead{($\log$ erg s$^{-1}$)} & \colhead{($\log$ dyne s)} & \colhead{($\log$ dyne) \vspace{-1.0em}}\\
\colhead{(1)} & \colhead{(2)} & \colhead{(3)} & \colhead{(4)} & \colhead{(5)} & \colhead{(6)} & \colhead{(7)} & \colhead{(8)}
}
\startdata
{\bf NGC~4151} \\ \hline
12.8	 & 	106.6	 & 	0.03	$\pm$	0.01	 & 	0.04	$\pm$	0.01	 & 	50.57	$\pm$	0.19	 & 	38.18	$\pm$	0.19	 & 	43.85	$\pm$	0.19	 & 	31.46	$\pm$	0.19	 \\ 
21.4	 & 	178.3	 & 	0.07	$\pm$	0.01	 & 	0.14	$\pm$	0.02	 & 	51.31	$\pm$	0.19	 & 	39.15	$\pm$	0.19	 & 	44.36	$\pm$	0.19	 & 	32.20	$\pm$	0.19	 \\ 
29.9	 & 	249.1	 & 	0.14	$\pm$	0.02	 & 	0.42	$\pm$	0.05	 & 	51.94	$\pm$	0.19	 & 	39.92	$\pm$	0.19	 & 	44.85	$\pm$	0.19	 & 	32.82	$\pm$	0.19	 \\ 
38.4	 & 	319.9	 & 	0.20	$\pm$	0.03	 & 	0.78	$\pm$	0.10	 & 	52.31	$\pm$	0.19	 & 	40.40	$\pm$	0.19	 & 	45.11	$\pm$	0.19	 & 	33.19	$\pm$	0.19	 \\ 
47.0	 & 	391.5	 & 	0.25	$\pm$	0.03	 & 	1.17	$\pm$	0.16	 & 	52.58	$\pm$	0.18	 & 	40.75	$\pm$	0.18	 & 	45.28	$\pm$	0.18	 & 	33.46	$\pm$	0.18	 \\ 
55.5	 & 	462.3	 & 	0.32	$\pm$	0.04	 & 	1.81	$\pm$	0.24	 & 	52.84	$\pm$	0.18	 & 	41.09	$\pm$	0.18	 & 	45.47	$\pm$	0.18	 & 	33.72	$\pm$	0.18	 \\ 
64.1	 & 	534.0	 & 	0.45	$\pm$	0.06	 & 	2.91	$\pm$	0.39	 & 	53.11	$\pm$	0.18	 & 	41.42	$\pm$	0.18	 & 	45.68	$\pm$	0.18	 & 	33.99	$\pm$	0.18	 \\ 
72.6	 & 	604.8	 & 	0.41	$\pm$	0.06	 & 	3.01	$\pm$	0.43	 & 	53.18	$\pm$	0.18	 & 	41.54	$\pm$	0.18	 & 	45.70	$\pm$	0.18	 & 	34.06	$\pm$	0.18	 \\ 
81.2	 & 	676.4	 & 	0.33	$\pm$	0.05	 & 	2.72	$\pm$	0.45	 & 	53.18	$\pm$	0.16	 & 	41.59	$\pm$	0.16	 & 	45.65	$\pm$	0.16	 & 	34.06	$\pm$	0.16	 \\ 
89.7	 & 	747.2	 & 	0.27	$\pm$	0.05	 & 	2.44	$\pm$	0.42	 & 	53.18	$\pm$	0.15	 & 	41.63	$\pm$	0.15	 & 	45.60	$\pm$	0.15	 & 	34.06	$\pm$	0.15	 \\ 
98.3	 & 	793.5	 & 	0.19	$\pm$	0.04	 & 	1.80	$\pm$	0.41	 & 	53.07	$\pm$	0.10	 & 	41.55	$\pm$	0.10	 & 	45.47	$\pm$	0.10	 & 	33.95	$\pm$	0.10	 \\ 
106.8	 & 	771.1	 & 	0.10	$\pm$	0.04	 & 	0.89	$\pm$	0.33	 & 	52.75	$\pm$	0.03	 & 	41.22	$\pm$	0.03	 & 	45.17	$\pm$	0.03	 & 	33.64	$\pm$	0.03	 \\ 
115.3	 & 	748.8	 & 	0.08	$\pm$	0.04	 & 	0.69	$\pm$	0.32	 & 	52.63	$\pm$	0.13	 & 	41.09	$\pm$	0.13	 & 	45.06	$\pm$	0.13	 & 	33.51	$\pm$	0.13	 \\ 
123.9	 & 	726.1	 & 	0.09	$\pm$	0.04	 & 	0.77	$\pm$	0.35	 & 	52.66	$\pm$	0.12	 & 	41.11	$\pm$	0.12	 & 	45.10	$\pm$	0.12	 & 	33.55	$\pm$	0.12	 \\ 
132.4	 & 	703.8	 & 	0.07	$\pm$	0.04	 & 	0.57	$\pm$	0.38	 & 	52.52	$\pm$	0.39	 & 	40.95	$\pm$	0.39	 & 	44.98	$\pm$	0.39	 & 	33.41	$\pm$	0.39	 \\ \hline
{\bf NGC~1068} \\ \hline
14.4	 & 	1142.7	 & 	0.61	$\pm$	0.08	 & 	4.92	$\pm$	0.71	 & 	53.90	$\pm$	0.18	 & 	42.31	$\pm$	0.17	 & 	46.14	$\pm$	0.18	 & 	34.55	$\pm$	0.17	 \\ 
28.8	 & 	648.4	 & 	0.54	$\pm$	0.04	 & 	2.50	$\pm$	0.16	 & 	53.36	$\pm$	0.24	 & 	41.52	$\pm$	0.24	 & 	45.85	$\pm$	0.24	 & 	34.01	$\pm$	0.24	 \\ 
57.6	 & 	1029.9	 & 	1.24	$\pm$	0.09	 & 	9.04	$\pm$	0.65	 & 	54.12	$\pm$	0.23	 & 	42.48	$\pm$	0.24	 & 	46.40	$\pm$	0.23	 & 	34.77	$\pm$	0.24	 \\ 
72.0	 & 	1018.1	 & 	0.93	$\pm$	0.17	 & 	6.75	$\pm$	1.13	 & 	53.98	$\pm$	0.15	 & 	42.34	$\pm$	0.15	 & 	46.28	$\pm$	0.15	 & 	34.64	$\pm$	0.15	 \\ 
100.8	 & 	1509.9	 & 	0.70	$\pm$	0.05	 & 	7.53	$\pm$	0.48	 & 	54.20	$\pm$	0.24	 & 	42.73	$\pm$	0.25	 & 	46.32	$\pm$	0.24	 & 	34.86	$\pm$	0.25	 \\ 
129.6	 & 	915.7	 & 	0.23	$\pm$	0.06	 & 	1.48	$\pm$	0.43	 & 	53.28	$\pm$	0.08	 & 	41.59	$\pm$	0.04	 & 	45.62	$\pm$	0.08	 & 	33.93	$\pm$	0.04	 \\ \hline
{\bf Mrk~3} \\ \hline
0.0	 & 	6.7	 & 	0.05	$\pm$	0.01	 & 	0.01	$\pm$	0.01	 & 	48.32	$\pm$	0.10	 & 	33.84	$\pm$	0.06	 & 	42.80	$\pm$	0.10	 & 	28.32	$\pm$	0.06	 \\ 
65.1	 & 	1299.3	 & 	1.79	$\pm$	0.43	 & 	3.65	$\pm$	0.89	 & 	54.48	$\pm$	0.09	 & 	42.29	$\pm$	0.08	 & 	46.66	$\pm$	0.09	 & 	34.47	$\pm$	0.08	 \\ 
123.8	 & 	742.8	 & 	2.78	$\pm$	0.20	 & 	3.24	$\pm$	0.16	 & 	54.18	$\pm$	0.24	 & 	41.75	$\pm$	0.26	 & 	46.61	$\pm$	0.24	 & 	34.18	$\pm$	0.26	 \\ 
136.8	 & 	201.0	 & 	9.08	$\pm$	3.65	 & 	2.86	$\pm$	1.15	 & 	53.56	$\pm$	0.07	 & 	40.56	$\pm$	0.07	 & 	46.56	$\pm$	0.07	 & 	33.56	$\pm$	0.07	 \\ 
188.9	 & 	416.5	 & 	11.92	$\pm$	0.52	 & 	7.79	$\pm$	0.34	 & 	54.31	$\pm$	0.26	 & 	41.63	$\pm$	0.26	 & 	46.99	$\pm$	0.26	 & 	34.31	$\pm$	0.26	 \\ 
260.6	 & 	9.8	 & 	14.61	$\pm$	1.88	 & 	0.22	$\pm$	0.03	 & 	51.14	$\pm$	0.19	 & 	36.83	$\pm$	0.19	 & 	45.45	$\pm$	0.19	 & 	31.14	$\pm$	0.19	 \\ \hline
{\bf Mrk~573} \\ \hline
19.2	 & 	106.7	 & 	0.06	$\pm$	0.01	 & 	0.02	$\pm$	0.01	 & 	50.82	$\pm$	0.18	 & 	37.78	$\pm$	0.13	 & 	44.10	$\pm$	0.18	 & 	31.06	$\pm$	0.13	 \\ 
57.5	 & 	342.0	 & 	0.31	$\pm$	0.04	 & 	0.28	$\pm$	0.05	 & 	52.56	$\pm$	0.19	 & 	40.02	$\pm$	0.14	 & 	45.32	$\pm$	0.19	 & 	32.78	$\pm$	0.14	 \\ 
95.8	 & 	580.7	 & 	0.70	$\pm$	0.09	 & 	1.08	$\pm$	0.19	 & 	53.37	$\pm$	0.19	 & 	41.06	$\pm$	0.14	 & 	45.90	$\pm$	0.19	 & 	33.60	$\pm$	0.14	 \\ 
134.1	 & 	667.8	 & 	0.89	$\pm$	0.11	 & 	1.58	$\pm$	0.28	 & 	53.59	$\pm$	0.19	 & 	41.35	$\pm$	0.14	 & 	46.07	$\pm$	0.19	 & 	33.82	$\pm$	0.14	 \\ 
172.4	 & 	689.8	 & 	1.07	$\pm$	0.14	 & 	1.97	$\pm$	0.35	 & 	53.70	$\pm$	0.19	 & 	41.47	$\pm$	0.14	 & 	46.17	$\pm$	0.19	 & 	33.93	$\pm$	0.14	 \\ 
210.7	 & 	781.3	 & 	1.61	$\pm$	0.20	 & 	3.35	$\pm$	0.60	 & 	53.99	$\pm$	0.19	 & 	41.81	$\pm$	0.14	 & 	46.40	$\pm$	0.19	 & 	34.22	$\pm$	0.14	 \\ 
249.1	 & 	744.7	 & 	1.43	$\pm$	0.18	 & 	2.84	$\pm$	0.51	 & 	53.90	$\pm$	0.19	 & 	41.70	$\pm$	0.14	 & 	46.33	$\pm$	0.19	 & 	34.13	$\pm$	0.14	 \\ 
287.4	 & 	637.7	 & 	1.61	$\pm$	0.20	 & 	2.74	$\pm$	0.49	 & 	53.81	$\pm$	0.19	 & 	41.54	$\pm$	0.14	 & 	46.31	$\pm$	0.19	 & 	34.04	$\pm$	0.14	 \\ 
325.7	 & 	463.0	 & 	1.35	$\pm$	0.17	 & 	1.67	$\pm$	0.30	 & 	53.46	$\pm$	0.19	 & 	41.05	$\pm$	0.14	 & 	46.09	$\pm$	0.19	 & 	33.69	$\pm$	0.14	 \\ 
364.0	 & 	355.1	 & 	1.48	$\pm$	0.19	 & 	1.41	$\pm$	0.25	 & 	53.27	$\pm$	0.19	 & 	40.75	$\pm$	0.14	 & 	46.02	$\pm$	0.19	 & 	33.50	$\pm$	0.14	 \\ 
402.3	 & 	267.3	 & 	1.12	$\pm$	0.14	 & 	0.80	$\pm$	0.14	 & 	52.90	$\pm$	0.19	 & 	40.26	$\pm$	0.14	 & 	45.78	$\pm$	0.19	 & 	33.13	$\pm$	0.14	 \\ 
440.6	 & 	224.0	 & 	1.13	$\pm$	0.14	 & 	0.67	$\pm$	0.12	 & 	52.75	$\pm$	0.19	 & 	40.03	$\pm$	0.14	 & 	45.70	$\pm$	0.19	 & 	32.98	$\pm$	0.14	 \\ 
478.9	 & 	204.8	 & 	1.17	$\pm$	0.15	 & 	0.64	$\pm$	0.11	 & 	52.69	$\pm$	0.19	 & 	39.93	$\pm$	0.14	 & 	45.68	$\pm$	0.19	 & 	32.92	$\pm$	0.14	 \\ 
517.3	 & 	86.3	 & 	1.95	$\pm$	0.25	 & 	0.45	$\pm$	0.08	 & 	52.16	$\pm$	0.19	 & 	39.02	$\pm$	0.14	 & 	45.52	$\pm$	0.19	 & 	32.39	$\pm$	0.14	 \\ 
555.6	 & 	28.0	 & 	2.75	$\pm$	0.35	 & 	0.21	$\pm$	0.04	 & 	51.33	$\pm$	0.19	 & 	37.71	$\pm$	0.14	 & 	45.19	$\pm$	0.19	 & 	31.56	$\pm$	0.14	 \\ 
593.9	 & 	13.6	 & 	2.90	$\pm$	0.37	 & 	0.11	$\pm$	0.02	 & 	50.72	$\pm$	0.19	 & 	36.78	$\pm$	0.14	 & 	44.89	$\pm$	0.19	 & 	30.95	$\pm$	0.14	 \\ \hline
{\bf Mrk~78} \\ \hline
127.4	 & 	100.0	 & 	0.64	$\pm$	0.22	 & 	0.05	$\pm$	0.02	 & 	51.80	$\pm$	0.02	 & 	38.20	$\pm$	0.12	 & 	45.10	$\pm$	0.02	 & 	31.50	$\pm$	0.12	 \\ 
254.9	 & 	200.0	 & 	2.82	$\pm$	0.99	 & 	0.45	$\pm$	0.22	 & 	53.05	$\pm$	0.02	 & 	39.75	$\pm$	0.17	 & 	46.05	$\pm$	0.02	 & 	32.75	$\pm$	0.17	 \\ 
509.7	 & 	400.0	 & 	5.25	$\pm$	1.84	 & 	1.66	$\pm$	0.82	 & 	53.92	$\pm$	0.02	 & 	40.92	$\pm$	0.17	 & 	46.62	$\pm$	0.02	 & 	33.62	$\pm$	0.17	 \\ 
764.6	 & 	600.0	 & 	5.06	$\pm$	1.77	 & 	2.40	$\pm$	1.19	 & 	54.26	$\pm$	0.02	 & 	41.43	$\pm$	0.17	 & 	46.78	$\pm$	0.02	 & 	33.96	$\pm$	0.17	 \\ 
1019.5	 & 	800.0	 & 	10.38	$\pm$	3.63	 & 	6.56	$\pm$	3.25	 & 	54.82	$\pm$	0.02	 & 	42.12	$\pm$	0.17	 & 	47.22	$\pm$	0.02	 & 	34.52	$\pm$	0.17	 \\ 
1274.3	 & 	1000.0	 & 	11.80	$\pm$	4.12	 & 	9.32	$\pm$	4.60	 & 	55.07	$\pm$	0.02	 & 	42.47	$\pm$	0.17	 & 	47.37	$\pm$	0.02	 & 	34.77	$\pm$	0.17	 \\ 
1529.2	 & 	1200.0	 & 	6.97	$\pm$	2.44	 & 	6.61	$\pm$	3.27	 & 	55.00	$\pm$	0.02	 & 	42.48	$\pm$	0.17	 & 	47.22	$\pm$	0.02	 & 	34.70	$\pm$	0.17	 \\ 
1784.0	 & 	971.4	 & 	6.12	$\pm$	2.14	 & 	4.70	$\pm$	2.32	 & 	54.76	$\pm$	0.02	 & 	42.15	$\pm$	0.17	 & 	47.07	$\pm$	0.02	 & 	34.46	$\pm$	0.17	 \\ 
2038.9	 & 	742.9	 & 	7.39	$\pm$	2.59	 & 	4.34	$\pm$	2.15	 & 	54.61	$\pm$	0.02	 & 	41.88	$\pm$	0.17	 & 	47.04	$\pm$	0.02	 & 	34.31	$\pm$	0.17	 \\ 
2293.8	 & 	514.3	 & 	15.35	$\pm$	5.36	 & 	6.24	$\pm$	3.08	 & 	54.61	$\pm$	0.02	 & 	41.72	$\pm$	0.17	 & 	47.20	$\pm$	0.02	 & 	34.31	$\pm$	0.17	 \\ 
2548.6	 & 	384.0	 & 	7.17	$\pm$	2.51	 & 	2.18	$\pm$	1.08	 & 	54.02	$\pm$	0.02	 & 	41.01	$\pm$	0.17	 & 	46.74	$\pm$	0.02	 & 	33.72	$\pm$	0.17	 \\ 
2803.5	 & 	352.0	 & 	4.84	$\pm$	1.69	 & 	1.35	$\pm$	0.66	 & 	53.78	$\pm$	0.02	 & 	40.72	$\pm$	0.16	 & 	46.53	$\pm$	0.02	 & 	33.48	$\pm$	0.16	 \\ 
3058.4	 & 	320.0	 & 	11.05	$\pm$	3.86	 & 	2.80	$\pm$	1.38	 & 	54.05	$\pm$	0.02	 & 	40.96	$\pm$	0.17	 & 	46.85	$\pm$	0.02	 & 	33.75	$\pm$	0.17	 \\ 
3313.2	 & 	288.0	 & 	17.88	$\pm$	6.25	 & 	4.07	$\pm$	2.02	 & 	54.17	$\pm$	0.02	 & 	41.03	$\pm$	0.17	 & 	47.01	$\pm$	0.02	 & 	33.87	$\pm$	0.17	 \\ 
3568.1	 & 	256.0	 & 	8.24	$\pm$	2.88	 & 	1.67	$\pm$	0.83	 & 	53.73	$\pm$	0.02	 & 	40.54	$\pm$	0.17	 & 	46.62	$\pm$	0.02	 & 	33.43	$\pm$	0.17	 \\ 
3823.0	 & 	224.0	 & 	2.83	$\pm$	0.99	 & 	0.50	$\pm$	0.25	 & 	53.15	$\pm$	0.01	 & 	39.90	$\pm$	0.17	 & 	46.10	$\pm$	0.01	 & 	32.85	$\pm$	0.17	 \\  \hline
{\bf Mrk~34} \\ \hline
67.5	 & 	191.8	 & 	0.84	$\pm$	0.13	 & 	0.12	$\pm$	0.03	 & 	52.49	$\pm$	0.17	 & 	39.15	$\pm$	0.11	 & 	45.50	$\pm$	0.17	 & 	32.17	$\pm$	0.11	 \\ 
202.4	 & 	2347.9	 & 	2.26	$\pm$	0.35	 & 	4.02	$\pm$	0.88	 & 	55.09	$\pm$	0.17	 & 	42.84	$\pm$	0.11	 & 	47.02	$\pm$	0.17	 & 	34.77	$\pm$	0.11	 \\ 
337.4	 & 	1976.6	 & 	4.17	$\pm$	0.64	 & 	6.24	$\pm$	1.36	 & 	55.21	$\pm$	0.17	 & 	42.89	$\pm$	0.11	 & 	47.21	$\pm$	0.17	 & 	34.89	$\pm$	0.11	 \\ 
472.4	 & 	1789.2	 & 	9.18	$\pm$	1.42	 & 	12.45	$\pm$	2.72	 & 	55.47	$\pm$	0.17	 & 	43.10	$\pm$	0.11	 & 	47.51	$\pm$	0.17	 & 	35.15	$\pm$	0.11	 \\ 
607.3	 & 	1358.2	 & 	2.94	$\pm$	0.45	 & 	3.03	$\pm$	0.66	 & 	54.73	$\pm$	0.17	 & 	42.25	$\pm$	0.11	 & 	46.90	$\pm$	0.17	 & 	34.41	$\pm$	0.11	 \\ 
742.3	 & 	1108.4	 & 	2.51	$\pm$	0.39	 & 	2.11	$\pm$	0.46	 & 	54.49	$\pm$	0.17	 & 	41.91	$\pm$	0.11	 & 	46.74	$\pm$	0.17	 & 	34.17	$\pm$	0.11	 \\ 
877.2	 & 	46.0	 & 	1.15	$\pm$	0.18	 & 	0.04	$\pm$	0.01	 & 	51.38	$\pm$	0.17	 & 	37.43	$\pm$	0.10	 & 	45.02	$\pm$	0.17	 & 	31.06	$\pm$	0.10	 \\ 
1012.2	 & 	113.0	 & 	4.06	$\pm$	0.63	 & 	0.35	$\pm$	0.08	 & 	52.71	$\pm$	0.17	 & 	39.15	$\pm$	0.11	 & 	45.96	$\pm$	0.17	 & 	32.39	$\pm$	0.11	 \\ 
1147.1	 & 	368.6	 & 	13.20	$\pm$	2.04	 & 	3.69	$\pm$	0.81	 & 	54.25	$\pm$	0.17	 & 	41.20	$\pm$	0.11	 & 	46.99	$\pm$	0.17	 & 	33.93	$\pm$	0.11	 \\ 
1282.1	 & 	533.5	 & 	13.53	$\pm$	2.09	 & 	5.47	$\pm$	1.20	 & 	54.58	$\pm$	0.17	 & 	41.69	$\pm$	0.11	 & 	47.16	$\pm$	0.17	 & 	34.26	$\pm$	0.11	 \\ 
1417.0	 & 	350.7	 & 	23.01	$\pm$	3.55	 & 	6.11	$\pm$	1.34	 & 	54.45	$\pm$	0.17	 & 	41.37	$\pm$	0.11	 & 	47.21	$\pm$	0.17	 & 	34.13	$\pm$	0.11	 \\ 
1552.0	 & 	407.1	 & 	18.39	$\pm$	2.84	 & 	5.67	$\pm$	1.24	 & 	54.48	$\pm$	0.17	 & 	41.47	$\pm$	0.11	 & 	47.17	$\pm$	0.17	 & 	34.16	$\pm$	0.11	 \\ 
1687.0	 & 	440.4	 & 	16.17	$\pm$	2.50	 & 	5.40	$\pm$	1.18	 & 	54.49	$\pm$	0.17	 & 	41.52	$\pm$	0.11	 & 	47.15	$\pm$	0.17	 & 	34.18	$\pm$	0.11	 \\ 
1821.9	 & 	517.0	 & 	24.38	$\pm$	3.77	 & 	9.55	$\pm$	2.09	 & 	54.81	$\pm$	0.17	 & 	41.91	$\pm$	0.11	 & 	47.40	$\pm$	0.17	 & 	34.49	$\pm$	0.11	 \\ 
1956.9	 & 	538.7	 & 	19.68	$\pm$	3.04	 & 	8.03	$\pm$	1.76	 & 	54.75	$\pm$	0.17	 & 	41.87	$\pm$	0.11	 & 	47.32	$\pm$	0.17	 & 	34.44	$\pm$	0.11	 \\ 
\enddata
\vspace{-0.75em}
\tablecomments{Numerical results for the mass and energetic quantities as functions of radial distance that are shown in Figures~\ref{mdots} -- \ref{results}. The columns are (1) deprojected distance from the nucleus, (2) mass-weighted mean velocity of the outflowing kinematic components, (3) logarithmic gas mass, (4) mass outflow rates, (5) kinetic energies, (6) kinetic energy outflow rates, (7) momenta, and (8) momenta flow rates. The value at each distance is the quantity contained within the annuli of width $\delta r$ as shown in Figure~\ref{imaging}. Values for the last two rows of Mrk~78 are lower limits because the \hst image does not fully cover the NLR emission.}
\end{deluxetable*}

\clearpage

\end{document}